\documentclass[crcready]{iosart2x}

\usepackage{dcolumn}

\newcolumntype{d}[1]{D{.}{.}{#1}}

\usepackage[ruled,vlined]{algorithm2e}
\usepackage{tikz}
\usepackage{caption}
\usepackage{listings}
\usepackage{tablefootnote}
\lstdefinelanguage{SAS}{morestring=[b]"}
\usepackage[bottom]{footmisc}
\usepackage{lscape}
\usepackage{latexsym}
\usepackage{graphicx}
\usepackage{multirow}
\usepackage{stmaryrd}
\usepackage{url}
\usepackage{breakurl}
\usepackage{float}
\usepackage{wrapfig}
\usepackage{verbatim}
\usepackage{stackengine}
\usepackage[fleqn]{amsmath}
\allowdisplaybreaks[4]
\usepackage{amsfonts}

\let\endproof\relax
\usepackage{amsthm}
\theoremstyle{definition}
\newtheorem{defn}{Definition}
\newtheorem{xmpl}{Example}
\newtheorem {rem}{Remark}
\usepackage[nodisplayskipstretch]{setspace}
\usepackage{array}
\newcolumntype{h}{>{\setbox0=\hbox\bgroup}c<{\egroup}@{}}
\makeatletter
\newcommand{\thickhline}{%
    \noalign {\ifnum 0=`}\fi \hrule height 1pt
    \futurelet \reserved@a \@xhline
}
\newcolumntype{"}{@{\hskip\tabcolsep\vrule width 1pt\hskip\tabcolsep}}
\makeatother

\usepackage{lscape}

\def\ojoin{\setbox0=\hbox{$\bowtie$}%
  \rule[-.02ex]{.25em}{.4pt}\llap{\rule[\ht0]{.25em}{.4pt}}}
\def\leftouterjoin{\mathbin{\ojoin\mkern-5.8mu\bowtie}}
\def\rightouterjoin{\mathbin{\bowtie\mkern-5.8mu\ojoin}}

\usepackage{framed}
\usepackage{adjustbox}
\usepackage{lipsum}
\usepackage{subcaption}
\usepackage{mathtools}
\usepackage{cuted}
\usepackage[colorinlistoftodos]{todonotes}
\usepackage{enumitem}

\newcommand{\s}{\textit{SETL\textsubscript{PROG}}}
\newcommand{\scon}{\textit{SETL\textsubscript{CONSTRUCT}}}
\newcommand{\sauto}{\textit{SETL\textsubscript{AUTO}}}

\newcommand{\ql}{SPARQL}
\newcommand{\ab}{ABox}
\newcommand{\tb}{TBox}
\newcommand{\dab}{$DAB$}
\newcommand{\sd}{SDW}

\newcommand{\mv}{S2TMAP}

\newcommand{\dad}{Danish Agricultural dataset}
\newcommand{\dbd}{Danish Business dataset}

\newcommand{\dl}{Definition Layer}
\newcommand{\exl}{Execution Layer}
\newcommand{\el}{ETL Layer}

\newcommand{\sC}{\texttt{owl:subClassOf}}
\newcommand{\eC}{\texttt{owl:equivalentClass}}
\newcommand{\eP}{\texttt{owl:equivalentProperty}}
\newcommand{\sP}{\texttt{owl:subPropertyOf}}
\newcommand{\inv}{\texttt{owl:inverseOf}}
\newcommand{\TP}{\texttt{owl:TransitiveProperty}}
\newcommand{\SP}{\texttt{owl:SymmetricProperty}}
\newcommand{\sA}{\texttt{owl:sameAs}}
\newcommand{\FP}{\texttt{owl:FunctionalProperty}}
\newcommand{\IFP}{\texttt{owl:InverseFunctionalProperty}}
\newcommand{\dom}{\texttt{rdfs:domain}}
\newcommand{\rng}{\texttt{rdfs:range}}
\def\louterjoin{\mathrel{\sqsupset\mkern-2.5mu\bowtie}}
\usepackage{hyperref}

\begin{document}
\makeatletter
\def\put@numberlines@box{}

\begin{frontmatter} 

%
\title{High-Level ETL for Semantic Data Warehouses---Full Version\thanks{This journal paper is submitted to Semantic Web – Interoperability, Usability, Applicability an IOS Press Journal }}
\runningtitle{High-Level ETL for Semantic Data Warehouses}

\author[A,B]{\fnms{Rudra Pratap} \snm{Deb Nath}\thanks{Corresponding author. E-mail: rudra@cs.aau.dk.}}, 
\author[B]{\fnms{Oscar} \snm{Romero}},
\author[A]{\fnms{Torben Bach} \snm{Pedersen}} and
\author[A]{\fnms{Katja} \snm{Hose}}
\runningauthor{R. Deb Nath et al.}
\address[A]{Department of Computer Science, Aalborg University, Denmark\\ 
E-mail: \{rudra,tbp,khose\}@cs.aau.dk}	
\address[B]{Department of Service and Information System Engineering, Universitat Politècnica de Catalunya, Spain\\
E-mail:  \{rudra,oromero\}@essi.upc.edu}
\vspace{-0.2in}	

\begin{abstract}
The popularity of the Semantic Web (SW) encourages organizations to organize and publish semantic data using the RDF model. This growth poses new requirements to Business Intelligence (BI) technologies to enable On-Line Analytical Processing (OLAP)-like analysis over semantic data. 
The incorporation of semantic data into a Data Warehouse (DW) is not supported by the traditional Extract-Transform-Load (ETL) tools because they do not consider semantic issues in the integration process. 
In this paper, we propose a layer-based integration process and a set of high-level RDF-based ETL constructs required to define, map, extract, process, transform, integrate, update, and load  (multidimensional) semantic data. Different to other ETL tools, we automate the ETL data flows by creating metadata at the schema level. Therefore, it relieves ETL developers from the burden of manual mapping at the ETL operation level. We create a prototype, named Semantic ETL Construct (\scon), based on the innovative ETL constructs proposed here. To evaluate \scon, we create a multidimensional semantic DW by integrating a Danish Business dataset and an EU Subsidy dataset using  it and compare it with the previous programmable framework \s\ in terms of productivity, development time and performance. The evaluation shows that 1) \scon\ uses 92\% fewer Number of Typed Characters (NOTC) than \s, and \sauto\ (the extension of \scon\ for generating ETL execution flow automatically) further reduces the Number of Used Concepts (NOUC) by another 25\%; 2) using \scon, the development time is almost cut in half compared to \s, and is cut by another 27\% using \sauto; 3) \scon\ is scalable and has similar performance compared to \s.

\end{abstract}

\begin{keyword}
\kwd{RDF}
\kwd{Layer-based Semantic Data Integration}
\kwd{ETL constructs}
\kwd{Semantic ETL}
\kwd{Semantic Data Warehosue}
\end{keyword}
\end{frontmatter}

\section{Introduction}
\label{sec:introduction}

Semantic Web technologies enable adding a semantic layer over the data; thus, the data can be processed and effectively retrieved by both humans and machines. The Linked Data (LD) principles are the set of standard rules to publish and connect data using semantic links~\cite{hartig2019linked}. With the growing popularity of the SW and LD, more and more organizations natively manage data using SW standards, such as Resource Description Framework (RDF), RDF-Schema (RDFs), the Web Ontology Language (OWL), etc.~\cite{deb2020setlbi}. Moreover, one can easily convert data given in another format (database, XML, JSON, etc.) into RDF format using an RDF wrapper. As a result, a lot of semantic datasets are now available in different data portals, such as DataHub\footnote{\url{https://datahub.io/}}, Linked Open Data Cloud~\footnote{\url{https://lod-cloud.net/}} (LOD), etc. Most SW data provided by international and governmental organizations include facts and figures, which give rise to new requirements for Business Intelligence (BI) tools to enable analyses in the style of Online Analytical Processing (OLAP) over those semantic data~\cite{kalampokis2015challenges}. 

OLAP is a well-recognized technology to support decision making by analyzing data integrated from multiple sources. The integrated data are stored in a Data Warehouse (DW), typically structured following the Multidimensional Model (MD) that represents data in terms of facts and dimensions to enable OLAP queries. The integration process for extracting data from different sources, translating it according to the underlying semantics of the DW, and loading it into the DW is known as Extract-Transform-Load (ETL). One way to enable OLAP analysis over semantic data is by extracting those data and translating them according to the DW's format using a traditional ETL process. \cite{nebot2012building} outlines such a type of semi-automatic method to integrate semantic data into a traditional RDBMS-centric MD DW. However, the process does not maintain all the semantics of data as they are conveying in the semantic sources; hence the integrated data no more follow the SW data principles defined in~\cite{heath2011linked}. The semantics of the data in a semantic data source is defined by 1) using Internationalized Resource Identifier (IRIs) to uniquely identify resources globally, 2) providing common terminology, 3) semantically linking with published information, and 4) providing further knowledge (e.g., logical axioms) to allow reasoning~\cite{berlanga2011semantic}. 

Therefore, considering semantic issues in the integration process should be emphasized. Moreover, initiatives such as Open Government Data\footnote{\url{https://opengovdata.org/}} encourage organizations to publish their data using standards and non-proprietary formats~\cite{varga2016dimensional}. The integration of semantic data into a DW raises the challenges of schema derivation, semantic heterogeneity, semantic annotation, linking as well as the schema and data management system over traditional DW technologies and ETL tools. The main drawback of a state-of-the-art Relational Database Management System (RDBMS)-based DW is that it is strictly schema dependent and less flexible to evolving business requirements. To cover new business requirements, every step of the development cycle needs to be updated to cope with the new requirements. This update process is time-consuming as well as costly and is sometimes not adjustable with the current setup of the DW; hence it introduces the need for a novel approach. The limitations of traditional ETL tools to process semantic data sources are: (1) they do not fully support semantic-aware data, (2) they are entirely schema dependent (i.e., cannot handle data expressed without pre-defined schema), (3) they do not focus on meaningful semantic relationships to integrate data from disparate sources, and (4) they neither support to capture the semantics of data nor support to derive new information by active inference and reasoning on the data. 

Semantic Web technologies address the problems described above, as they allow adding semantics at both data and schema level in the integration process and publish data in RDF using the LD principles. On the SW, the RDF model is used to manage and exchange data, and RDFS and OWL are used in combination with the RDF data model to define constraints that data must meet. Moreover, QB~\cite{ciferri2013cube} and QB4OLAP~\cite{etcheverry2015modeling} vocabularies can be used to define data with MD semantics. \cite{nath2017setl} refers to an MD DW that is semantically annotated both at the schema and data level as a Semantic DW (SDW). An SDW is based on the assumption that the schema can evolve and be extended without affecting the existing structure. Hence, it overcomes the problems triggered by the evolution of an RDBMS-based data warehousing system. In  \cite{nath2017setl}, we proposed SETL (throughout this present paper, we call it \s), a programmable semantic ETL framework that semantically integrates both semantic and non-semantic data sources. In \s, an ETL developer has to create hand-code specific modules to deal with semantic data. Thus, there is a lack of a well-defined set of basic ETL constructs that allow developers having a higher level of abstraction and more control in creating their ETL process. In this paper, we propose a strong foundation for an RDF-based semantic integration process and a set of high-level ETL constructs that allows defining, mapping, processing, and integrating semantic data. The unique contributions of this paper are: 

\begin{enumerate}
\item We structure the integration process into two layers: \dl\ and \exl. Different to SETL or other ETL tools, here, we propose a new paradigm: the ETL flow transformations are characterized once and for all at the \dl\ instead of independently within each ETL operation (in the \exl). This is done by generating a mapping file that gives an overall view of the integration process. This mapping file is our primary metadata source, and it will be fed (by either the ETL developer or the automatic ETL execution flow generation process) to the ETL operations, orchestrated in the ETL flow (Execution Layer), to parametrize themselves automatically. Thus, we are unifying the creation of the required metadata to automate the ETL process in the Definition layer. We propose an OWL-based Source-To-target Mapping (S2TMAP) vocabulary to express the source-to-target mappings.
\item We provide a set of high-level ETL constructs for each layer. The \dl\ includes constructs for target schema\footnote{Here, we use the terms ``target" and ``MD SDW" interchangeably.} definition, source schema derivation, and source-to-target mappings generation. The \exl\ includes a set of high-level ETL operations for semantic data extraction, cleansing, joining, MD data creation, linking, inferencing, and for dimensional data update. 
\item We propose an approach to automate the ETL execution flows based on metadata generated in the \dl.
\item We create a prototype \scon, based on the innovative ETL constructs proposed here. \scon\ allows creating ETL flows by dragging, dropping, and connecting the ETL operations. In addition, it allows creating ETL data flows automatically (we call it \sauto). 
\item We perform a comprehensive experimental evaluation by producing an MD SDW that integrates an EU farm Subsidy dataset and a Danish Business dataset. The evaluation shows that \scon\ improves considerably over \textit{SET-} \textit{L\textsubscript{PROG}} in terms of productivity, development time, and performance. In summary: 1) \scon\ uses 92\% fewer NOTC than \s, and \sauto\ further reduces NOUC by another 25\%; 2) using\scon, the development time is almost cut in half compared to \s, and is cut by another 27\% using \sauto; 3) \scon\ is scalable and has similar performance compared to \s. 

\end{enumerate}
The remainder of the paper is organized as follows. We discuss the terminologies and the notations used throughout the paper in Section~\ref{sec:preliminarydefinitions}. Section~\ref{sec:usecase} explains the structure of the datasets we use as a use case. Section~\ref{sec:overview} gives the overview of an integration process. The description of the \dl\ and \exl\ constructs are given in Sections~\ref{sec:definitionLayer} and ~\ref{sec:executionLayer}, respectively. Section~\ref{sec:aeefg} presents the automatic ETL execution flow generation process. 
In Section~\ref{sec:evaluation}, we create an MD SDW for the use case using \scon\ and compare the process with \s\ using different metrics. The previous research related to our study is discussed in Section~\ref{sec:relatedwork}. Finally, we conclude and give pointers to future work in Section~\ref{sec:conclusion}. 


\section{Preliminary Definitions}
\label{sec:preliminarydefinitions}
In this section, we provide the definitions of the notions and terminologies used throughout the paper. 
\subsection{RDF Graph} An RDF graph is represented as a set of statements, called RDF triples. The three parts of a triple are subject, predicate, and object, respectively, and a triple represents a relationship between its subject and object described by its predicate. Each triple, in the RDF graph, is represented  as $subject\xrightarrow[]{predicate} object$, where subject and object of the triple are the nodes of the graph, and the label corresponds to the predicate of the triple.  Given $I, B,$ and $L$ are the sets of IRIs, blank nodes, and literals, and $(I\cap B \cap L) = \emptyset $, an RDF triple is $(s, p, o)$, where $s \in (I \cup B)$, $p \in I$ , and $ o \in (I \cup B \cup L)$. An RDF graph $G$ is a set of RDF triples, where $G\subseteq (I \cup B) \times I \times (I \cup B \cup L) $~\cite{harth2014linked}. 

\subsection{Semantic Data Source} We define a semantic data source as a Knowledge Base (KB) where data are semantically defined. A KB is composed of two components, \tb\ and \ab. The \tb\ introduces terminology, the vocabulary of a domain, and the \ab\ is the assertions of the \tb. 
  The \tb\ is formally defined as a 3-tuple: $TBox=(C, P, A^O)$, where $C$, $P$, and $A^O$ are the sets of concepts, properties, and terminological axioms, respectively~\cite{baader2003description}.  Generally, a concept\footnote{In this paper, we use the terms ``concept" and ``class" interchangeably.} provides a general framework for a group of instances that have similar properties. A property either relates the instances of concepts or associates the instances of a concept to literals. Terminological axioms are used to describe the domain's concepts, properties, and the relationships and constraints among them. 
 In this paper, we consider a KB as an RDF graph; therefore, the components of the KB are described by a set of RDF triples. Some standard languages such as Resource Description Framework Schema (RDFS) and Web Ontology Language (OWL) provide standard terms to define the formal semantics of a \tb. In RDFS, the core classes \texttt{rdfs:Class}, and \texttt{rdf:Property} are used to define the concepts and properties of a \tb; one can distinguish between instances and classes by using the \texttt{rdf:type} property, express concept and property taxonomies by using \texttt{rdfs:subClassOf} and \texttt{rdfs:subPropertyOf}, and specify the domain and range of properties by using the \texttt{rdfs:domain} and \texttt{rdfs:range} properties. Similarly, OWL uses \texttt{owl:Class} to define concepts and either \texttt{owl:Dat-\\aTypeProperty} or \texttt{owl:ObjectProperty} for properties. In addition to \texttt{rdfs:subClassOf}, it uses \texttt{owl:equivalentClass} and \texttt{owl:disjoi-\\ntWith} constructs for class axioms to give additional characteristics of classes. Property axioms define additional characteristics of properties. In addition to supporting RDFS constructs 
for property axioms, OWL provides \texttt{owl:equivalentProeprty} and \texttt{owl:inverseOf} to relate different properties, provides \texttt{owl:FunctionalProperty} and \texttt{owl:InverseFunctionalProperty} for imposing global cardinality constraints, and  supports \texttt{owl:-\\SymmetricProperty} and \texttt{owl:Transitivity-\\Property} for characterizing the relationship type of properties. As a KB can be defined by either language or both, we generalize the definition of $C$ and $P$ in a \tb\ T as $C(T) =\{c|\; type(c) \in \mathbb{P}(\{ \texttt{rdfs:Class},\texttt{owl:Class}\})\}$ and $P(T) =\{ p|\\ \; type(p) \in \mathbb{P}(\{\texttt{rdf:Property},\texttt{owl:ObjectProp}\\ \texttt{erty}, \texttt{owl:DatatypeProperty}\})\}$,  respectively, where $type(x)$ returns the set of classes of $x$, i.e., ($x$ \texttt{rdf:type} $?type(x)$)--- it returns the set of the objects of the triples whose subjects and  predicates are $x$ and \texttt{rdf:type}, respectively--- and $\mathbb{P}(s)$ is the power set of $s$.

\subsection{Semantic Data Warehouse} \label{sec:sdw} A semantic data warehouse (SDW) is a DW with the semantic annotations. We also considered it as a KB. Since the DW is represented with Multidimensional (MD) model for enabling On-Line Analytical Processing (OLAP) queries, the KB for an SDW needs to be defined with MD semantics. In the MD model, data are viewed in an n-dimensional space, usually known as a data cube, composed of facts (the cells of the cube) and dimensions (the axes of the cube). Therefore, it allows users to analyze data along several dimensions of interest. For example, a user can analyze sales of products according to time and store (dimensions). Facts are the interesting things or processes to be analyzed (e.g., sales of products) and the attributes of the fact are called measures (e.g., quantity, amount of sales), usually represented as numeric values. A dimension is organized into hierarchies, composed of several levels, which permit users to explore and aggregate measures at various levels of detail. For example, the \textit{location} hierarchy ($ municipality \rightarrow region \rightarrow state \rightarrow country$) of the \textit{store} dimension allows to aggregate the sales at various levels of detail. 

\begin{figure*}[hpt]
  \centering
  \includegraphics[width=1\linewidth]{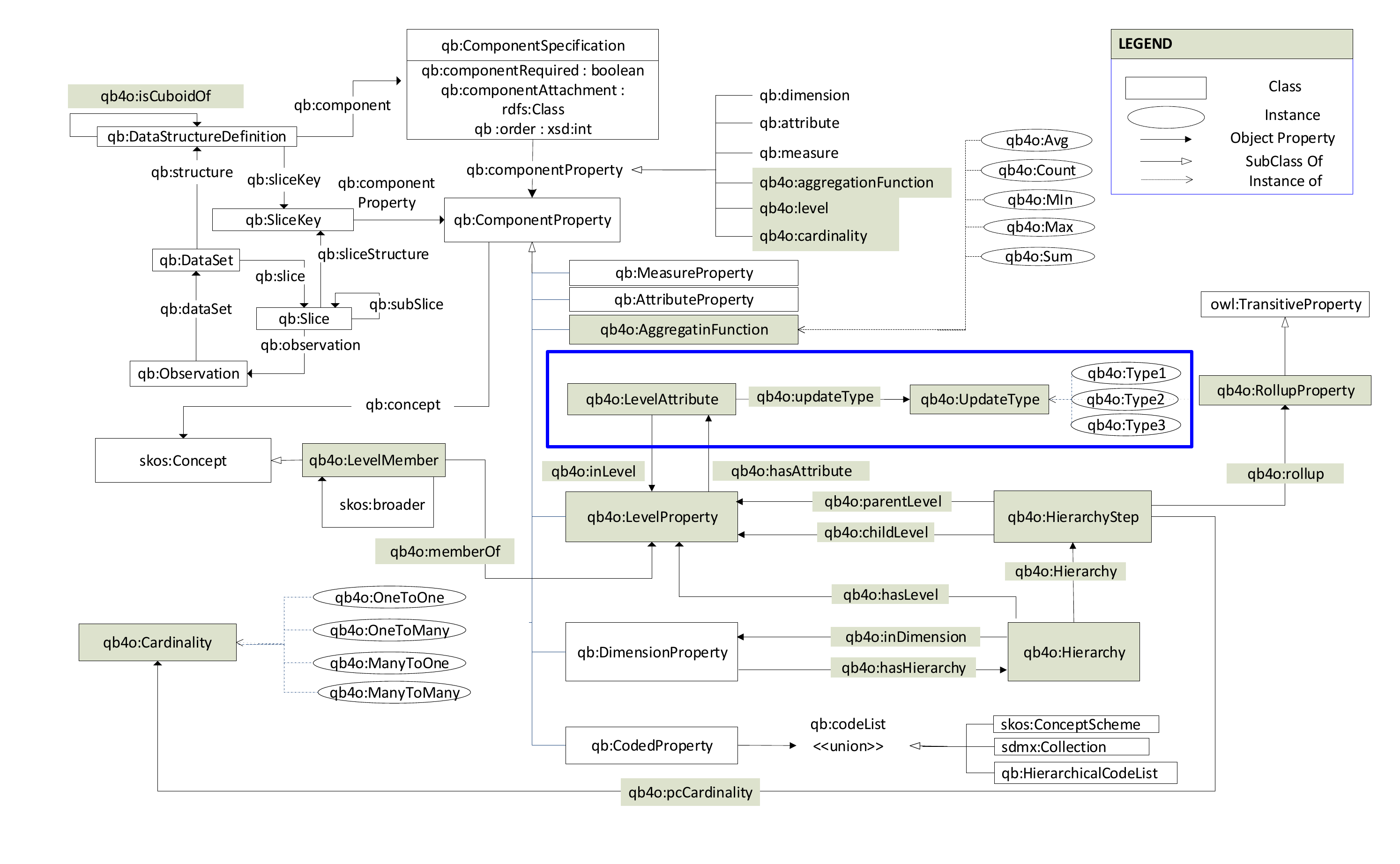}
  \caption{QB4OLAP vocabulary.}
  \label{fig:qb13}
\end{figure*}

We use the QB4OLAP vocabulary to describe the multidimensional semantics over a KB~\cite{etcheverry2015modeling}. QB4OLAP is used to annotate the \tb\  with MD components and is based on the RDF Data Cube (QB) which is the W3C standard to publish MD data on the Web~\cite{cyganiak2014rdf}. The QB is mostly used for analyzing statistical data and does not adequately support OLAP MD constructs.  Therefore, in this paper, we choose QB4OLAP. Figure~\ref{fig:qb13} depicts the ontology of QB4OLAP~\cite{varga2016dimensional}. The terms prefixed with ``qb:'' are from the original QB vocabulary, and QB4OLAP terms are prefixed with ``qb4o:'' and displayed with gray background.  Capitalized terms represent OWL classes, and non-capitalized terms represent OWL properties. Capitalized terms in italics represent classes with no instances. The blue-colored square in the figure represents our extension of QB4OLAP ontology. 

In QB4OLAP, the concept \texttt{qb:DataSet} is used to define a dataset of observations. The structure of the dataset is defined using the concept \texttt{qb:DataStruc-\\tureDefinition}. The structure can be a cube (if it is defined in terms of dimensions and measures) or a cuboid (if it is defined in terms of lower levels of the dimensions and measures). The property \texttt{qb4o:isCuboidOf} is used to relate a cuboid to its  corresponding cube. To define dimensions, levels and hierarchies, the concepts \texttt{qb4o:DimensionProp-\\erty}, \texttt{qb4o:LevelProperty}, and \texttt{qb4o:Hiera-\\rchy} are used. A dimension can have one or more hierarchies. The relationship between a dimension and its hierarchies are connected via the \texttt{qb4o:hasHiera-\\rchy} property or its inverse property \texttt{qb4o:inHier-\\archy}. Conceptually, a level may belong to different hierarchies; therefore, it may have one or more parent levels. Each parent and child pair has a cardinality constraint (e.g., 1-1, n-1, 1-n, and n-n.)~\cite{varga2016dimensional}. To allow this kind of complex nature, hierarchies in QB4OLAP are defined as a composition of pairs of levels, which are represented using the concept \texttt{qb4o:HierarchyStep}. Each hierarchy step (pair) is connected to its component levels using the properties \texttt{qb4o:parentLevel} and \texttt{qb4o:childLevel}. A rollup relationship between two levels are defined by creating a property which is an instance of the concept \texttt{qb4o:RollupProperty}; each hierarchy step is linked to a rollup relationship with the property \texttt{qb4o:rollup} and the cardinality constraint of that relationship is connected to the hierarchy step using the \texttt{qb4o:pcCardinality} property.  A hierarchy step is attached to the hierarchies it belongs to using the property \texttt{qb4o:inHierarchy} \cite{etcheverry2015modeling}. The concept \texttt{qb4o:LevelAttributes} is used to define attributes of levels. 
We extend this QB4OLAP ontology (the blue-colored box in the figure) to enable different types of dimension updates (Type 1, Type 2, and Type 3) to accommodate dimension update in an SDW, which are defined by Ralph Kimball in~\cite{kimball1996data}. To define the update-type of a level attribute in the TBox level, we introduce the \texttt{qb4o:UpdateType} class whose instances are \texttt{qb4o:Type1}, \texttt{qb4o:Type1} and \texttt{qb4o:Type3}. A level attribute is connected to its update-type by the property \texttt{qb4o:updateType}. The level attributes are linked to its corresponding levels using the property \texttt{qb4o:hasAttribute}. 
We extend the definition of $C$ and $P$ of a \tb, T for an SDW as  
\begin{multline}
C(T)=\{c|\; type(c) \in \mathbb{P}(\{\texttt{rdfs:Class},\texttt{owl:} \\\textit{Class}, \texttt{qb:DataStructureDefinition},\\ \texttt{qb:DataSet}, \texttt{qb:DimensionProperty},\\\texttt{qb4o:LevelProperty},\texttt{qb4o:Hierarchy},\\\texttt{qb4o:HierarchyStep}\})\}
\label{eq:concept}
\end{multline}
\vspace{-0.2 cm}
\begin{multline}
P(T)=\{p|\; type(p) \in \mathbb{P}( \{\texttt{rdf:Property},\\\texttt{owl:ObjectProperty}, \texttt{owl:Datatype}\\\texttt{Property}, \texttt{qb4o:LevelAttribute},\\\texttt{qb:MeassureProperty},\texttt{qb4o:Rollup}\\\texttt{Property}\})\}
\label{eq:property}
\end{multline}

\section{A Use Case}
\label{sec:usecase}

\begin{figure*}[t]
\centering
\includegraphics[scale=0.7]{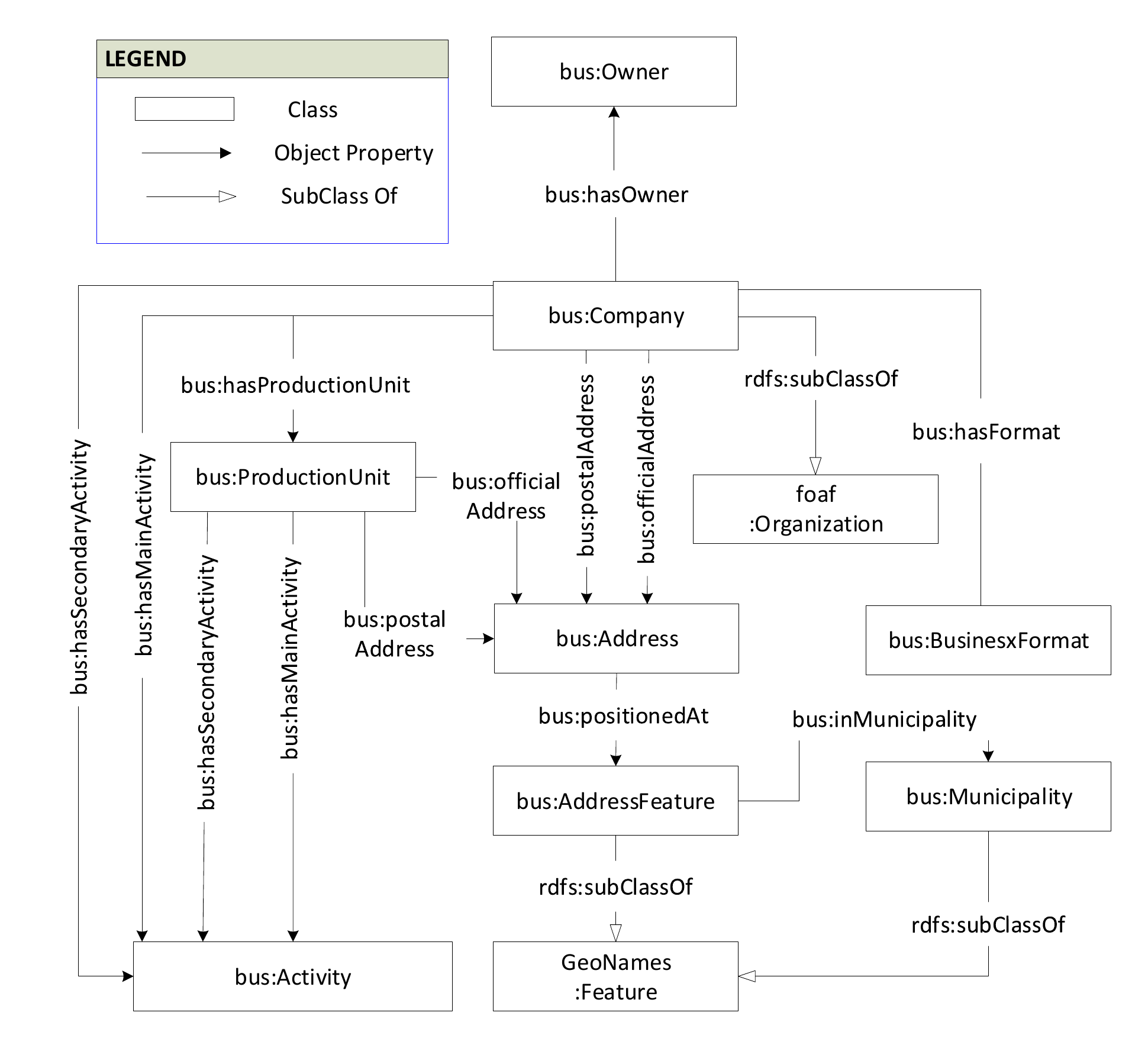}
 
\centering
\caption{The ontology of the \dbd. Due to the large number of datatype properties, they are not included.}
\label{fig:business}
\end{figure*} 

We create a Semantic Data Warehouse (SDW) by integrating two data sources, namely, a Danish Agriculture and Business knowledge base and an EU Farm Subsidy dataset. Both data sources are described below. 
\paragraph{Description of Danish Agriculture and Business knowledge base} The Danish Agriculture and Business knowledge base integrates a \dad\ and a \dbd. The knowledge base can be queried through the \ql\ endpoint \url{http://extbi.lab.aau.dk:8080/sparql/}. In our use case, we only use the business related information from this knowledge base and call it the Danish Business dataset (DBD). The relevant portion of the ontology of the knowledge base is illustrated in  Figure \ref{fig:business}. Generally, in an ontology, a concept provides a general description of the properties and behavior for the similar type of resources; an object property relates among the instances of concepts; a data type property is used to associate the instances of a concept to literals.

We start the description from the concept \texttt{bus:Ow-\\ner}. This concept contains information about the owners of companies, the type of the ownership, and the start date of the company ownership. A company is connected to its owner through the \texttt{bus:hasOwner} property. The \texttt{bus:Company} concept is related to \texttt{bus:BusinessFormat} and \texttt{bus:Production-\\Unit} through the \texttt{bus:hasProductionUnit} and \texttt{bus:hasFormat} properties. Companies and their production units have one or more main and secondary activities. Each company and each production unit has a postal address and an official address. Each address is positioned at an address feature, which is in turn contained within a particular municipality. 
\begin{figure}[h!]
\centering
\includegraphics[scale=0.5]{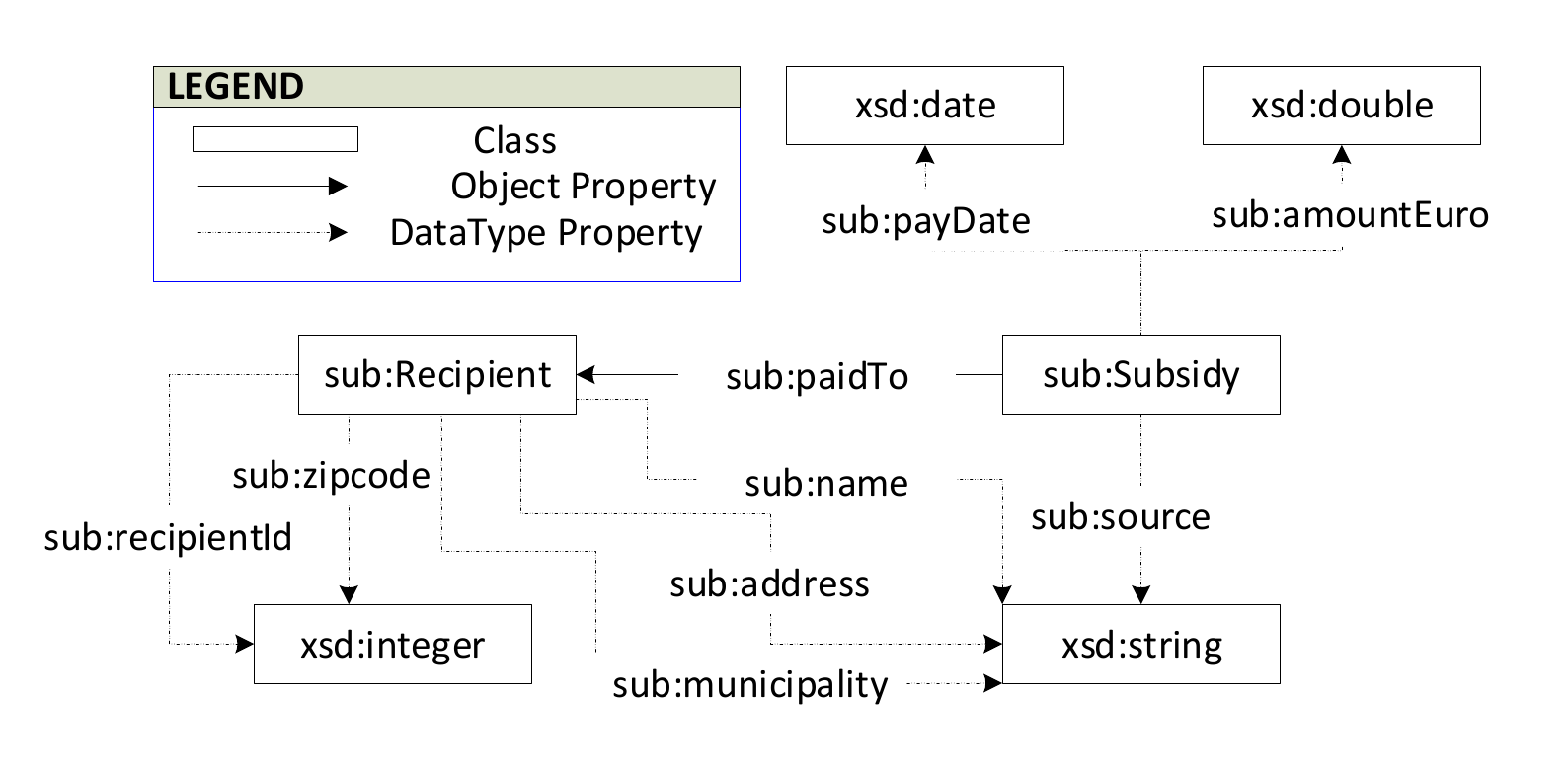}

\caption{The ontology of the Subsidy dataset. Due to the large number of datatype properties, all are not included.}
\label{fig:subsidy}
\end{figure}

\begin{figure*}[h!]
\centering
\includegraphics[scale=0.6]{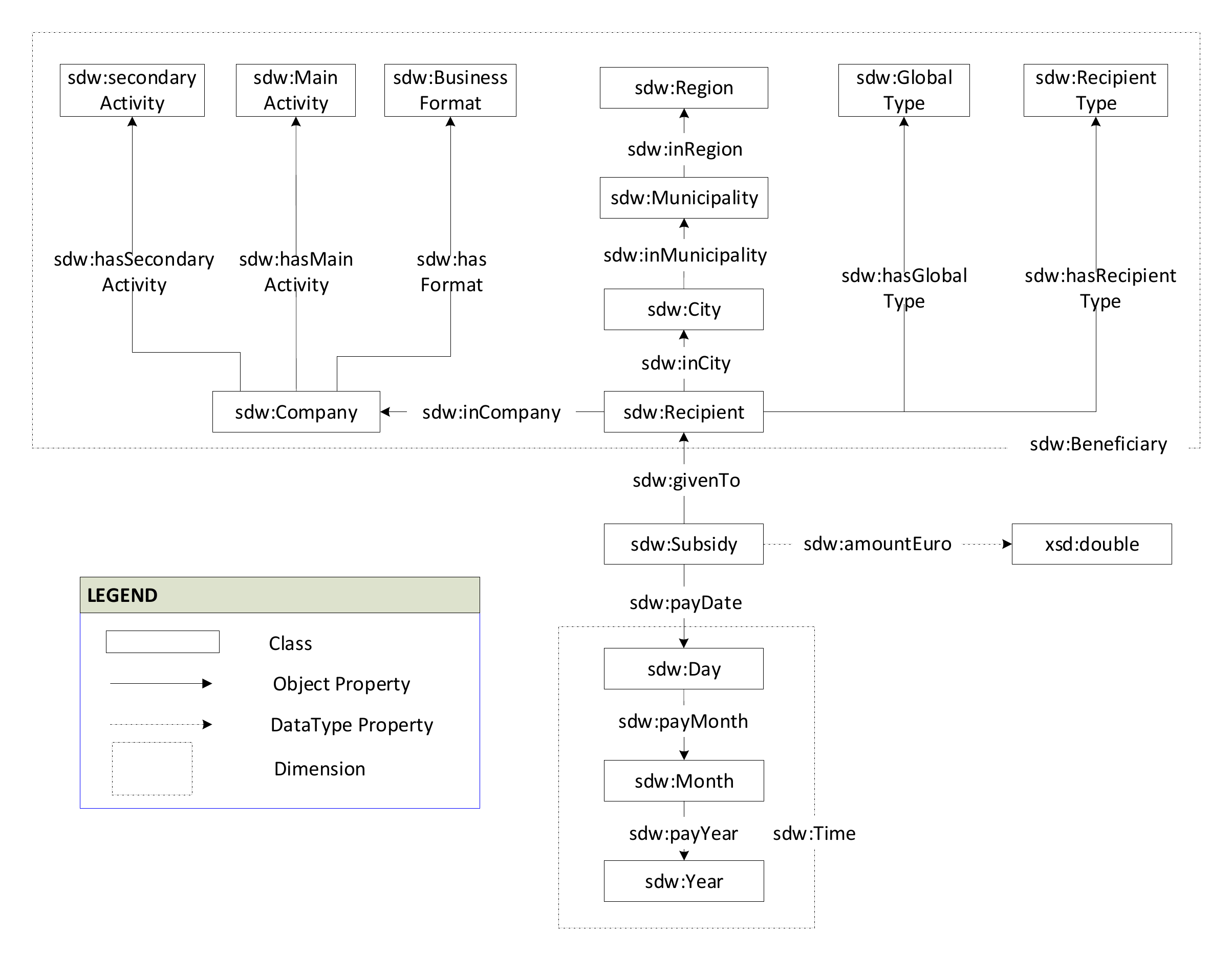}
 
\centering
\caption{The ontology of the MD SDW. Due to the large number, data properties of the dimensions are not shown.}
\label{fig:target}
\end{figure*}

\paragraph{Description of the EU subsidy dataset}Every year, the European Union provides subsidies to the farms of its member countries. We collect EU Farm subsidies for Denmark from \url{https://data.farmsubsidy.org/Old/}. The dataset contains two MS Access database tables: Recipient and Payment. The Recipient table contains the information of recipients who receive the subsidies, and the Payment table contains the amount of subsidy given to the recipients. We create a semantic version of the dataset using \s\ framework~\cite{nath2015towards}. We call it the Subsidy dataset. At first, we manually define an ontology, to describe the schema of the dataset, and the mappings between the ontology and datatabase tables. Then, we populate the ontology with the instances of the source database files. Figure~\ref{fig:subsidy} shows the ontology of the Subsidy dataset. 
\begin{xmpl}
Listing~\ref{list:instance} shows the example instances of (\texttt{bus:Company}) the Danish Business dataset and (\texttt{sub:Recipient} and \texttt{sub:Subsidy}) the EU Subsidy dataset. 
\begin{spacing}{1}
\begin{lstlisting}[caption={: Example instances of the DBD and the Subsidy dataset. }, label=list:instance,
   basicstyle=\tiny\tt, showstringspaces=false, frame=single,language=SQL,morekeywords={PREFIX, a}]
### Business Dataset
PREFIX rdf: <http://www.w3.org/1999/02/22-rdf-syntax-ns#>
PREFIX bus: <http://extbi.lab.aau.dk/ontology/business/>
PREFIX company: <http://extbi.lab.aau.dk/ontology/
                                      business/Company#>
PREFIX activity: <http://extbi.lab.aau.dk/ontology/
                                      business/Activity#>
PREFIX businessType: <http://extbi.lab.aau.dk/ontology/
                                      business/BusinessType#>
PREFIX owner: <http://extbi.lab.aau.dk/ontology/
                                      business/Owner#>
## Examples of bus:Company instances
company:10058996 rdf:type bus:Company;
   bus:companyId 10058996;
   bus:name "Regerupgard v/Kim Jonni Larsen";
   bus:mainActivity activity:11100;
   bus:secondaryActivity activity:682040;
   bus:hasFormat businessType:Enkeltmandsvirksomhed:
   bus:hasOwner  owner:4000175029_10058996;
   bus:ownerName "Kim Jonni Larsen";
   bus:officialaddress "Valsomaglevej 117, Ringsted".
company:10165164 rdf:type bus:Company;
   bus:companyId 10165164;
   bus:name "Idomlund 1 Vindmollelaug I/S";
   bus:mainActivity activity:351100;
   bus:hasFormat businessType:Interessentskab;
   bus:hasOwner  owner:4000170495_10165164;
   bus:ownerName "Anders Kristian Kristensen";
   bus:officialaddress "Donskaervej 31,Vemb".
---------------------------------------------------------
---------------------------------------------------------
### Subsidy Dataset
PREFIX rdf: <http://www.w3.org/1999/02/22-rdf-syntax-ns#>
PREFIX sub: <http://extbi.lab.aau.dk/ontology/subsidy/>
PREFIX recipient: <http://extbi.lab.aau.dk/ontology/
                                    subsidy/Recipient#>
PREFIX subsidy: <http://extbi.lab.aau.dk/ontology/
                                    subsidy/Subsidy#>
## Example of sub:Recipient instances.
recipient:291894 rdf:type sub:Recipient;
   sub:name "Kristian Kristensen";
   sub:address "Donskaervej 31,Vemb";
   sub:municipality "Holstebro";
   sub:recipientID 291894;
   sub:zipcode 7570.             
## Example of sub:Subsidy instances.                 
subsidy:10615413 rdf:type sub:Subsidy;
   sub:paidTo recipient:291894;
   sub:amountEuro "8928.31";
   sub:payDate "2010-05-25";
\end{lstlisting}
\end{spacing}
\end{xmpl}

\paragraph{Description of the Semantic Data Warehouse} Our goal is to develop an MD Semantic Data Warehouse (SDW) by integrating the Subsidy and the DBD datasets. The \texttt{sub:Recipient} concept in the Subsidy dataset contains the information of recipient id, name, address, etc. From \texttt{bus:Company} in the DBD, we can extract information of an owner of a company who received the EU farm subsidies. Therefore, we can integrate both DBD and Subsidy datasets. The ontology of the MD SDW to store EU subsidy information corresponding to the Danish companies is shown in Figure~\ref{fig:target}, where the concept \texttt{sdw:Subsidy} represents the facts of the SDW. The SDW has two dimensions, namely \texttt{sdw:Benificiary} and \texttt{sdw:Time}. The dimensions are shown by a box with dotted-line in Figure~\ref{fig:target}. Here, each level of the dimensions are represented by a concept, and the connections among levels are represented through object properties.








\section{Overview of the Integration Process}
\label{sec:overview}
\begin{figure*}[h!]
\centering
\includegraphics[scale=0.6]{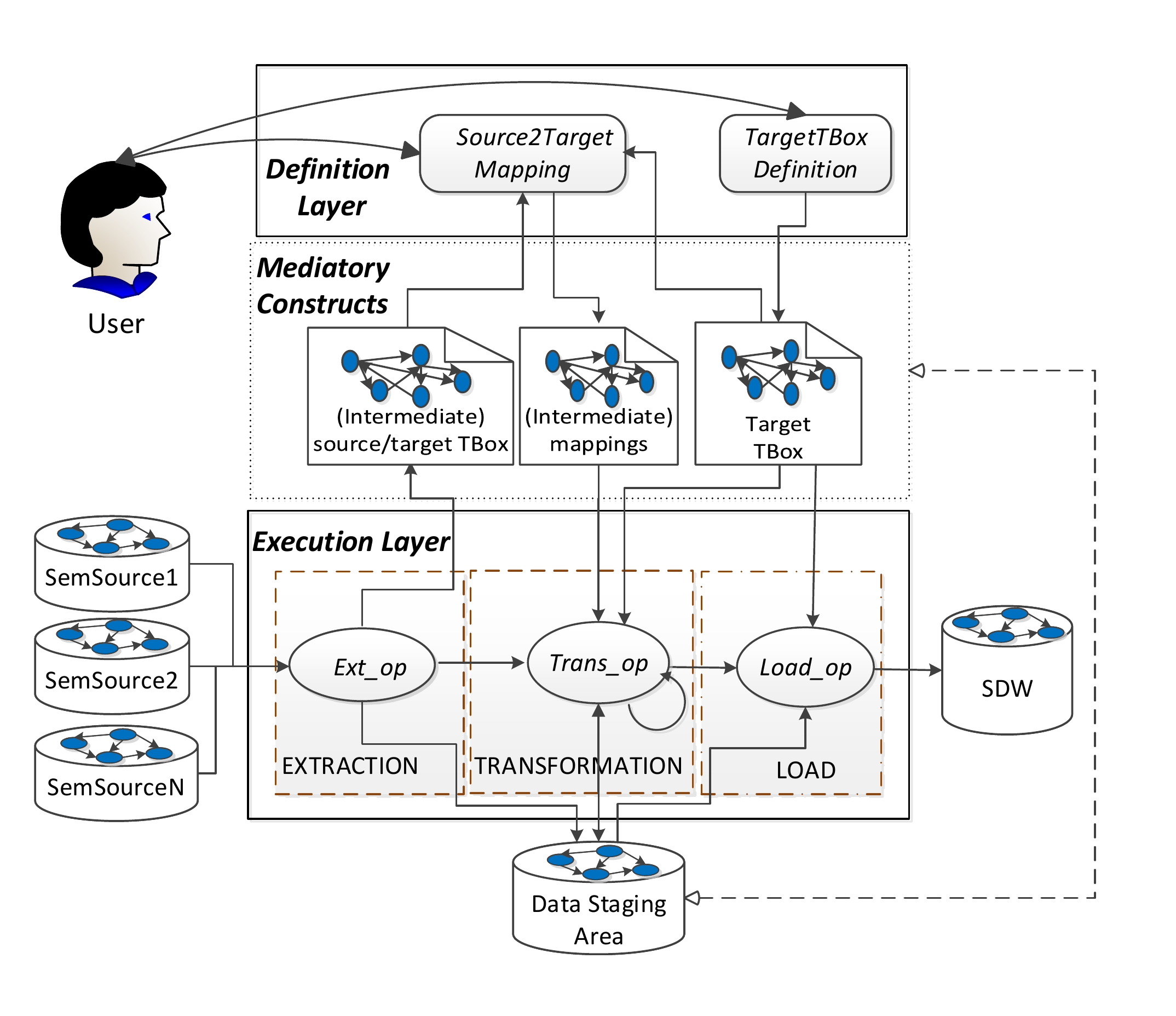}
 
\centering
\caption{The overall semantic data integration process. Here, the round-corner rectangle, data stores, dotted boxes, ellipses, and arrows indicate the  tasks, semantic data sources, the phases of the ETL process, ETL operations and flow directions.}
\label{fig:FSA}
\end{figure*}

\begin{table*}[h!]
\caption{\label{tab:operations} Summary of the ETL operations.}
\resizebox{1\textwidth}{!}{%
\begin{tabular}{|l|l|l|l|}
\hline
\textbf{\begin{tabular}[c]{@{}l@{}}Operation \\ Category\end{tabular}} & \textbf{\begin{tabular}[c]{@{}l@{}}Operation\\ Name\end{tabular}}                  & \textbf{\begin{tabular}[c]{@{}l@{}}Compatible\\ Successors\end{tabular}}                                                                                                                                                 & \textbf{Objectives}                                                                                                                                                       \\ \hline
Extraction                                                             & GraphExtractor                                                                     & \begin{tabular}[c]{@{}l@{}}GraphExtractor,\\ TBoxExtraction,\\ TransformationOnLiteral,\\ JoinTransformation,\\ LevelMemberGenerator,\\ ObservationGenerator,\\ DataChangeDetector,\\ UpdateLevel,\\ Loader\end{tabular} & \begin{tabular}[c]{@{}l@{}}It retrieves an RDF graph in \\ terms of RDF triples from \\ semantic data sources.\end{tabular}                                               \\ \hline
\multirow{9}{*}{Transformation}                                        & TBoxExtraction                                                                     &                                                                                                                                                                                                                          & \begin{tabular}[c]{@{}l@{}}It derives a TBox from a \\ given ABox.\end{tabular}                                                                                           \\ \cline{2-4} 
                                                                       & TransformationOnLiteral                                                            & \begin{tabular}[c]{@{}l@{}}TransformationOnLiteral,\\ JoinTransformation,\\ LevelMemberGenerator,\\ ObservationGenerator,\\ Loader\end{tabular}                                                                          & \begin{tabular}[c]{@{}l@{}}It transforms the source data\\ according to the expressions \\ described in the source-to-target\\ mapping.\end{tabular}                      \\ \cline{2-4} 
                                                                       & JoinTransformation                                                                 & \begin{tabular}[c]{@{}l@{}}TransformationOnLiteral,\\ JoinTransformation,\\ LevelMemberGenerator,\\ ObservationGenerator,\\ Loader\end{tabular}                                                                          & \begin{tabular}[c]{@{}l@{}}It joins two data sources and \\ transforms the data according to \\ the expressions described in the\\ source-to-target mapping.\end{tabular} \\ \cline{2-4} 
                                                                       & \begin{tabular}[c]{@{}l@{}}LevelMemberGenerator\\ (QB4OLAP construct)\end{tabular} & Loader                                                                                                                                                                                                                   & \begin{tabular}[c]{@{}l@{}}It populates levels of the target\\ with the input data.\end{tabular}                                                                          \\ \cline{2-4} 
                                                                       & \begin{tabular}[c]{@{}l@{}}ObservationGenerator\\ (QB4OLAP construct)\end{tabular} & Loader                                                                                                                                                                                                                   & \begin{tabular}[c]{@{}l@{}}It populates facts of the target\\ with the input data.\end{tabular}                                                                           \\ \cline{2-4} 
                                                                       & DataChangeDetector                                                                 & \begin{tabular}[c]{@{}l@{}}LevelMemberGenerator,\\ UpdateLevel\end{tabular}                                                                                                                                              & \begin{tabular}[c]{@{}l@{}}It returns the differences between\\ the new source dataset and the old\\ one.\end{tabular}                                                    \\ \cline{2-4} 
                                                                       & UpdateLevel                                                                        & Loader                                                                                                                                                                                                                   & \begin{tabular}[c]{@{}l@{}}It reflects the changes occurred in \\ the source data to the target level.\end{tabular}                                                       \\ \cline{2-4} 
                                                                       & MaterializeInference                                                               & Loader                                                                                                                                                                                                                   & \begin{tabular}[c]{@{}l@{}}It enriches the SDW by material-\\ izing the inferred triples.\end{tabular}                                                                    \\ \cline{2-4} 
                                                                       & ExternalLinking                                                                    & Loader                                                                                                                                                                                                                   & \begin{tabular}[c]{@{}l@{}}It links internal resources with\\ external KBs.\end{tabular}                                                                                  \\ \hline
Load                                                                   & Loader                                                                             &                                                                                                                                                                                                                          & It loads the data into the SDW.                                                                                                                                           \\ \hline
\end{tabular}%
}
\end{table*}
In this paper, we assume that all given data sources are semantically defined and the goal is to develop an SDW. The first step of building an SDW is to design its TBox. There are two approaches to design the TBox of an SDW, namely source-driven  and demand-driven~\cite{vaisman2014data}. In the former, the SDW's TBox is obtained by analyzing the sources. Here, ontology alignment techniques~\cite{li2008rimom} can be used to semi-automatically define the SDW. Then, designers can identify the multidimensional constructs from the integrated TBox and annotate them with the QB4OLAP vocabulary. In the latter, SDW designers first identify and analyze the needs of business users as well as decision makers, and based on those requirements, they define the target TBox with multidimensional semantics using the QB4OLAP vocabulary. How to design a Target TBox is orthogonal to our approach. Here, we merely provide an interface to facilitate creating it regardless of whatever approach was used to design it.  

After creating the TBox of the SDW, the next step is to create the ETL process. ETL is the backbone process by which data are entered into the SDW and the main focus of this paper. The ETL process is composed of three phases: extraction, transformation, and load. A phase is a sub-process of the ETL which provides a meaningful output that can be fed to the next phase as an input. Each phase includes a set of operations. The extraction operations extract data from the data sources and make it available for further processing as intermediate results. The transformation operations are applied on intermediate results, while the load operations load the transformed data into the DW. The intermediate results can be either materialized in a data staging area or kept in memory. A data staging area (temporary) persists data for cleansing, transforming, and future use. It may also  prevents the loss of extracted or transformed data in case of the failure of the loading process.  

As we want to separate the metadata needed to create ETL flows from their execution, we introduce a two-layered integration process, see Figure~\ref{fig:FSA}. In the \dl, a single source of metadata truth is defined. This includes: the target SDW, semantic representation of the source schemas, and a source to target mapping file. Relevantly, the metadata created represents the ETL flow at the schema level. In the \exl, ETL data flows based on high-level operations are created. This layer executes the ETL flows for instances (i.e., at the data level). Importantly, each ETL operation will be fed the metadata created to parameterize themselves automatically. Additionally, the \exl\ automatically checks the correctness of the flow created, by checking the compatibility of the output and input of consecutive operators. Overall the data integration process requires the following  four steps in the detailed order. 
\begin{enumerate}
\item Defining the target TBox with MD semantics using QB and QB4OLAP constructs. In addition, the TBox can be enriched with RDFS/OWL classes and properties. However, we do not validate the correctness of the added semantics beyond the MD model. This step is done at the Definition Layer. 

\item Extracting source TBoxes from the given sources. This step is done at the Definition Layer.
\item Creating mappings among source and target constructs to characterize ETL flows. The created mappings are expressed using the S2TMAP vocabulary proposed. This step is also done at the Definition Layer.
\item Populating the ABox of the SDW implementing ETL flows. This step is done at the Execution Layer.
\end{enumerate} 
Figure~\ref{fig:FSA} illustrates the whole integration process and how the constructs of each layer communicate with each other. Here, we introduce two types of constructs: tasks and operations. On the one hand, a task requires developer interactions with the interface of the system to produce an output. Intuitively, one may consider the tasks output as the required metadata to automate operations.  On the other hand, from the given metadata, an operation produces an output. The \dl\ consists of two tasks (\textit{TargetTBoxDefinition} and  \textit{SourceToTargetMapping}) and one operation (\textit{TBoxExtraction}). These two tasks respectively address the first and third steps of the integration process mentioned above, while the \textit{TBoxExtraction} operation addresses the second step. This is the only operation shared by both layers (see the input of \textit{SourceToTargetMapping} in Figure~\ref{fig:FSA}). Therefore, the \dl\ creates three types of metadata: target TBox (created by \textit{TargetTBoxDefinition}), source TBoxes (create by \textit{TBoxExtraction}), and source-to-target mappings (created by  \textit{SourceToTargetMapping}). The Execution Layer covers the fourth step of the integration process and includes a set of operations to create data flows from the sources to the target. Figure~\ref{fig:FSA}  shows constructs (i.e., the Mediatory Constructs) used by the  ETL task/operations to communicate between them. These mediatory constructs store the required metadata created to automate the process. In the figure, $Ext_{op}$, $Trans_{op}$, and $Load_{op}$ are the set of extraction, transformation, and load operations used to create the flows at the instance level. The ETL data flows created in the Execution Layer are automatically validated by checking the compatibility of the operations. Precisely, if an operation $O_1$'s output is accepted by $O_2$, then we say $O_2$ is compatible with $O_1$ and express it as $O_1 \rightarrow O_2$.


Since traditional ETL tools (e.g., PDI) do not have ETL operations supporting the creation of an SDW, we propose a set of ETL operations for each phase of the ETL to process semantic data sources. The operations are categorized based on their functionality. Table~\ref{tab:operations} summarizes each operation with its corresponding category name, compatible successors, and its objectives. 
Next, we present the details of each construct of the layers presented in Figure~\ref{fig:FSA}. 

\section{The Definition Layer}
\label{sec:definitionLayer}
\subsection{Definition Layer}
This layer contains two  tasks (\textit{TargetTBoxDefinition} and \textit{Source2TargetMapping}) and one operation \texttt{TBoxExtraction}. The semantics of the tasks are described below. 

\paragraph{\textit{TargetTBoxDefinition}} \label{para:ttd} The objective of this task is to define a target \tb\ with MD semantics. There are two main components of the MD structure: dimensions and cubes. To formally define the schema of these components, we use the notation from~\cite{ciferri2013cube} with some modifications.
\theoremstyle{definition}
\begin{defn}\label{def:dimension} 
A \textbf{dimension schema} can formally be defined as a 5-tuple $(D_{name}, \mathcal{L}, \rightarrow, \mathcal{H}, \mathcal{F_R})$ where (a) $D_{name}$ is the name of the dimension;
(b) $\mathcal{L}$  is a set of level tuples $( L_{name}, L_A ) $ such that $L_{name}$ is the name of a level and $L_A$ is the set of attributes describing the level $L_{name}$. There is a unique bottom level (the finest granularity level) $L_b$, and unique top level (the coarsest one)  denoted $L_{All}$, such that $( L_{All}, \emptyset ) \in \mathcal{L} $;  
 (c) $\rightarrow$ is a strict partial order on $\mathcal{L}$. The poset $(\mathcal{L}, \rightarrow)$ can be represented as a directed graph where each node represents an aggregation level $L\in \mathcal{L}$, and every pair of levels $(L_i, L_j)$ in the poset are represented by a directed edge from the finer granularity level $L_i$  to the coarser granularity level $L_j$, which means that $L_i$ rolls up to $L_j$ or $L_j$ drills down to $L_i$.  Each distinct path between the $L_b$ and $L_{All}$ is called a hierarchy; (d) $\mathcal{H}$ is a set of hierarchy tuples $( H_{name}, H_{L} )$ where $H_{name}$ is the name of a hierarchy and $H_{L} \subseteq \mathcal{L}$  is the set of levels composing the hierarchy; (e) $\mathcal{F_R}$ is a set of roll-up relations. A roll-up relation is a mathematical relation that relates between the members of two adjacent levels in the hierarchy, i.e, $ RUP_{(L_i,L_j)} \in \mathcal{F_R} = \{(lm_i, lm_j)\;|\; lm_i \in L_i \;and \;lm_j \in L_j\; and \;(L_i, L_j) \in (\mathcal{L}, \rightarrow)  \}$. 
\end{defn} 

\begin{xmpl}
Figure~\ref{fig:target} shows that our use case MD SDW has two dimensions: \texttt{sdw:Time} and \texttt{sdw:Ben\\eficiary}. The dimension schema of \texttt{sdw:Time} is formally defined as follows: 

\begin{enumerate}
\item $D_{name} =  \texttt{sdw:Time}$,
\item $\mathcal{L} = \{(\texttt{sdw:Day},\langle \texttt{sdw:dayId} ,  \texttt{sdw:} \\ \texttt{dayName} \rangle ), (\texttt{sdw:Month},\langle \texttt{sdw:monthId},\\  \texttt{sdw:monthName} \rangle ), (\texttt{sdw:Year}, \\
\langle \texttt{sdw:yearId},   \texttt{sdw:yearName} \rangle ) \}$, 
\item $(\mathcal{L},\rightarrow) =\{(\texttt{sdw:Day}, \texttt{sdw:Month}),\\ (\texttt{sdw:Month}, \texttt{sdw:Year}), \\(\texttt{sdw:Year}, \texttt{sdw:All})\}$,
\item $\mathcal{H}= \{\texttt{sdw:TimeHierarchy},\{ \texttt{sdw:Day}, \\ \texttt{sdw:Month}, \texttt{sdw:Year}, \texttt{sdw:All} \}\}, $ and
\item $\mathcal{F_R}= \{\\
RUP_{(\texttt{sdw:Day}, \texttt{sdw:Month})}=\texttt{sdw:payMonth}, \\
RUP_{(\texttt{sdw:Month}, \texttt{sdw:Year})}=
\texttt{sdw:payYear},\\
RUP_{(\texttt{sdw:Year}, \texttt{sdw:All})}= \texttt{sdw:payAll}\}$.
\end{enumerate}

\end{xmpl}

\begin{defn} 
\label{def:cube}
A \textbf{cube schema} is a tuple $(C_{name}, \mathcal{D}_{l_b}, \mathcal{M},$ $\mathcal{F_A})$, where  (a) $C_{name}$ is the name of the cube; (b) $\mathcal{D}_{l_b}$ is a finite set of bottom levels of dimensions, with $|\mathcal{D}_{l_b}|=n$, corresponding to $n$ bottom levels of $n$ dimension schemas different from each other; (c) $\mathcal{M}$ is a finite set of attributes called measures, and each measure $m \in \mathcal{M}$ has an associated domain $Dom(m)$; and (d) $\mathcal{F_A}$ is a mathematical relation that relates each measure to one or more aggregate function in $\mathcal{A}= \{SUM, MAX, AVG ,MIN, COUNT.\}$, i.e.,  $\mathcal{F_A} \subseteq \mathcal{M} \times \mathcal{A}$.
\end{defn}

\begin{xmpl}
The cube schema of our use case, shown in Figure~\ref{fig:target}, is formally defined as follows:
\begin{enumerate}
\item $C_{name}=\texttt{sdw:Subsidy}$,
\item $D_L=\{\texttt{sdw:Day}, \texttt{sdw:Recipient} \}$,
\item $\mathcal{M}= \{\texttt{sdw:amounteuro}\},$ and
\item $\mathcal{F_A}=\{(\texttt{sdw:amounteuro}, SUM),\\ (\texttt{sdw:amounteuro}, AVG)\})$.
\end{enumerate}

\end{xmpl} 

In Section~\ref{sec:sdw},  we discussed how the QB4OLAP vocabulary is used to define different constructs of an SDW. Listing~\ref{list:tboxdefinition} represents the \texttt{sdw:Time} dimension and \texttt{sdw:Subsidy} cube in QB4OLAP.  
\begin{spacing}{0.8}
\begin{lstlisting}[caption={: QB4OLAP representation of \texttt{sdw:Time} dimension and \texttt{sdw:Subsidy} cube.}, label=list:tboxdefinition,
   basicstyle=\tiny\tt,showstringspaces=false,frame=single,language=SQL,morekeywords={PREFIX, a}]
PREFIX sdw: <http://extbi.lab.aau.dk/ontology/sdw/>
PREFIX rdf: 	http://www.w3.org/1999/02/22-rdf-syntax-ns#
PREFIX rdfs: 	http://www.w3.org/2000/01/rdf-schema#
PREFIX qb: <http://purl.org/linked-data/cube#>
PREFIX qb4o: <http://purl.org/qb4olap/cubes#>

## Time Dimension
sdw:Time rdf:type qb:DimensionProperty;
         rdfs:label "Time Dimension";
         qb4o:hasHierarcy sdw:TimeHierarchy.

# Dimension Hierarchies
sdw:TimeHierarchy rdf:type qb4o:Hierarchy;
         rdfs:label "Time Hierarchy";
         qb4o:inDimension sdw:Time;
         qb4o:hasLevel  sdw:Day, sdw:Month, sdw:Year.

# Hierarchy levels 
sdw:Day rdf:type qb4o:LevelProperty;
        rdfs:label "Day Level";
        qb4o:hasAttribute sdw:dayId, sdw:dayName.
sdw:Month rdf:type qb4o:LevelProperty;
        rdfs:label "Month Level";
        qb4o:hasAttribute sdw:monthId, sdw:monthName.
sdw:Year rdf:type qb4o:LevelProperty;
        rdfs:label "Year Level";
        qb4o:hasAttribute sdw:yearId, sdw:yearName.
sdw:All rdf:type qb4o:LevelProperty;
        rdfs:label "ALL".

# Level attributes
sdw:dayId rdf:type qb4o:LevelAttribute;
       rdfs:label "day ID";
       qb4o:updateType qb4o:Type2;
       rdfs:range xsd:String.
sdw:monthId rdf:type qb4o:LevelAttribute;
       rdfs:label "Month ID";
       qb4o:updateType qb4o:Type2;
       rdfs:range xsd:String.
sdw:yearId rdf:type qb4o:LevelAttribute;
       rdfs:label "year ID";
       qb4o:updateType qb4o:Type2;
       rdfs:range xsd:String.
sdw:dayName rdf:type qb4o:LevelAttribute;
       rdfs:label "day Name";
       qb4o:updateType qb4o:Type1;
       rdfs:range xsd:String.
sdw:monthName rdf:type qb4o:LevelAttribute;
       rdfs:label "Month Name";
       qb4o:updateType qb4o:Type1;
       rdfs:range xsd:String.
sdw:yearName rdf:type qb4o:LevelAttribute;
       rdfs:label "year Name";
       qb4o:updateType qb4o:Type1;		 
       rdfs:range xsd:String.
		 
#rollup relations
sdw:payMonth rdf:type qb4o:RollupProperty.	 
sdw:payYear rdf:type qb4o:RollupProperty.
sdw:payAll rdf:type qb4o:RollupProperty.

# Hierarchy Steps
_:ht1 rdf:type qb4o:HierarchyStep;
      qb4o:inHierarchy sdw:TimeHierarchy;
      qb4o:childLevel sdw:Day;
      qb4o:parentLevel sdw:Month;
      qb4o:pcCardinality qb4o:OneToMany;
      qb4o:rollup sdw:payMonth.
_:ht2 rdf:type qb4o:HierarchyStep;
      qb4o:inHierarchy sdw:TimeHierarchy;
      qb4o:childLevel sdw:Month;
      qb4o:parentLevel sdw:Year;
      qb4o:pcCardinality qb4o:OneToMany;
      qb4o:rollup sdw:payYear.
_:ht2 rdf:type qb4o:HierarchyStep;
      qb4o:inHierarchy sdw:TimeHierarchy;
      qb4o:childLevel sdw:Year;
      qb4o:parentLevel sdw:All;
      qb4o:pcCardinality qb4o:OneToMany;
      qb4o:rollup sdw:payAll.

## Subsidy Cube 
sdw:amounteuro rdf:type qb:MeasureProperty;
      rdfs:label "subsidy amount"; rdfs:range xsd:Double.
sdw:SubsidyStructure rdf:type qb:DataStructureDefinition;
      qb:component[qb4o:level sdw:Recipient];
      qb:component[qb4o:level sdw:Day];
      qb:component[qb:measure sdw:amounteuro; 
           qb4o:aggregateFunction qb4o:sum, qb4o:avg].           
# Subsidy Dataset
sdw:SubsidyMD rdf:type qb:Dataset;
      rdfs:label "Subsidy dataset";
      qb:structure sdw:SubsidyStructure;                     
                     

\end{lstlisting}
\end{spacing}

\paragraph{\textit{TBoxExtraction}} \label{para:tbe} After defining a target TBox, the next step is to extract source TBoxes. Typically, in a semantic source, the TBox and ABox of the source are provided. Therefore, no external extraction task/operation is required. However, sometimes, the source contains only the ABox, and no TBox. In that scenario, an extraction process is required to derive a TBox from the ABox. We formally define the process as follows.
\begin{defn}

The TBox extraction operation from a given ABox, \textit{ABox} is defined as $f_{ABox2TBox}(ABox) \rightarrow  TBox $. The derived \textit{TBox}  is defined in terms of the following TBox constructs: a set of concepts \textit{C}, a set of concept taxonomies \textit{H}, a set of properties \textit{P}, and the sets of property domains \textit{D} and ranges \textit{R}. The following steps describe the process to derive each TBox element for \textit{TBox}.
\begin{enumerate}
\item \textit{C}: By checking the unique  objects of the triples in \textit{ABox} where \texttt{rdf:type} is used as a predicate, \textit{C} is identified.
\item \textit{H}: The taxonomies among concepts are identified by checking the instances they share among themselves.  Let $C_1$ and $C_2$ be two concepts. One of the following taxonomic relationships holds between them: 1) if $C_1$ contains all instances of $C_2$, then we say $C_2$ is a subclass of $C_1$ ($C_2$ \texttt{rdfs:subClassOf} $C_1$); 2) if they do not share any instances, they are disjoint ($C_1$ \texttt{owl:disjointWith} $C_2$); and 3) if $C_1$ and $C_2$ are both a subclass of each other, then they are equivalent ($C_1$ \texttt{owl:equivalentClass} $C_2$). 
\item $P, D, R$: By checking the unique predicates of the triples, $P$ is derived. A property $p\in P$ can relate resources with either resources or literals. If the objects of the triples where $p$ is used as predicates are IRIs, then $p$ is an object property; the domain of $p$ is the set of the types of the subjects of those triples, and the range of $p$ is the types of the objects of those triples. If the objects of the triples where $p$ is used as predicates are literals, then  $p$ is a datatype property; the domain of $p$ is the set of types the subjects of those triples, and the range is the set of data types of the literals.    
\end{enumerate}
\end{defn}

\paragraph{\textit{SourceToTargetMapping}} \label{para:s2t}

Once the target and source TBoxes are defined, the next task is to characterize the ETL flows at the \dl\ by creating source-to-target mappings. Because of the heterogeneous nature of source data, mappings among sources and the target should be done at the TBox level. In principle, mappings are constructed between sources and the target; however, since mappings can get very complicated, we allow to create a sequence of \textit{SourceToTargetMapping} definitions  whose subsequent input is generated by the preceding operation. The communication between these operations is by means of a materialized intermediate mapping definition and its meant to facilitate the creation of very complex flows (i.e., mappings) between source and target. 

A source-to-target mapping is constructed between a source and a target TBox, and it consists of a set of concept-mappings. A concept-mapping defines i) a relationship (equivalence, subsumption, supersumption, or join) between a source and the corresponding target concept, ii) which source instances are mapped (either all or a subset defined by a filter condition), iii) the rule to create the IRI for target concept instances, iv) the source and target ABox locations, v) the common properties between two concepts if their relationship is join, vi) the sequence of ETL operations required to process the concept-mapping,  and vii) a set of property-mappings for each property having the target concept as a domain. A property-mapping defines how a target property is mapped from either a source property or an expression over properties. Definition~\ref{defn:s2t} formally defines a source-to-target mapping.  

\begin{figure*}[h!]
\centering
\includegraphics[scale=0.75]{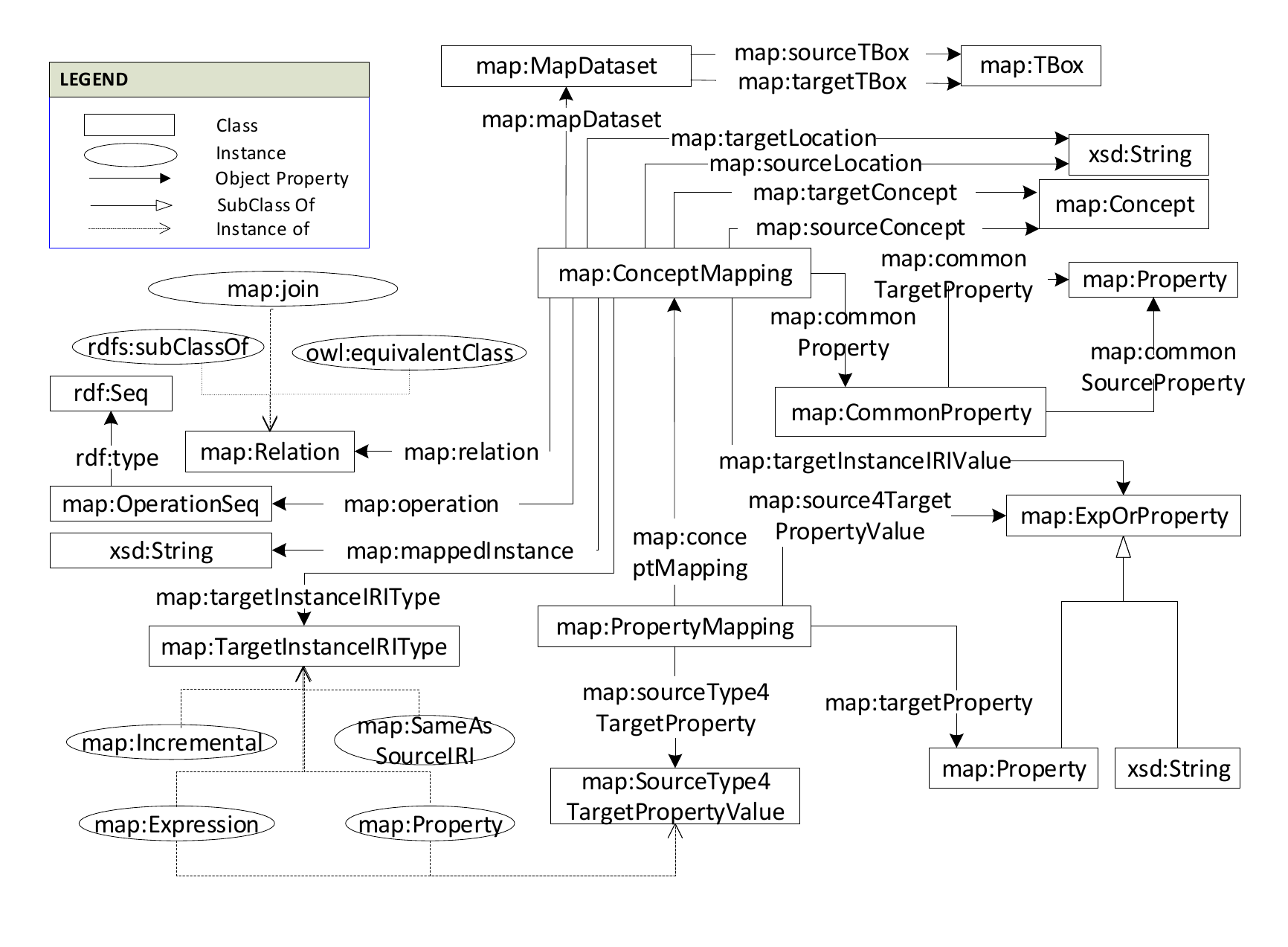}
\centering
\caption{Graphical overview of key terms and their relationship to the \mv\ vocabulary.}
\label{fig:mappingvocabulary}
\end{figure*}
\begin{defn}
\label{defn:s2t}
Let $T_S$ and $T_T$ be a  source \tb\ and a target \tb. We formally define a source-to-target mapping as a set of concept-mappings, wherein each concept-mapping is defined with a 10-tuple formalizing the elements discussed above (i-vii):

$SourceToTargetMapping(T_S,T_T) = \{(c_{s},  relation,\\  c_{t}, loc_{c_{s}}, loc_{c_{t}}, mapIns, p_{map}, tin_{iri}, p_{com}, op)\}.$

The semantics of each concept-mapping tuple is given below. 
\begin{enumerate}
\item[-]  $ c_{s} \in  \mathcal{C}(T_S)$ and $ e_{t_i}  \in  \mathcal{C}(T_T)$ are a source and a target concept respectively, where $\mathcal{C}(T)$ defined in Equation~\ref{eq:concept}.
\item[-]  $ relation \in \{\equiv , \sqsubseteq ,\sqsupseteq, \bowtie \}$ represents the relationship between the source and target  concept. The relationship can  be either $equivalence$ $(e_{s_i} \equiv e_{t_i}$), $supersumption$  $(e_{s_i} \sqsupseteq e_{t_i})$, $subsumption$  $(e_{s_i} \sqsubseteq e_{t_i})$, or $join\; (e_{s_i} \bowtie e_{t_i})$. A join relationship exists between two sources when there is a need to populate a target element (a level, a (QB) dataset, or a concept) from multiple sources. Since a concept-mapping represents a binary relationship, to join $n$ sources, an ETL process requires $n-1$ join concept-mappings. A concept-mapping with a join relationship requires two sources (i.e., the concept-mapping source and target concepts) as  input and updates the target concept according to the join result. Thus, for multi-way joins, the output of a concept-mapping is a source concept of the next concept-mapping. Note that, a join relationship can be natural join ($\bowtie$), right-outer join ($\rightouterjoin$), or left-outer join ($\leftouterjoin$). 

\item[-]  $loc_{c_{s}}$ and $loc_{c_{t}}$ are the locations of source and target concept ABoxes.

\item[-] $ mapIns  \in (\{All\} \cup FilterCondition)$ indicates which instances of the source concept to use to populate the target concept; it can either be all source instances or a subset of source instances defined by a filter condition.  

\item[-] $p_{map}=\{(p_{c_s},p_{c_t})\}$ is a set of property-mappings across the properties of $ c_{s} $ and $ c_{t}$. $p_{c_s}$ can be a property from $property(c_{s})$ or an expression over the elements of $exp(property(c_{s})\:\cup\: property(c_{t}))$ and $p_{c_t}$ is a property from $property$ \\$(c_{t})$. Here, $property(c)$ returns the union of the set of properties which are connected with concept $c$ either using the \texttt{rdfs:domain} or \texttt{qb4olap:inLevel} properties, or the set of rollup properties related to $c$. An expression allows to apply arithmetic operations and/or some high-level functions for manipulating strings, data types, numbers, dates defined as standard SPARQL functions in~\cite{harris2013sparql} over the properties. 
\item[-] $tin_{iri}$ indicates how the unique IRIs of target instances are generated. The IRIs can be either the same as the source instances, or created using a property of $ c_{s}$, or using an expression from $exp(property(c_{s}))$, or in an incremental way.
\item[-] $p_{com} = \{(scom_i, tcom_i)| scom_i\in (property (e_{s_i}),$ \\ $ tcom_i\in property (e_{t_i})) \}$ is a set of common property pairs. In each pair, the first element is a source property and the second one is a target property. $p_{com}$ is required when the relationship between the source and target concept is a join.
\item[-] $op$ is an ETL operation or a sequence of ETL operations required to implements the mapping element in the ABox level. When $op$ is a sequence of ETL operations, the location of the input ABox location for the first operation in the sequence is $loc_{c_{s}}$; the subsequent operations in the sequence take the output of their preceding operation as the input ABox. This generation of intermediate results is automatically handled by the automatic ETL generation process described in Section~\ref{sec:aeefg}.

\end{enumerate}
\end{defn}
In principle, an SDW is populated from multiple sources, and a source-to-target ETL flow requires more than one intermediate concept-mapping definitions. Therefore, a complete ETL process requires a set of source-to-target mappings. We say a mapping file is a set of source-to-target mappings. Definition~\ref{defn:mappingfile} formally defines a mapping file. 

\begin{defn}
\label{defn:mappingfile}
$Mapping\; file = $ \\ $\bigcup_{i\in S}{SourceToTargetMapping(T_i, T_j)}$, where $S$ is the set of all sources and intermediate results  schemas, and $j$ the set of all intermediate results and the target schemas.
\end{defn}

To implement the source-to-target mappings formally defined above, we propose an OWL-based mapping vocabulary: Source-to-Target Mapping (\mv). Figure~\ref{fig:mappingvocabulary} depicts the mapping vocabulary. A mapping between a source and a target TBox is represented as an instance of the class \texttt{map:MapDataset}. The source and target TBoxes are defined by instantiating \texttt{map:TBox}, and these TBoxes are connected to the mapping dataset using the properties \texttt{map:sourceTBox} and \texttt{map:targetTBox}, respectively. A concept-mapping (an instance of \texttt{map:-\\ConceptMapping}) is used to map between a source and a target concepts (instances of \texttt{map:Concept}). A concept-mapping is connected to a mapping dataset using the \texttt{map:mapDataset} property. The source and target ABox locations of the concept-mapping are defined through the \texttt{map:sourceLocation} and \texttt{map:targetLocation} properties. The relationship between the concepts can be either \textit{\texttt{rdfs:subC-\\lassOf}}, or \textit{\texttt{map:join}}, or \textit{\texttt{owl:equivalentCla-\\ss}}, and it is connected to the concept-mapping via the \texttt{map:relation} property. The sequence of ETL operations, required to implement the concept-mapping at the ABox level, is defined through an RDF sequence. To express joins, the source and target concept in a concept-mapping represent the concepts to be joined, and the join result is stored in the target concept as an intermediate result. In a concept-mapping,  we, via \texttt{map:commonProperty}, identify the join attributes with a blank node (instance of \texttt{map:CommonProperty}) that has, in turn, two properties identifying the source and target join attributes; i.e., \texttt{map:commonSourceProperty} and \texttt{map:commonTargetProperty}.  Since a join can be defined on multiple attributes, we may have multiple blank node definitions. The type of target instance IRIs is stated using the property \texttt{map:TargetInst-\\anceIRIType}. If the type is either \textit{\texttt{map:Property}} or \textit{\texttt{map:Expression}}, then the property or expression, to be used to generate the IRIs,  is given by \texttt{map:targetInstanceIRIvalue}. 

To map at the property stage, a property-mapping (an instance of \texttt{map:PropertyMapping}) is used.  The association between a property-mapping and a concept-mapping is defined by \texttt{map:conceptMap-\\ping}. The target property of the property-mapping is stated using \texttt{map:targetProperty}, and that target property can be mapped with either a source property or an expression. The source type of target property is determined through \texttt{map:sourceType4Ta-\\rgetProperty} property, and the value is defined by \texttt{map:source4TargetPropertyValue}.      

\begin{xmpl}\label{xmpl:mapping}
Listing~\ref{list:mapping} represents a snippet of the mapping file of our use case MD SDW and the source datasets. 
In the \exl, we show how the different segments of this mapping file will be used by each ETL operation. 
\begin{spacing}{0.9}
\begin{lstlisting}[caption={: An \mv\ representation of the mapping file of our use case. }, label=list:mapping,
   basicstyle=\tiny\tt, showstringspaces=false, frame=single, keywordstyle=\bfseries, language=SQL, morekeywords={PREFIX, a, :Recipient_Company}]
PREFIX onto: <http://extbi.lab.aau.dk/ontology/>
PREFIX bus:  <http://extbi.lab.aau.dk/ontology/business/>
PREFIX sub:  <http://extbi.lab.aau.dk/ontology/subsidy/>
PREFIX rdf:  <http://www.w3.org/1999/02/22-rdf-syntax-ns#>
PREFIX rdfs: <http://www.w3.org/2000/01/rdf-schema#>
PREFIX map:  <http://extbi.lab.aau.dk/ontology/s2tmap/>
PREFIX sdw:  <http://extbi.lab.aau.dk/sdw>
PREFIX : <http://extbi.lab.aau.dk/ontology/s2map/example#>
## MapDataset 
:mapDataset1 rdf:type map:Dataset;
  rdfs:label "Map-dataset for business and subsidy ontology";
  map:sourceTBox "/map/businessTBox.ttl";  
  map:targetTBox "/map/subsidyTBox.ttl".
:mapDataset2 rdf:type map:Dataset;
  rdfs:label "Map-dataset for subsidy and subsidyMD ontology";
  map:sourceTBox "/map/subsidyTBox.ttl";
  map:targetTBox "/map/subsidyMDTBox.ttl".
##ConceptMapping: Joining  Recipient and Company
:Recipient_Company rdf:type map:ConceptMapping;
  rdfs:label "join-transformation between 
          bus:Company and sub:Recipient";
  map:mapDataset :mapDataset1;
  map:sourceConcept bus:Company;
  map:targetConcept sub:Recipient;
  map:sourceLocation "/map/dbd.nt";
  map:targetLocation "/map/subsidy.nt";
  map:relation map:rightOuterjoin;				  
  map:mappedInstance "All";
  map:targetInstanceIRIUniqueValueType map:SameAsSourceIRI;
  map:operation _:opSeq;
  map:commonProperty _:cp1, _:cp2.
_:opSeq  rdf:type rdf:Seq;
  rdf:_1 map:joinTransformation.
_:cp1 map:sourceCommonProperty bus:ownerName;
  map:targetCommonProperty sub:name.
_:cp2 map:sourceCommonProperty bus:officialAddress;
  map:targetCommonProperty sub:address.   

#concept-mapping: Populating the sdw:Recipient level
:Recipient_RecipientMD   rdf:type map:ConceptMapping;
  rdfs:label "Level member generation";
  map:mapDataset :mapDataset2;
  map:sourceConcept sub:Recipient;
  map:targetConcept sdw:Recipient;
  map:sourceLocation "/map/subsidy.nt";
  map:targetLocation "/map/sdw";
  map:relation owl:equivalentClass;				  
  map:mappedInstance "All";
  map:targetInstanceIRIValueType map:Property;
  map:targetInstanceIRIValue sub:recipientID;
  map:operation _:opSeq1.
_:opSeq1 rdf:type rdf:Seq;
  rdf:_1 map:LevelMemberGenerator;
  rdf:_2 map:Loader.
#concept-mapping: Populating the cube dataset
:Subsidy_SubsidyMD rdf:type map:ConceptMapping;
  rdfs:label "Observation generation";
  map:mapDataset :mapDataset2;
  map:sourceConcept sub:Subsidy;
  map:targetConcept sdw:SubsidyMD;
  map:sourceLocation "/map/subsidy.nt";
  map:targetLocation "/map/sdw";
  map:relation owl:equivalentClass;				  
  map:mappedInstance "All";
  map:targetInstanceIRIUniqueValueType map:Incremental;
  map:operation _:opSeq2.
_:opSeq2 rdf:type rdf:Seq;
  rdf:_1 map:GraphExtractor;
  rdf:_2 map:TransformationOnLiteral;
  rdf:_3 map:ObservationGenerator;
  rdf:_4 map:Loader.
## property-mapping under :Recipient_Company  
:companyID_companyID rdf:type map:PropertyMapping;
  rdfs:label "property-mapping for companyID";
  map:conceptMapping :Recipient_Company;
  map:targetProperty sub:companyId;
  map:sourceType4TargetPropertyValue map:Property;
  map:source4TargetPropertyValue bus:companyId.
:businessType_businessType rdf:type map:PropertyMapping;
  rdfs:label "property-mapping for business type";
  map:conceptMapping :Recipient_Company;
  map:targetProperty sub:businessType;
  map:sourceType4TargetPropertyValue map:Property;
  map:source4TargetPropertyValue bus:hasFormat.
:address_city rdf:type map:PropertyMapping;
  rdfs:label "property-mapping for city";
  map:conceptMapping :Recipient_Company;
  map:targetProperty sub:cityId;
  map:sourceType4TargetPropertyValue map:Expression;
  map:source4TargetPropertyValue STRAFTER(sub:address,",").
:name_name rdf:type map:PropertyMapping;
  rdfs:label "property-mapping for name";
  map:conceptMapping :Recipient_Company;
  map:targetProperty sub:name;
  map:sourceType4TargetPropertyValue map:Property;
  map:source4TargetPropertyValue sub:name.
# property-mappings under :Recipient_RecipientMD
:companyId_company rdf:type map:PropertyMapping;
  rdfs:label "property-mapping for companyId";
  map:conceptMapping :Recipient_RecipientMD;
  map:targetProperty sdw:hasCompany;
  map:sourceType4TargetPropertyValue map:Property;
  map:source4TargetPropertyValue sub:companyId;
:cityId_city rdf:type map:PropertyMapping;
  rdfs:label "property-mapping for cityId";
  map:conceptMapping :Recipient_RecipientMD;
  map:targetProperty sdw:inCity;
  map:sourceType4TargetPropertyValue map:Property;
  map:source4TargetPropertyValue sub:city;
:name_name rdf:type map:PropertyMapping;
  rdfs:label "property-mapping for name";
  map:conceptMapping :Recipient_RecipientMD;
  map:targetProperty sdw:name;
  map:sourceType4TargetPropertyValue map:Property;
  map:source4TargetPropertyValue sub:name
# property-mappings under :Subsidy_SubsidyMD 
:Recipient_recipientId rdf:type map:PropertyMapping;
  rdfs:label "property-mapping for recipient in sdw:Subsidy";
  map:conceptMapping :Subsidy_SubsidyMD;
  map:targetProperty sdw:Recipient;
  map:sourceType4TargetPropertyValue map:Property;
  map:source4TargetPropertyValue sub:paidTo.
:hasPayDate_Day rdf:type map:PropertyMapping;
  rdfs:label "property-mapping for Day of sdw:SubsidyMD";
  map:conceptMapping :Subsidy_SubsidyMD;
  map:targetProperty sdw:Day;
  map:sourceType4TargetPropertyValue map:Expression;
  map:source4TargetPropertyValue 
  "CONCAT(STR(DAY(sub:payDate)),"/",
   STR(MONTH(sub:payDate)),"/",STR(YEAR(sub:payDate)))".
:amountEuro_amountEuro rdf:type map:PropertyMapping;
  rdfs:label "property-mapping for amountEuro measure";
  map:conceptMapping :Subsidy_SubsidyMD;
  map:targetProperty sdw:amountEuro;
  map:sourceType4TargetPropertyValue map:Property;
  map:source4TargetPropertyValue sub:amountEuro.
\end{lstlisting}
\end{spacing}
\end{xmpl}
\begin{figure*}[h!]
\includegraphics[scale=0.6]{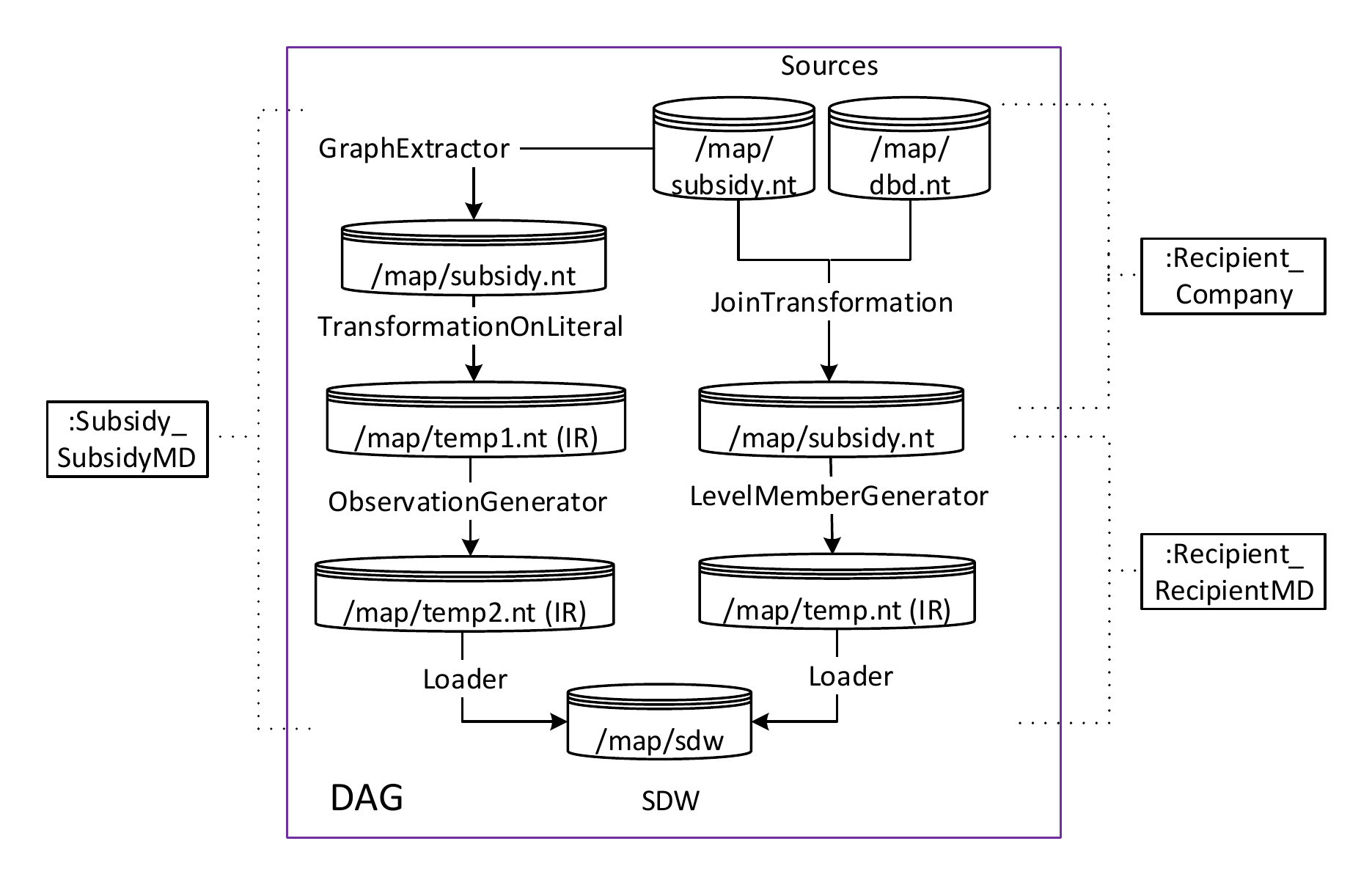}
\caption{The conceptual presentation of Listing~\ref{list:mapping}.}
\label{fig:dag}
\end{figure*}

A mapping file is a Directed Acyclic Graph (DAG). Figure~\ref{fig:dag} shows the DAG representation of Listing~\ref{list:mapping}. In this figure, the sources, intermediate results and the SDW are denoted as nodes of the DAG and edges of the DAG represent the operations. The dotted-lines shows the parts of the ETL covered by concept-mappings, represented by a rectangle.

\section{The Execution Layer}
\label{sec:executionLayer}
In the Execution Layer, ETL data flows are constructed to populate an MD SDW. Table~\ref{tab:operations} summarizes the set of ETL operations.  In the following, we present an overview of each operation category-wise. Here, we give the intuitions of the ETL operations in terms of definitions and examples. To reduce the complexity and length of the paper, we place the formal semantics of the ETL operations in Appendix~\ref{sec:operations}. In this section, we only present the signature of each operation. That is, the main inputs required to execute the operation. As an ETL data flow is a sequence of operations and an operation in the sequence communicates with its preceding and subsequent operations by means of materialized intermediate results, all the operations presented here have side effects\footnote{An operation has a  side effect if it modifies some state variable value(s) outside its local environment (\url{https://en.wikipedia.org/wiki/Side_effect_(computer_science)})} instead of returning output.  

Developers can use either of the two following options: (i) The recommended option is that given a TBox construct \textit{aConstruct} (a concept, a level, or a QB dataset) and a mapping file \textit{aMappings} generated in the \dl, the automatic ETL execution flow generation process will automatically extract the parameter values from \textit{aMappings} (see Section~\ref{sec:aeefg} for a detailed explanation of the automatic ETL execution flow generation process). (ii) They can manually set input parameters at the operation level.  
 In this section, we follow the following order to present each operation: 1) we first give a high-level definition of the operation; 2) then, we define how the automatic ETL execution flow generation process parameterizes the operation from the mapping file, and 3) finally, we present an example showing how developers can manually parameterize the operation. When introducing the operations and referring to their automatic parametrization, we will refer to \textit{aMappings} and \textit{aConstruct} as defined here. Note that each operation is bound to exactly one concept-mapping at a time in the mapping file (discussed in  Section~\ref{sec:aeefg}).

\subsection{Extraction Operations} Extraction is process of data retrieval from the sources. Here, we introduce two extraction operations for semantic sources: (i) \textit{GraphExtractor} - to form/extract an RDF graph  from a semantic source and (ii) \textit{TBoxExtraction} - to derive a TBox from a semantic source as described in Section~\ref{sec:definitionLayer}. As such, \textit{TBoxExtraction} is the only operation in the Execution Layer generating metadata stored in the Mediatory Constructs (see Figure~\ref{fig:FSA}).
\paragraph{\textit{GraphExtractor(Q, G, outputPattern, tABox)}} Since the data integration process proposed in this paper uses RDF as the canonical model, we extract/generate RDF triples from the sources with this operation. \textit{GraphExtractor} is functionally equivalent to SPARQL CONSTRUCT queries~\cite{kostylev2015construct}. 

If the ETL execution flow is generated automatically, the automatic ETL execution flow generation process first identifies the concept-mapping \textit{cm} from \textit{aMappings} where \textit{aConstruct} appears (i.e., in \textit{aMapping}, $cm\xrightarrow[]{map:targetConcept} aConstruct$) and the operation to process \textit{cm} is \textit{GraphExtractor} (i.e., \textit{GraphExtractor} is an element in the operation sequence defined by the \texttt{map:operation} property). Then, it parametrizes \textit{GraphExtractor} as follows:  1) $G$ is  the location of the source ABox defined by the property \texttt{map:sourceLocation} of $cm$; 2) $Q$ and \textit{outputPattern} are internally built based on the \texttt{map:mappedI-\\nstance} property, which defines whether all instances (defined by "All") or a subset of the instances (defined by a filter condition) will be extracted; 3) \textit{tABox} is the location of target ABox defined by the property \texttt{map:targetLocation}. A developer can also manually set the parameters. From the given inputs, \textit{GraphExtractor} operation performs a pattern matching operation over the given source $G$ (i.e., finds a map function binding the variables in query pattern $Q$ to constants in $G$), and then, for each binding, it creates triples according to the triple templates in \textit{outputPattern}. Finally, the operation stores the output in the path \textit{tABox}.

\begin{xmpl}
Listing~\ref{list:instance} shows the example instances of the Danish Business Dataset (DBD). To extract all instances of \texttt{bus:Company} from the dataset, we use the \textit{GraphExtractor(Q, G, outputPattern, tABox)} operation, where
 
\begin{enumerate}
\item \textit{Q=((?ins,\texttt{rdf:type},\texttt{bus:Company})\\AND (?ins,?p,?v))},\footnote{To make it easily distinguishable, here, we use comma instead of space to separate the components of a triple pattern and an RDF triple.}
\item \textit{G}="/map/dbd.ttl",
\item \textit{outputPattern= (?ins,?p,?v)},
\item \textit{tABox}="/map/com.ttl"\footnote{We present the examples in Turtle format for reducing the space and better understanding. In practice, our system prefers N-Triples format to support scalability.}.
\end{enumerate}
Listing~\ref{list:graphextract} shows the output of this operation. 

\begin{lstlisting}[caption={: Example of \textit{GraphExtractor}.}, label=list:graphextract,
   basicstyle=\tiny\tt, showstringspaces=false, frame=single,language=SQL,morekeywords={PREFIX, a}]
company:10058996 rdf:type bus:Company;
    bus:name "Regerupgard v/Kim Jonni Larsen";
    bus:mainActivity activity:11100;
    bus:secondaryActivity activity:682040;
    bus:hasFormat businessType:Enkeltmandsvirksomhed;
    bus:hasOwner  owner:4000175029_10058996;
    bus:ownerName "Kim Jonni Larsen";
    bus:address "Valsomaglevej 117, Ringsted".
company:10165164 rdf:type bus:Company;
    bus:name "Idomlund 1 Vindmollelaug I/S";
    bus:mainActivity activity:351100;
    bus:hasFormat businessType:Interessentskab;
    bus:hasOwner  owner:4000170495_10165164;
    bus:ownerName "Anders Kristian Kristensen";
    bus:address "Donskaervej 31,Vemb".
\end{lstlisting}
\end{xmpl}
\textit{TBoxExtraction} is already described in Section~\ref{para:tbe}, therefore, we do not repeat it here. 

\subsection{Transformation Operations}
\label{subsec:transformation}
Transformation operations transform the extracted data according to the semantics of the SDW. Here, we define the following semantic-aware ETL transformation operations: \textit{TransformationOnLiteral, JoinTransformation, LevelMemberGenerator, ObservationGenerator, ChangedDataCapture, UpdateLevel, External linking,} and \textit{MaterializeInference}. The following describe each operation. 
\paragraph{TransformationOnLiteral(sConstruct, tConstruct, sTBox, sABox propertyMappings, tABox)} As described in the \textit{SourceToTargetMapping} task, a property (in a property-mapping) of a target construct (i.e., a level, a QB dataset, or a concept) can be mapped to either a source concept property or an expression over the source properties. An expression allows arithmetic operations, datatype (string, number, and date) conversion and processing functions, and group functions (sum, avg, max, min, count) as defined in SPARQL~\cite{harris2013sparql}. This operation generates the instances of the target construct by resolving the source expressions mapped to its properties.

If the ETL execution flow is generated automatically, the automatic ETL execution flow generation process first identifies the concept-mapping $cm$ from \textit{aMappings}, where \textit{aConstruct} appears and the operation to process \textit{cm} is \textit{TransformationOnLiteral}. Then, the process parametrizes \textit{TransformationOnLiteral} as follows: 1) \textit{sConstruct} and \textit{tConstruct} are defined by \texttt{map:sourceConcept} and \texttt{map:targetConce-\\pt}; 2) \textit{sTBox} is the target TBox of \textit{cm}'s map-dataset, defined by the property \texttt{map:sourceTBox}; 3) \textit{sABox} is the location of the source ABox defined by \texttt{map:so-\\urceLocation}; 3) \textit{propertyMappings} is the set of property-mappings defined under $cm$; 4) \textit{tABox} is the location of the target ABox defined by \texttt{map:target-\\Location}. A developer can also manually set the parameters. From the given inputs, this operation transforms (or directly returns) the \textit{sABox} triple objects  according to the expressions (defined through \texttt{map:source4TargetPropertyValue}) in \textit{propertyMappings} and stores the triples in \textit{tABox}. This operation first creates a SPARQL SELECT query based on the expressions defined in \textit{propertyMappings}, and then, on top of the SELECT query, it forms a SPARQL CONSTRUCT query to generate the transformed ABox for \textit{tConstruct}. 

\begin{xmpl}
\label{xmple:transformationonliteral}
Listing~\ref{list:transformationonliteral} (lines 16-19) shows the transformed instances after applying the operation \textit{TransformationOnLiteral(sConstruct, tConstruct, sTBox, sABox, PropertyMappings, tABox)}, where
\begin{enumerate}
\item \textit{sConstruct} = \textit{tConstruct}=\texttt{sub:Subsidy},
\item \textit{sTBox}="/map/subsidyTBox.ttl",
\item \textit{sABox}= source instances of \texttt{sub:Subsidy} (lines 47-50 in Listing~\ref{list:instance}),
\item \textit{propertyMappings} = lines 2-14 in Listing~\ref{list:transformationonliteral},
\item \textit{tABox}="/map/temp1.ttl".

\end{enumerate}

\begin{lstlisting}[caption={: Example of \textit{TransformationOnLiteral}.}, label=list:transformationonliteral,
   basicstyle=\tiny\tt, showstringspaces=false, frame=single,language=SQL,morekeywords={PREFIX, a}]
## Property-mappings input
:hasPayDate_Day rdf:type map:PropertyMapping;
 map:targetProperty sub:hasPayDate;
 map:sourceType4TargetPropertyValue map:Expression;
 map:source4TargetPropertyValue "CONCAT(STR(DAY(hasPayDate)),
 "/", STR(MONTH(hasPayDate)),"/",STR(YEAR(hasPayDate)))".
:Recipient_recipientId rdf:type map:PropertyMapping;
 map:targetProperty sub:hasRecipient;
 map:sourceType4TargetPropertyValue map:Property;
 map:source4TargetPropertyValue sub:hasRecipient.
:amountEuro_amountEuro rdf:type map:PropertyMapping;
 map:targetProperty sub:amountEuro;
 map:sourceType4TargetPropertyValue map:Expression;
map:source4TargetPropertyValue "xsd:integer(sub:amountEuro)".
## sub:Subsidy instances after TransformationOnLiteral.                 
subsidy:10615413 rdf:type sub:Subsidy;
 sub:hasRecipient recipient:291894;
 sub:amountEuro 8928;
 sub:hasPayData "25/25/2010".
\end{lstlisting}
\end{xmpl}
\paragraph{\textit{JoinTransformation(sConstruct, tConstruct, sTBox, tTBox, sABox, tABox, comProperty, propertyMappings)}} A TBox construct (a concept, a level, or a QB dataset) can be populated from multiple sources. Therefore, an operation is necessary to join and transform data coming from different sources. Two constructs of the same or different sources can only be joined if they share some common properties. This operation joins a source and a target constructs based on their common properties and produce the instances of the target construct by resolving the source expressions mapped to target properties. To join \textit{n} sources, an ETL process requires \textit{n-1} \textit{JoinTransformation} operations. 

If the ETL execution flow is generated automatically, the automatic ETL execution flow generation process first identifies the concept-mapping \textit{cm} from \textit{aMappings}, where \textit{aConstruct} appears and the operation to process $cm$ is \textit{JoinTransformation}. Then it parameterizes \textit{JoinTransformation} as follows: 1) \textit{sConstruct} and \textit{tConstruct} are defined by the \texttt{map:sourc-\\eConcept} and \texttt{map:targetConcept} properties; 2) \textit{sTBox} and \textit{tTBox} are the source and target TBox of \textit{cm}'s map-dataset, defined by \texttt{map:sourceTBox} and \texttt{map:targetTBox}; 3) \textit{sABox} and \textit{tABox} are defined by the \texttt{map:sourceLocation} and \texttt{map:target-\\Location} properties; 4) \textit{comProperty} is defined by \texttt{map:commonProperty}; 5) \textit{propertyMappings} is the set of property-mappings defined under \textit{cm}.

A developer can also manually set the parameters. Once it is parameterized, \textit{JoinTransformation} joins two constructs based on \textit{comProperty}, transforms their data based on the expressions (specified through \texttt{map:source4TargetPropertyValue}) defined in \textit{propertyMappings}, and updates \textit{tABox} based on the join result. It creates a SPARQL SELECT query joining two constructs using either \texttt{AND} or \texttt{OPT} features, and on top of that query, it forms a SPARQL CONSTRUCT query to generate the transformed \textit{tABox}. 
 
\begin{xmpl}
\label{xmple:joinTransformation}
The recipients in \texttt{sdw:Recipient} need to be enriched with their company information available in the Danish Business dataset. Therefore, a join operation is necessary between \texttt{sub:Recipient} and \texttt{bus:Company}. The concept-mapping of this join is described in Listing~\ref{list:mapping} at lines 19-37. They are joined by two concept properties: recipient names and their addresses (lines 31, 34-37). We join and transform \texttt{bus:Company} and  \texttt{sub:Recipient} using \textit{JoinTransformation(sConstruct, tConstruct, sTBox, tTBox, sABox, tABox, comProperty, propertyMappings)}, where
\begin{enumerate}
\item \textit{sConstruct}= \texttt{bus:Company},
\item \textit{tConstruct}= \texttt{sub:Recipient},
\item \textit{sTBox}= "/map/businessTBox.ttl",
\item \textit{tTBox}= "/map/subsidyTBox.ttl",
\item \textit{sABox}= source instances of \texttt{bus:Company} (lines 13-29 in Listing~\ref{list:instance}),
\item \textit{tABox}= source instances of \texttt{sub:Recipient} (lines 40-45 in Listing~\ref{list:instance}),
\item \textit{comProperty} = lines 31, 34-37 in Listing~\ref{list:mapping},
\item \textit{propertyMappings} = lines 73-96 in Listing~\ref{list:mapping}.
\end{enumerate}
Listing~\ref{list:join} shows the output of the \textit{joinTransformation} operation.
\begin{lstlisting}[caption={: Example of \textit{JoinTransformation}.}, label=list:join,
   basicstyle=\tiny\tt, showstringspaces=false, frame=single,language=SQL,morekeywords={PREFIX, a}]
## Example of sub:Recipient instances.
recipient:291894 rdf:type sub:Recipient;
   sub:name "Kristian Kristensen";
   sub:cityId "Vemb";
   sub:companyId company:10165164;
   sub:businessType businessType:Interessentskab.
                 
\end{lstlisting}
\end{xmpl} 

\paragraph{\textit{LevelMemberGenerator(sConstruct, level, sTBox, sABox, tTBox, iriValue, iriGraph, propertyMappings, tABox)}} In QB4OLAP, dimensional data are physically stored in levels.  A level member, in an SDW, is described by a unique IRI and its semantically linked properties (i.e., level attributes and rollup properties). This operation generates data for a dimension schema defined in Definition~\ref{def:dimension}. 

If the ETL execution flow is generated automatically, the automatic process first identifies the concept-mapping \textit{cm} from \textit{aMappings}, where \textit{aConstruct} appears and the operation to process \textit{cm} is \textit{LevelMemberGenerator}. Then it parameterizes \textit{LevelMemberGenerator} as follows: 1) \textit{sConstruct} is the source construct defined by \texttt{map:sourceConcept}; 2) \textit{level} is the target level\footnote{A level is termed as a level property in QB4OLAP, therefore, throughout this paper, we use both the term ``level" and ``level property"  interchangeably.} defined by the \texttt{map:targetConcept} property; 3) \textit{sTBox} and \textit{tTBox} are the source and target TBoxes of \textit{cm}'s map dataset, defined by the properties \texttt{map:sourceTBox} and \texttt{map:targetTBox}; 4) \textit{sABox} is the source ABox defined by the property \texttt{map:sourceLocation}; 5) \textit{iriValue} is a rule\footnote{A rule can be either a source property, an expression or incremental, as described in Section~\ref{para:s2t}.} to create IRIs for the level members and it is defined defined by the \texttt{map:TargetInstanceIriValue} property; 6) \textit{iriGraph} is the IRI graph\footnote{The IRI graph is an RDF graph that keeps a triple for each resource in the SDW with their corresponding source IRI.} within which to look up IRIs, given by the developer in the automatic ETL flow generation process; 7) \textit{propertyMappings} is the set of property-mappings defined under \textit{cm}; 8) \textit{tABox} is the target ABox location defined by \texttt{map:targetLocation}. 

A developer can also manually set the paramenters. Once it is parameterized, \textit{LevelMemberGenerator} operation generates QB4OLAP-compliant triples for the level members of \textit{level} based on the semantics encoded in \textit{tTBox} and stores them in \textit{tABox}. 

\begin{xmpl}
Listing~\ref{list:mapping} shows a concept-mapping (lines 40-54) describing how
to populate \texttt{sdw:Recipient} from \texttt{sub:Recipient}. Listing~\ref{list:levelmembergenerator} shows the level
member created by the \textit{LevelMemberGenerator(level, tTBox, sABox, iriValue, iriGraph,  propertyMappings, tABox)} operation, where

\begin{enumerate}
\item \textit{sConstruct}= \texttt{sub:Recipient},
\item \textit{level}= \texttt{sdw:Recipient},
\item \textit{sTBox}= "/map/subsidyTBox.ttl",
\item \textit{sABox}= "/map/subsidy.ttl", shown in Example~\ref{xmple:joinTransformation},
\item \textit{tTBox} = "/map/subsidyMDTBox.ttl",
\item \textit{iriValue} = \texttt{sub:recipientID},
\item \textit{iriGraph} = "/map/provGraph.nt", 
\item \textit{propertyMappings}= lines 98-115 in Listing~\ref{list:mapping},
\item \textit{tABox}="/map/temp.ttl".

\end{enumerate} 

\begin{lstlisting}[caption={: Example of \textit{LevelMemberGenerator}.}, label=list:levelmembergenerator,
   basicstyle=\tiny\tt, showstringspaces=false, frame=single,language=SQL,morekeywords={PREFIX, a}]
PREFIX sdw: <http://extbi.lab.aau.dk/ontology/sdw/>
PREFIX recipient: <http://extbi.lab.aau.dk/ontology
                                  /sdw/Recipient#>
PREFIX company: <http://extbi.lab.aau.dk/ontology
                                    /sdw/Company#>
PREFIX city: <http://extbi.lab.aau.dk/ontology
                                      /sdw/City#>
## Example of a recipient level member.
recipient:291894 rdf:type qb4o:LevelMember;
                 qb4o:memberOf sdw:Recipient.
                 sdw:name "Kristian Kristensen";
                 sdw:inCity city:Vemb;
                 sdw:hasCompany company:10165164. 
\end{lstlisting}
\end{xmpl}
\paragraph{\textit{ObservationGenerator(sConstruct, dataset, sTBox, sABox, tTBox, iriValue, iriGraph, propertyMappings, tABox)}} In QB4OLAP, an observation represents a fact. A fact is uniquely identified by an IRI, which is defined by a combination of several members from different levels and contains values for different measure properties. This operation generates data for a cube schema defined in Definition~\ref{def:cube}. 

If the ETL execution flow is generated automatically, the way used by the automatic ETL execution flow generation process to extract values for the parameters of \textit{ObservationGenerator} from \textit{aMappings} is analogous to \textit{LevelMemberGenerator}. Developers can also manually set the parameters. Once it is parameterized, the operation generates QB4OLAP-compliant triples for observations of the QB dataset\textit{dataset} based on the semantics encoded in \textit{tTBox} and stores them in \textit{tABox}.

\begin{xmpl} Listing~\ref{list:observationgenerator} (lines 21-25) shows a QB4OL-AP-compliant observation create by the \textit{ObservationGenerator(sConstruct, dataset, sTBox, sABox, tTBox, iriValue, iriGraph, propertyMappings, tABox)} operation, where 
\begin{enumerate}
\item \textit{sConstruct}=\texttt{sub:Subsidy},
\item \textit{dataset}= \texttt{sdw:SubsidyMD},
\item \textit{sTBox}="/map/subsidyTBox.ttl"
\item \textit{sABox}= "/map/subsidy.ttl", shown in Example~\ref{xmple:transformationonliteral},
\item \textit{tTBox} = "/map/subsidyMDTBox.ttl", shown in Listing~\ref{list:tboxdefinition},
\item \textit{iriValue} = "Incremental",
\item \textit{iriGraph} = "/map/provGraph.nt", 
\item \textit{propertyMappings}= lines 8-19 in Listing~\ref{list:observationgenerator},
\item \textit{tABox}= "/map/temp2.ttl".
\end{enumerate} 
\begin{lstlisting}[caption={: Example of \textit{ObservationGenerator}. }, label=list:observationgenerator,
   basicstyle=\tiny\tt, showstringspaces=false, frame=single,language=SQL,morekeywords={PREFIX, a}]
PREFIX sdw: <http://extbi.lab.aau.dk/ontology/sdw/>
PREFIX subsidy: <http://extbi.lab.aau.dk/ontology
                                   /sdw/Subsidy#>
PREFIX recipient: <http://extbi.lab.aau.dk/ontology
                                    /sdw/Recipient#>
PREFIX day: <http://extbi.lab.aau.dk/ontology/sdw/Day#>  
## Property-Mappings
:recipientId_Recipient rdf:type map:PropertyMapping;
  map:targetProperty sdw:Recipient;
  map:sourceType4TargetPropertyValue map:Property;
  map:source4TargetPropertyValue sub:hasRecipient.
:hasPayDate_Day rdf:type map:PropertyMapping;
  map:targetProperty sdw:Day;
  map:sourceType4TargetPropertyValue map:Property;
  map:source4TargetPropertyValue sub:hasPaydate.
:amountEuro_amountEuro rdf:type map:PropertyMapping;
  map:targetProperty sdw:amountEuro;
  map:sourceType4TargetPropertyValue map:Property;
  map:source4TargetPropertyValue sub:amountEuro.	 
## Example of observations              
subsidy:_01 rdf:type qb4o:Observation;
  inDataset sdw:SubsidyMD;
  sdw:hasRecipient recipient:291894;
  sdw:amountEuro "8928.00";
  sdw:hasPayData day:25/25/2010.
\end{lstlisting}
\end{xmpl}

\paragraph{\textit{ChangedDataCapture(nABox, oABox, flag)}}In a real-world scenario changes occur in a semantic source both at the schema and instance level. Therefore, an SDW needs to take action based on the changed schema and instances. The adaption of the SDW TBox with the changes of source schemas is an analytical task and requires the involvement of domain experts, therefore, it is out of the scope of this paper. Here, only the changes at the instance level are considered. 

If the ETL execution flow is generated automatically, the automatic process first identifies the concept-mapping \textit{cm} from \textit{aMappings}, where \textit{aConstruct} appears and the operation to process \textit{cm} is \textit{ChangedDataCapture}. Then, it takes \texttt{map:sourceLocation} and \texttt{map:targetLocation} for \textit{nABox} (new dimensional instances in a source) and \textit{oABox} (old dimensional instances in a source), respectively, to parameterize this operations. \textit{flag} depends on the next operation in the operation sequence.

Developers can also manually set the parameters. From the given inputs, \textit{ChangedDataCapture} outputs either 1) a set of new instances (in the case of SDW evolution, i.e., \textit{flag=0}) or 2) a set of updated triples ---the existing triples changed over time--- (in the case of SDW update, i.e., \textit{flag=1}) and overwrites \textit{oABox}. This is done by means of the set difference operation. This operation must then be connected to either \textit{LevelMemberGenerator} to create the new level members or \textit{updateLevel} (described below) to reflect the changes in the existing level members.

\begin{xmpl} Suppose Listing~\ref{list:join} is the old ABox of \texttt{sub:Recipient} and the new ABox is at lines 2-10 in Listing~\ref{list:cdc}. This operation outputs either 1) the new instance set (in this case, lines 12-15 in the listing) or 2) the updated triples (in this case, line 17).

\begin{lstlisting}[caption={: Example of \textit{ChangedDataCapture}.}, label=list:cdc,
   basicstyle=\tiny\tt, showstringspaces=false, frame=single,language=SQL,morekeywords={PREFIX, a}]
## New snapshot of sub:Recipient instances.
recipient:291894 rdf:type sub:Recipient;
                 sub:name "Kristian Jensen";
                 sub:cityId "Vemb";
                 sub:companyId company:10165164;
                 businessType:Interessentskab.
recipient:301894 rdf:type sub:Recipient;
                 sub:name "Jack";
                 sub:cityId "Aalborg";
                 sub:companyId company:100000.
## New instances to be inserted
recipient:301894 rdf:type sub:Recipient;
                 sub:name "Jack";
                 sub:cityId "Aalborg";
                 sub:companyId company:100000.            
## Update triples
recipient:291894 sub:name "Kristian Jensen".                 
                 
\end{lstlisting}

\end{xmpl}
\paragraph{\textit{UpdateLevel(level, updatedTriples, sABox tTBox, tABox, propertyMappings, iriGraph)}} Based on the triples updated in the source ABox \textit{sABox} for the level \textit{level} (generated by \textit{ChangedDataCapture}), this operation updates the target ABox \textit{tABox} to reflect the changes in the SDW according to the semantics encoded in the target TBox \textit{tTBox} and \textit{level}  property-mappings \textit{propertyMappings}. Here, we address three update types (Type1-update, Type2-update, and Type3-update), defined by Ralph Kimball in~\cite{kimball1996data} for a traditional DW, in an SDW environment. The update types are defined in \textit{tTBox} for each level attribute of \textit{level} (as discussed in Section~\ref{sec:sdw}). As we consider only instance level updates, only the objects of the source updated triples are updated. To reflect a source updated triple in \textit{level}, the level member using the triple to describe itself, will be updated.  In short, the level member is updated in the following three ways: 1) A Type1-update simply overwrites the old object with the current object. 2) A Type2-update creates a new version for the level member (i.e., it keeps the previous version and creates a new updated one). It adds the validity interval for both versions. Further, if the level member is a member of an upper level in the hierarchy of a dimension, the changes are propagated downward in the hierarchy, too. 3) A Type3-update overwrites the old object with the new one. Besides, it adds an additional triple for each changed object to keep the old object. The subject of the additional triple is the instance IRI, the object of the triple is the old object, and the predicate is \textit{concat(oldPredicate, "oldValue")}.

If the ETL execution flow is generated automatically, this operation first identifies the concept-mapping \textit{cm} from \textit{aMappings}, where \textit{aConstruct} appears and the operation to process \textit{cm} is \textit{UpdateLevel}. Then it parameterizes \textit{LevelMemberGenerator} as follows: 1) \textit{level} is defined by the \texttt{map:targetConcept}; 2) sABox is old source data; 3) \textit{updatedTriples} is the source location defined by \texttt{map:sourceLocation}; 4) \textit{tTBox} is the target TBox of \textit{cm}'s map-dataset, defined by the property \texttt{map:targetTBox}; 5) \textit{tABox} is defined by \texttt{map:targetLocation}; 6) \textit{propertyMappings} is the set of property-mappings defined under \textit{cm}; 7) \textit{iriGraph} is given by developer in the automatic ETL flow generation process.

\begin{xmpl}
Listing~\ref{list:update} describes how different types of updates work by considering two members of the \texttt{sdw:Recipient} level (lines 2-11). As the name of the second member (lines 7-11) is changed to ``Kristian Jensen" from ``Kristian Kristensen", as found in Listing~\ref{list:cdc}. A Type1-update simply overwrites the existing name (line 20). A Type2-update creates a new version (lines 39-46). Both old and new versions contain validity interval (lines 35-37) and (lines 44-46). A Type3-Update overwrites the old name (line 56) and adds a new triple to keep the old name (line 57). 
\begin{lstlisting}[caption={: Example of different types of updates.}, label=list:update,
   basicstyle=\tiny\tt, showstringspaces=false, frame=single,language=SQL,morekeywords={PREFIX, a}]
## sdw:Recipient Level
recipient:762921 rdf:type qb4o:LevelMember;
                 qb4o:member sdw:Recipient;
                 sdw:name "R. Nielsen";
                 sdw:cityId city:Lokken;
                 sdw:hasCompany company:10165164.            
recipient:291894 rdf:type qb4o:LevelMember;
                 qb4o:memberOf sdw:Recipient.
                 sdw:name "Kristian Kristensen";
                 sdw:cityId city:Vemb;
                 sdw:hasCompany company:10165164. 
## After type1 update
recipient:762921 rdf:type qb4o:LevelMember;
                 qb4o:member sdw:Recipient;
                 sdw:name "R. Nielsen";
                 sdw:cityId city:Lokken;
                 sdw:hasCompany company:10165164.            
recipient:291894 rdf:type qb4o:LevelMember;
                 qb4o:memberOf sdw:Recipient.
                 sdw:name "Kristian Jensen";
                 sdw:cityId city:Vemb;
                 sdw:hasCompany company:10165164. 
##After type2 update
recipient:762921 rdf:type qb4o:LevelMember;
                 qb4o:member sdw:Recipient;
                 sdw:name "R. Nielsen";
                 sdw:cityId city:Lokken;
                 sdw:hasCompany company:10165164.            

recipient:291894 rdf:type qb4o:LevelMember;
                 qb4o:memberOf sdw:Recipient.
                 sdw:name "Kristian Kristensen";
                 sdw:cityId city:Vemb;
                 sdw:hasCompany company:10165164.
                 sdw:fromDate ``0000-00-00";
                 sdw:toDate ``2017-09-25";
                 sdw:status ``Expired".

recipient:291894\_2017\_09\_26 rdf:type qb4o:LevelMember;
                            qb4o:memberOf sdw:Recipient.
                            sdw:name "Kristian Jensen";
                            sdw:cityId city:Vemb;
                            sdw:hasCompany company:10165164; 
                            sdw:fromDate ``2017-09-26";
                            sdw:toDate ``9999-12-31";
                            sdw:status ``Current".
## After type3 update
recipient:762921 rdf:type qb4o:LevelMember;
                qb4o:member sdw:Recipient;
                sdw:name "R. Nielsen";
                sdw:cityId city:Lokken;
                sdw:hasCompany company:10165164.            

recipient:291894 rdf:type qb4o:LevelMember;
                 qb4o:memberOf sdw:Recipient.
                 sdw:name "Kristian Jensen";
                 sdw:name_oldValue "Kristian Kristensen";
                 sdw:cityId city:Vemb;
                 sdw:hasCompany company:10165164.

\end{lstlisting}    
\end{xmpl}

Besides the transformation operations discussed above, we define two additional transformation operations that cannot be run by the automatic ETL dataflows generation process. 
\paragraph{\textit{ExternalLinking (sABox, externalSource)}} This operation links the resources of \textit{sABox}  with the resources of an external source \textit{externalSource}. \textit{externalSource} can either be a SPARQL endpoint or an API.  For each resource $inRes \in$ \textit{sABox}, this operation extracts top k matching external resources either 1) submitting a query to \textit{externalSource} or 2) by sending a web service request embedding \textit{inRes} through the API (e.g., DBpedia lookup API). To find a potential link for each external resource \textit{exRes}, the Jaccard Similarity of the semantic bags of \textit{inRes} and \textit{exRes} is computed. The semantic bag of a resource consists of triples describing the resource~\cite{seddiqui2011augmentation,nath2012resolving,nath2014efficient}. The pair of the internal and external resources is considered as a match if the Jaccard Similarity exceeds a certain threshold. A triple with the \texttt{owl:sameAs} property is created to materialize the link in \textit{sABox} for each pair of matched internal and external resources.
    
\paragraph{\textit{MaterializeInference(ABox, TBox)}} This operation infers new information that has not been explicitly stated in an SDW. It analyzes the semantics encoded into the SDW and enriches the SDW with the inferred triples. A subset of the OWL 2 RL/RDF rules, which encodes part of the RDF-Based Semantics of OWL 2~\cite{polleres2013rdfs}, are considered here. The reasoning rules can be applied over the TBox \textit{TBox} and ABox \textit{ABox} separately, and then together. Finally, the resulting inference is asserted in the form of triples, in the same spirit as how SPARQL deals with inference.

\subsection{Load}
\paragraph{\textit{Loader(tripleSet, tsPath)}} An SDW is represented in the form of RDF triples and the triples are stored in a triple store (e.g., Jena TDB). Given a set of RDF triples \textit{triplesSet} and the path of a triple store \textit{tsPath}, this operation loads \textit{triplesSet} in the triple store.

If the ETL execution flow is generated automatically, this operation first identifies the concept-mapping \textit{cm} from \textit{aMappings}, where \textit{aConstruct} appears and the operation to process \textit{cm} is \textit{Loader}. Then, it takes values of \texttt{map:sourceLocation} and \texttt{map:targ-\\etLocation} for the parameters \textit{tripleSet} and \textit{tsPath}.

\section{Automatic ETL Execution Flow Generation}\label{sec:aeefg}
We can characterize ETL flows at the Definition Layer by means of the source-to-target mapping file; therefore, the ETL data flows in the Execution Layer can be generated automatically. This way, we can guarantee that the created ETL data flows are sound and relieve the developer from creating the ETL data flows manually. Note that this is possible because the Mediatory Constructs (see Figure~\ref{fig:FSA}) contain all the required metadata to automate the process.

\begin{algorithm}[h!]
\KwIn{TargetConstruct \textit{x}, MappingFile \textit{G}, IRIGraph $G_{IRI}$}
\KwOut{A set of ETL flows \textit{E}}
\Begin{

\nl $node_{target} \leftarrow$ \textsc{find}($x,G$) 

\nl $?c_{maps} \leftarrow $\textsc{findConMap}($(node_{target},G$)

\tcc{In $G:$ $\; node_{target} \xleftarrow{\texttt{map:targetConcept}}?c_{maps}$ ; $|?c_{maps}|\geq 0$}

\nl $E\leftarrow$ \textsc{createSet}()

\nl \If {$(?c_{maps}\ne \emptyset)$}
{

\nl \ForEach{$c \in ?c_{maps}$}{

\nl $s_c \leftarrow$ \textsc{createStack}()

\nl  $s_c \leftarrow$ \textsc{CreateAFlow}($c, s_c, G, G_{IRI}$)

\nl $E$.\textsc{add}($s_c$)
}
}
\nl \textbf{return} $E$

}
    \caption{\textsc{CreateETL} \label{alg:createETL}}

\end{algorithm}

\begin{algorithm}[h!]
\DontPrintSemicolon
\KwIn{ConceptMapping $c$, Stack $s_c$, MappingFile $G$, IRIGraph $G_{IRI}$}
\KwOut{Stack $s_c$}
\Begin{

\nl $ops \leftarrow$ \textsc{findOperations}($c, G$) 

\tcc{In $G:$ $\; c\xrightarrow{\texttt{map:operation}}ops$}





\nl $s_c$. \textsc{push}(\textsc{Parameterize}($ops,c, G, G_{IRI}$)) 

\tcc{Push parameterized operations in $s_c$}

\nl $scon \leftarrow$ \textsc{findSourceConcept}($c,G$) 

\tcc{In $G:$ $\; c\xrightarrow{\texttt{map:sourceConcept}}scon$}

\nl $ ?scon_{map} \leftarrow$ \textsc{findConMap}($scon,G$) 

\tcc{In $G:$ $\; scon \xleftarrow{\texttt{map:targetConcept}}?scon_{map}$; $|?scon_{map}|\leq 1$}

\nl \If {$(|?scon_{map}|= 1)$}
{
\nl \textsc{CreateAFlow}($nc\in ?scon_{map}, s_c, G$) 

\tcc{recursive call with $nc$}
}

\nl $s_c$.\textsc{push}(StartOp) 

\tcc{Push the ETL start operation to $s_c$}

\nl \textbf{return} $s_c$
}
    \caption{\textsc{CreateAFlow}  \label{alg:createAFlow}}

\end{algorithm}

\begin{algorithm}[h]
\KwIn{Seq \textit{ops}, ConceptMapping $cm$, MappingFile $G$, IRIGraph $G_{IRI}$}
\KwOut{Stack, $s_{op}$}
\Begin{
\nl $s_{op} \leftarrow$ \textsc{createStack}()

\nl \For {($i=1$ to \textsc{length}($ops$))}
{

\nl \If {$(i=1)$}
{
\nl \textsc{parameterizeOp}($op[i], G$, \textsc{loc}($cm$), $G_{IRI}$)

\nl $s_{op}$.\textsc{push}($op[i]$)

}
\nl \textsc{parameterizeOp}($op[i], G,$ \textsc{outputPath}($op[i-1]$), $G_{IRI}$)

\nl $s_{op}$.\textsc{push}($op[i]$)

}
\nl \textbf{return} $s_{op}$
}
    \caption{\textsc{Parameterize}  \label{alg:parameterize}}

\end{algorithm}

Algorithm~\ref{alg:createETL} shows the steps required to create ETL data flows to populate a target construct (a level, a concept, or a dataset). As inputs, the algorithm takes a target construct $x$, the mapping file $G$, and the IRI graph $G_{IRI}$, and it outputs a set of ETL data flows $E$. At first, it locates $x$ in $G$ (line 1) and gets the concept-mappings where $node_{target}$ participates (line 2). As a (final) target construct can be populated from multiple sources, it can  be connected to multiple concept-mappings, and for each concept-mapping, it creates an ETL data flow by calling the function\footnote{Here, functions used in the algorithms are characterized by a pattern over $G$, shown at its side as comments in the corresponding algorithm.} \textsc{CreateAFlow} (lines 5-8). 
Algorithm~\ref{alg:createAFlow} generates an ETL data flow for a concept-mapping $c$ and recursively does so if the current concept-mapping source element is connected to another concept-mapping as, until it reaches a source element. 
Algorithm~\ref{alg:createAFlow} recursively calls itself and uses a stack to preserve the order of the partial ETL data flows created for each concept-mapping. Eventually, the stack contains the whole ETL data flow between the source and target schema.

Algorithm~\ref{alg:createAFlow} works as follows. The sequence of operations in $c$ is pushed to the stack after parameterizing it (lines 1-2). Algorithm~\ref{alg:parameterize} parameterizes each operation in the sequence, as described in Section~\ref{sec:executionLayer} and returns a stack of  parameterized operations. As inputs, it takes the operation sequence, the concept-mapping, the mapping file, and the IRI graph. For each operation, it uses the \textsc{parameterizeOp}($op$, $G$, \textsc{loc($cm$), $G_{IRI}$}) function to automatically parameterize $op$ from $G$ (as all required parameters are available in $G$) and push the parameterized operation in a stack (line 2-7). Note that, for the first operation in the sequence, the algorithm uses the source ABox location of the concept-mapping (line 4) as an input, whereas for the remaining operations, it uses the output of the previous operation as input ABox (line 6). Finally, Algorithm~\ref{alg:parameterize} returns the stack of parameterized operations.
\begin{figure*}[h!]
\centering
\includegraphics[scale=0.45]{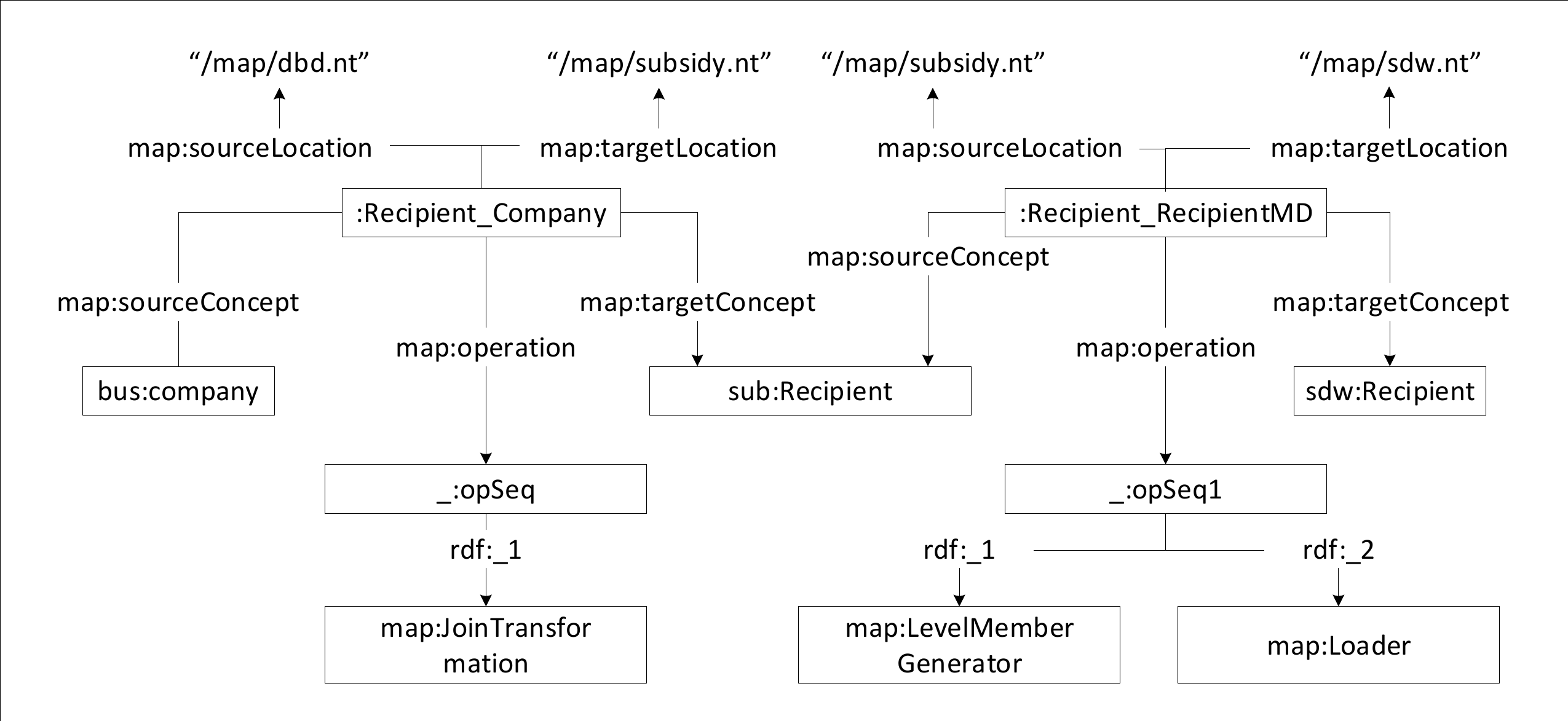}
\centering
\caption{The graph presentation of a part of Listing~\ref{list:mapping}.}
\label{fig:autoexample}
\end{figure*}

Then, Algorithm~\ref{alg:createAFlow} traverses to the adjacent concept-mapping of $c$ connected via $c$'s source concept (line 3-4). After that, the algorithm recursively calls itself for the adjacent concept-mapping (line 6). Note that, here, we set a restriction: except for the final target constructs, all the intermediate source and target concepts can be connected to at most one concept-mapping. This constraint is guaranteed when building the metadata in the \dl. Once there are no more intermediate concept-mappings, the algorithm pushes a dummy starting ETL operation (\textit{StartOp}) (line 7) to the stack and returns it. \textit{StartOp} can be considered as the root of the ETL data flows that starts the execution process. The stacks generated for each concept-mapping of $node_{target}$ are added to the output set $E$ (line 8 in Algorithm~\ref{alg:createETL}). The following section shows this process with an example of our use case. 
\subsection{Auto ETL Example} In this section, we show how to populate \texttt{sdw:Rec-\\ipient} level using \textsc{CreateETL} (Algorithm~\ref{alg:createETL}). As input, \textsc{CreateETL} takes the target construct \texttt{sdw:Recipient}, the mapping file Listing~\ref{list:mapping} and the location of IRI graph ``/map/provGraph.nt". Figure~\ref{fig:autoexample} presents a part (necessary to explain this process) of Listing~\ref{list:mapping} as an RDF graph. As soon as \texttt{sdw:Recipient} is found in the graph, the next task is to find the concept-mappings that are connected to \texttt{sdw:Recipient} through \texttt{map:targetConcept} (line 2). Here, $?c_{maps}$=\{\texttt{:Recipient\_Recipien-\\tMD}\}. For \texttt{:Recipient\_RecipientMD}, the algorithm creates an empty stack $s_c$ and calls \textsc{CreateAFlow} (\texttt{:Recipient\_RecipientMD}, $s_c$, Listing~\ref{list:mapping}) (Algorithm~\ref{alg:createAFlow}) at lines 6-7. 

\textsc{CreateAFlow} retrieves the sequence of operations (line 1) needed to process the concept-mapping, here: \textit{ops=(LevelMemberGenerator; Loader)} (see Figure~\ref{fig:autoexample}). Then, \textsc{CreateAFlow} parameterizes $ops$ by calling \textsc{Parameterize} (Algorithm~\ref{alg:parameterize}) and then pushes the parameterized operations to $s_c$ (line 2). \textsc{Parameterize} creates an empty stack $s_{op}$ (line 1) and for each operation in $ops$ it calls the \textsc{parameterize()} method to parameterize the operation using the concept-mapping information from Listing~\ref{list:mapping}, the source location of the concept-mapping, and the IRI graph, and then it pushes the parameterized operation in $s_{op}$ (lines 2-7). After the execution of the for loop in \textsc{Parameterize} (line 2-7), the value of the stack $s_{op}$ is (\textit{Loader}("/map/temp.nt", "/map/sdw"); \textit{LevelMemberGenerator}(\texttt{sub:Recipient},\texttt{sdw:Recipient},\\ "/map/subsidyTBox.ttl", "/map/subsidy.nt", "/map/subsidyMDTBox.ttl",  \texttt{sub:recipientID}, "/map/-\\provGraph.nt", propertyMappings\footnote{Here, lines 95-112 in Listing~\ref{list:mapping}.}, "/map/temp.nt")), which is returned to line 2 of \textsc{CreateAFlow}. Note that the output path of \textit{LevelMemberGenerator(..)}\footnote{(..) indicates that the operation is parameterized.} is used as the input source location of \textit{Loader(..)}.  \textsc{CreateAFlow} pushes the parameterized operations to $s_c$ (line 2), hence $s_c$= (\textit{LevelMemberGenerator(..); Loader(..)}). 

Then, \textsc{CreateAFlow} finds the source concept of \texttt{:Recipient\_RecipientMD} (line 3), which is \texttt{sub:Recipient}; retrieves the concept-mapping \texttt{:Recipient\_Company} of \texttt{sub:Recipient} from Listing~\ref{list:mapping} (line 4); and recursively calls itself for \texttt{:Recipient\_Company} (lines 5-6). The operation sequence of \texttt{:Recipient\_Company} (line 1) is (\textit{JoinTransformation}) (see Figure~\ref{fig:autoexample}), and the call of \textsc{Parameterize} at line 2 returns the parameterized \textit{JoinTransformation}(\texttt{bus:Company}, \texttt{Sub:Reicip-\\ient}, "/map/dbd.nt","/map/subsidy.nt", comProperty\footnote{lines 32,36-39 in Listing~\ref{list:mapping}.}, propertyMappings\footnote{lines 76-93 in Listing~\ref{list:mapping}.}) operation, which is pushed to the stack $s_c$, i.e., $s_c$= (\textit{JoinTransformation(...); LevelMemberGenerator(...); Loader(...)}). The source concept \texttt{bus:Company} is not connected to any other concept-mapping through the \texttt{map:targetConcept} property in the mapping file. Therefore, \textsc{CreateAFlow} skips lines 5 and 6. After that, it pushes the start operation (StartOp) in $s_c$, i.e., $s_c$= (\textit{StartOp, JoinTransformation(..); LevelMemberGenerator(..); Loader(..)}) and returns it to \textsc{CreateETL} (Algorithm~\ref{alg:createETL}) at line 7. Then, \textsc{CreateETL} adds it to the set of ETL data flows $E$ and returns it (line 9). Therefore, the ETL data flow to populate \texttt{sdw:Recipient} is \textit{StartOp $\Rightarrow$JoinTransformation(..) $\Rightarrow$LevelMemberGenerator(..) $\Rightarrow$Loader(..)}.


\section{Evaluation}
\label{sec:evaluation}
\begin{table*}[]
\centering
\caption{\label{tab:productivity1}Comparison among the productivity of \s, \scon, and \sauto\ for the \sd.}
\begin{adjustbox}{angle=90,width=1\textwidth}
\resizebox{\textwidth}{!}{%
\begin{tabular}{|l|l!{\vrule width 1.5pt}l|r|r!{\vrule width 1.5pt}l!{\vrule width 1.5pt}r|r|r|r!{\vrule width 1.5pt}r|r|r|r!{\vrule width 1.5pt}}
\hline
\multicolumn{1}{|r|}{\multirow{2}{*}{\textbf{Tool $\Rightarrow$}}}                                            & \multicolumn{1}{c|}{\multirow{2}{*}{\textbf{}}}             & \multicolumn{3}{c|}{\multirow{2}{*}{\textbf{\s}}}                                                                                                                                                        & \multirow{2}{*}{\textbf{}}                                                             & \multicolumn{4}{c|}{\textbf{\scon}}                                                        & \multicolumn{4}{c|}{\textbf{\sauto}}                                                       \\ \cline{7-14} 
\multicolumn{1}{|r|}{}                                                                          & \multicolumn{1}{c|}{}                                       & \multicolumn{3}{c|}{}                                                                                                                                                                                   &                                                                                        & \multirow{2}{*}{\textbf{NOUC}} & \multicolumn{3}{r|}{\textbf{Components of Each Concept}} & \multirow{2}{*}{\textbf{NOUC}} & \multicolumn{3}{r|}{\textbf{Components of Each Concept}} \\ \cline{1-6} \cline{8-10} \cline{12-14} 
\textbf{ETL Task  $\Downarrow$}                                                                               & \textbf{Sub Construct}                                      & \textbf{\begin{tabular}[c]{@{}l@{}}Procedures/\\ Data Structures\end{tabular}}                                                                                         & \textbf{NUEP} & \textbf{NOTC}  & \textbf{Task/Operation}                                                                &                                & \textbf{Clicks}  & \textbf{Selections}  & \textbf{NOTC}  &                                & \textbf{Clicks}  & \textbf{Selections}  & \textbf{NOTC}  \\ \noalign{\hrule height 1.5pt}
\multirow{10}{*}{\begin{tabular}[c]{@{}l@{}}TBox with\\ MD Semantics\end{tabular}}              & Cube Structure                                              & \multirow{10}{*}{\begin{tabular}[c]{@{}l@{}}Concept(),\\ Property(), \\ BlankNode(),\\ Namespaces(),\\ conceptPrope-\\ rtyBindings(),\\ createOntology()\end{tabular}} & 1             & 665            & \multirow{10}{*}{\begin{tabular}[c]{@{}l@{}}TargetTBox-\\ Definition\end{tabular}}     & 1                              & 9                & 9                    & 7              & 1                              & 9                & 9                    & 7              \\ \cline{2-2} \cline{4-5} \cline{7-14} 
                                                                                                & Cube Dataset                                                &                                                                                                                                                                        & 1             & 67             &                                                                                        & 1                              & 2                & 2                    & 18             & 1                              & 2                & 2                    & 18             \\ \cline{2-2} \cline{4-5} \cline{7-14} 
                                                                                                & Dimension                                                   &                                                                                                                                                                        & 2             & 245            &                                                                                        & 2                              & 3                & 4                    & 12             & 2                              & 3                & 4                    & 12             \\ \cline{2-2} \cline{4-5} \cline{7-14} 
                                                                                                & Hierarchy                                                   &                                                                                                                                                                        & 7             & 303            &                                                                                        & 7                              & 4                & 6                    & 18             & 7                              & 4                & 6                    & 18             \\ \cline{2-2} \cline{4-5} \cline{7-14} 
                                                                                                & Level                                                       &                                                                                                                                                                        & 16            & 219            &                                                                                        & 16                             & 4                & 5                    & 17             & 16                             & 4                & 5                    & 17             \\ \cline{2-2} \cline{4-5} \cline{7-14} 
                                                                                                & Level Attribute                                             &                                                                                                                                                                        & 38            & 228            &                                                                                        & 38                             & 4                & 3                    & 15             & 38                             & 4                & 3                    & 15             \\ \cline{2-2} \cline{4-5} \cline{7-14} 
                                                                                                & Rollup Property                                             &                                                                                                                                                                        & 12            & 83             &                                                                                        & 12                             & 2                & 1                    & 7              & 12                             & 2                & 1                    & 7              \\ \cline{2-2} \cline{4-5} \cline{7-14} 
                                                                                                & \begin{tabular}[c]{@{}l@{}}Measure \\ Property\end{tabular} &                                                                                                                                                                        & 1             & 82             &                                                                                        & 1                              & 4                & 3                    & 25             & 1                              & 4                & 3                    & 25             \\ \cline{2-2} \cline{4-5} \cline{7-14} 
                                                                                                & Hierarchy Step                                              &                                                                                                                                                                        & 12            & 315            &                                                                                        & 12                             & 6                & 6                    & 6              & 12                             & 6                & 6                    & 6              \\ \cline{2-2} \cline{4-5} \cline{7-14} 
                                                                                                & Prefix                                                      &                                                                                                                                                                        & 21            & 74             &                                                                                        & 21                             & 1                & 0                    & 48             & 21                             & 1                & 0                    & 48             \\ \noalign{\hrule height 1.5pt}
\multicolumn{3}{|l|}{\textbf{Subtotal (for the Target TBox)}}                                                                                                                                                                                                                                                                            & \textbf{1}    & \textbf{24509} & \textbf{}                                                                              & \textbf{101}                   & \textbf{382}     & \textbf{342}         & \textbf{1905}  & \textbf{101}                   & \textbf{382}     & \textbf{342}         & \textbf{1905}  \\ \noalign{\hrule height 1.5pt}
\multirow{3}{*}{\begin{tabular}[c]{@{}l@{}}Mapping \\ Generation\end{tabular}}                  & \begin{tabular}[c]{@{}l@{}}Mapping\\ Dataset\end{tabular}   & N/A                                                                                                                                                                     & 0             & 0              & \multirow{3}{*}{\begin{tabular}[c]{@{}l@{}}SourceTo-\\ Target-\\ Mapping\end{tabular}} & 5                              & 1                & 2                    & 6              & 3                              & 1                & 2                    & 6              \\ \cline{2-5} \cline{7-14} 
                                                                                                & Concept-mapping                                             & dict()                                                                                                                                                                 & 17            & 243            &                                                                                        & 23                             & 9                & 10                   & 0              & 18                             & 12               & 10                   & 15             \\ \cline{2-5} \cline{7-14} 
                                                                                                & Property-mapping                                            & N/A                                                                                                                                                                     & 0             & 0              &                                                                                        & 59                             & 2                & 2                    & 0              & 44                             & 2                & 2                    & 0              \\ \noalign{\hrule height 1.5pt}
\multicolumn{3}{|l|}{\textbf{subtotal (for the Mapping File)}}                                                                                                                                                                                                                                                                            & \textbf{1}    & \textbf{2052}  & \textbf{}                                                                              & \textbf{87}                    & \textbf{330}     & \textbf{358}         & \textbf{30}    & \textbf{65}                    & \textbf{253}     & \textbf{274}         & \textbf{473}   \\ \noalign{\hrule height 1.5pt}
\begin{tabular}[c]{@{}l@{}}Semantic Data \\ Extraction from \\ an RDF Dump File\end{tabular}    & -                                                           & query()                                                                                                                                                                & 17            & 40             & GraphExtractor                                                                         & 17                             & 5                & 0                    & 30             & 1                              & \multicolumn{3}{l|}{Auto Generation}                     \\ \hline
\begin{tabular}[c]{@{}l@{}}Semantic Data \\ Extraction through\\ a SPARQL Endpoint\end{tabular} & -                                                           & \begin{tabular}[c]{@{}l@{}}ExtractTriples-\\ FromEndpoint()\end{tabular}                                                                                               & 0             & 100            & GraphExtractor                                                                         & 0                              & 5                & 0                    & 30             & 0                              & \multicolumn{3}{l|}{Auto Generation}                     \\ \hline
Cleansing                                                                                       & -                                                           & Built-in Petl functions                                                                                                                                                & 15            & 80             & \begin{tabular}[c]{@{}l@{}}Transformation-\\ OnLiteral\end{tabular}                    & 5                              & 12               & 1                    & 0              & 1                              & \multicolumn{3}{l|}{Auto Generation}                     \\ \hline
Join                                                                                            & -                                                           & Built-in Petl functions                                                                                                                                                & 1             & 112            & \begin{tabular}[c]{@{}l@{}}Join-\\ Transformation\end{tabular}                         & 1                              & 15               & 1                    & 0              & 1                              & \multicolumn{3}{l|}{Auto Generation}                     \\ \hline
\begin{tabular}[c]{@{}l@{}}Level Member\\ Generation\end{tabular}                               & -                                                           & \begin{tabular}[c]{@{}l@{}}createDataTripleToFile(),\\ createDataTripleTo-\\ TripleStore()\end{tabular}                                                                & 16            & 75             & \begin{tabular}[c]{@{}l@{}}LevelMember-\\ Generator\end{tabular}                       & 16                             & 6                & 6                    & 0              & 1                              & \multicolumn{3}{l|}{Auto Generation}                     \\ \hline
\begin{tabular}[c]{@{}l@{}}Observation\\ Generation\end{tabular}                                & -                                                           & \begin{tabular}[c]{@{}l@{}}createDataTripleToFile(),\\ createDataTripleTo-\\ TripleStore()\end{tabular}                                                                & 1             & 75             & \begin{tabular}[c]{@{}l@{}}Observation-\\ Generator\end{tabular}                       & 1                              & 6                & 6                    & 0              & 1                              & \multicolumn{3}{l|}{Auto Generation}                     \\ \hline
\begin{tabular}[c]{@{}l@{}}Loading as \\ RDF Dump\end{tabular}                                  & -                                                           & \begin{tabular}[c]{@{}l@{}}insertTriplesIntoTDB(),\\ bulkLoadToTDB()\end{tabular}                                                                                      & 0             & 113            & Loader                                                                                 & 0                              & 1                & 2                    & 0              & 0                              & \multicolumn{3}{l|}{Auto Generation}                     \\ \hline
\begin{tabular}[c]{@{}l@{}}Loading to\\ a Triplestore\end{tabular}                             & -                                                           & \begin{tabular}[c]{@{}l@{}}insertTriplesIntoTDB(),\\ bulkLoadToTDB()\end{tabular}                                                                                      & 17            & 153            & Loader                                                                                 & 17                             & 1                & 2                    & 0              & 1                              & \multicolumn{3}{l|}{Auto Generation}                     \\ \hline
\begin{tabular}[c]{@{}l@{}}Auto ETL \\ Generation\end{tabular}                                  & -                                                           & -                                                                                                                                                                      & -             & -              & CreateETL                                                                              & 1                              & -                & -                    & -              & 1                              & 21               & 16                   & 0              \\ \noalign{\hrule height 1.5pt}
\multicolumn{3}{|l|}{\textbf{Total (for the Execution Layer Tasks)}}                                                                                                                                                                                                                                                                            & \textbf{1}    & \textbf{2807}  & \textbf{}                                                                              & \textbf{57}                    & \textbf{279}     & \textbf{142}         & \textbf{363}    & \textbf{58}                    & \textbf{21}     & \textbf{16}         & \textbf{0}   \\ \noalign{\hrule height 1.5pt}
\multicolumn{2}{|l|}{\textbf{\begin{tabular}[c]{@{}l@{}}Grand Total\\ (for the Whole  ETL Process)\end{tabular}}}                                                & \textbf{}    & \textbf{1}  &{\textbf{29358}}                                                                                                                                                                     & \multicolumn{2}{r|}{\textbf{245}}                                                                                       & \textbf{991}     & \textbf{826}         & \textbf{2298}  & \textbf{224}                   & \textbf{656}     & \textbf{632}         & \textbf{2378}  \\ \noalign{\hrule height 1.5pt}
\end{tabular}%
}
\end{adjustbox}
\end{table*}

We created a GUI-based prototype, named \textit{SETL\textsubscript{CON-}} \textit{\textsubscript{STRUCT}}~\cite{deb2020setlbi} based on the different high-level constructs described in Sections~\ref{sec:definitionLayer} and~\ref{sec:executionLayer}. We use Jena 3.4.0 to process, store, and query RDF data and Jena TDB as a triplestore. \scon\ is developed in Java 8. Like other traditional tools (e.g., PDI~\cite{casters2010pentaho}), \scon\ allows developers to create ETL flows by dragging, dropping, and connecting the ETL operations. The system is demonstrated in~\cite{deb2020setlbi}. On top of \scon, we implement the automatic ETL execution flow generation process discussed in Section~\ref{sec:aeefg}; we call it \sauto. The source code, examples, and developer manual for \scon\ and \sauto\ are available at \url{https://github.com/bi-setl/SETL}.

To evaluate \scon\ and \sauto, we create an MD SDW by integrating the Danish Business Dataset (DBD)~\cite{andersen2014publishing} and the Subsidy dataset (~\url{https://data.farmsubsidy.org/Old/}), described in Section~\ref{sec:usecase}. We choose this use case and these datasets for evaluation as there already exists an MD SDW, based on these datasets, that has been programatically created using \s\cite{nath2017setl}. 
Our evaluation focuses on three aspects: 1) productivity, i.e., to what extent \scon\ and \sauto\ ease the work of a developer when developing an ETL task to process semantic data, 2) development time, the time to develop an ETL process, and 3) performance, the time required to run the process. We run the experiments on a laptop with a 2.10 GHz Intel Core(TM) i7-4600U processor, 8 GB RAM, and  Windows 10.

\subsection{Productivity}
 \s\ requires Python knowledge to maintain and implement an ETL process. On the other hand, using \scon\ and \sauto,  a developer can create all the phases of an ETL process by interacting with a GUI. Therefore, they provide a higher level of abstraction to the developer that hides low-level details and requires no programming background. Table~\ref{tab:productivity1} summarizes the effort required by the developer to create different ETL tasks using \s, \scon, and \sauto. We measure the developer effort in terms of Number of Typed Characters (NOTC) for \s\ and in Number of Used Concepts (NOUC) for \scon\ and \sauto. Here, we define a concept as a GUI component of \scon\ that opens a new window to perform a specific action. A concept is composed of several clicks, selections, and typed characters. For each ETL task, Table~\ref{tab:productivity1} lists: 1) its sub construct, required procedures/data structures, number of the task used in the ETL process (NUEP), and NOTC for \s; 2) the required task/operation, NOUC, and  number of clicks, selections, and NOTC (for each concept) for \scon\ and \sauto. Note that NOTC depends on the user input. Here, we generalize it by using same user input for all systems. Therefore, the total NOTC is not the summation of the values of the respective column. 
 
To create a target TBox using \s, an ETL developer needs to use the following steps: 1) defining the TBox constructs by instantiating the \textit{Concept, Property,} and \textit{BlankNode} classes that play a meta modeling role for the constructs and 2) calling  \textit{conceptPropertyBinding()} and \textit{createTriples()}. Both procedures take the list of all concepts, properties, and blank nodes as parameters. The former one internally connects the constructs to each other, and the latter creates triples for the TBox. To create the TBox of our \sd\ using \s, we used 24,509 NOTC.
On the other hand, \scon\ and \sauto\ use the \textit{TargetTBoxDefinition} interface to create/edit a target TBox. To create the target TBox of our \sd\ in \scon\ and \sauto, we use 101 concepts that require only 382 clicks, 342 selections, and 1905 NOTC. Therefore, for creating a target TBox, \scon\ and \sauto\ reduce use 92\% fewer NOTC than \s.

To create source-to-target mappings, \s\ uses Python \textit{dictionaries} where the keys of each dictionary represent source constructs and values are the mapped target TBox constructs, and for creating the ETL process it uses 23 dictionaries, where the biggest dictionary used in the ETL process takes 253 NOTC. In total, \s\ uses 2052 NOTC for creating mappings for the whole ETL process. On the contrary, \scon\ and \sauto\ use a GUI. In total, \scon\ uses 87 concepts composed of 330 clicks, 358 selections, and 30 NOTC. Since \sauto\ internally creates the intermediate mappings, there is no need to create separate mapping datasets and concept-mappings for intermediate results. Thus, \sauto\ uses only 65 concepts requiring 253 clicks, 274 selections, and 473 NOTC. Therefore, \scon\ reduces NOTC of \s\ by 98\%. Although \sauto\ uses 22 less concepts than \scon, \scon\ reduces NOTC of \sauto\ by 93\%. This is because, in \sauto, we write the data extraction queries in the concept-mappings where in \scon\, we set the data extraction queries in the ETL operation level. 

To extract data from either an RDF local file or an SPARQL endpoint, \s\ uses the \textit{query()} procedure and the \textit{ExtractTriplesFromEndpoint()} class. On the other hand, \scon\ uses the \textit{GraphExtractor} operation. It uses 1 concept composed of 5 clicks and 20 NOTC for the local file and 1 concept with 5 clicks and 30 NOTC for the endpoint. \s\ uses different functions/procedures from the Petl Python library (integrated with \s) based on the cleansing requirements. In \scon, all data cleansing related tasks on  data sources are done using \textit{TransformationOnLiteral} (single source) and \textit{JoinTransformation} (for multi-source). \textit{TransformationOnLiteral} requires 12 clicks and 1 selection, and \textit{JoinTransformation} takes 15 clicks and 1 selection.

 To create a level member and observation, \s\ uses $createDataTripleToFile()$ and takes 125 NOTC. The procedure takes all the classes, properties, and blank nodes of a target TBox as input; therefore, the given TBox should be parsed for being used in the procedure. On the other hand, \scon\ uses the \textit{LevelMemberGenerator} and \textit{ObservationGenerator} operations, and each operation requires 1 concept, which takes 6 clicks and 6 selections.  \s\ provides procedures for either bulk or trickle loading to a file or an endpoint. Both procedures take 113 and 153 NOTC, respectively. For loading RDF triples either to a file or a triple store, \scon\ uses the \textit{Loader} operation, which needs 2 clicks and 2 selections. Therefore, \scon\ reduces NOTC for all transformation and loading tasks by 100\%. 

\sauto\ requires only a target TBox and a mapping file to generate ETL data flows through the \textit{CreateETL} interface, which takes only 1 concept composed of 21 clicks and 16 selections. Therefore, other \el\ tasks are automatically accomplished internally. In summary, \scon\ uses 92\% fewer NOTC than \s, and \sauto\ further reduces NOUC by another 25\%.

\subsection{Development Time} We compare the time used by \s, \\\scon, and \sauto\ to build the ETL processes for our use case SDW. As the three tools were developed within the same project scope and we master them, the first author conducted this test. We chose the running use case used in this paper and created a solution in each of the three tools and measured the development time. We used each tool twice to simulate the improvement we may obtain when we are knowledgeable about a given project. The first time it takes more time to analyze, think, and create the ETL process, and in the latter, we reduce the interaction time spent on analysis, typing, and clicking. Table~\ref{tab:develpedtime} shows the development time (in minutes) for main integration tasks used by the different systems to create the ETL processes. 
\begin{table}[h!]
\centering
\caption{\label{tab:develpedtime}ETL development time (in minutes) required for \s, \scon, and \sauto.}
\begin{adjustbox}{width=0.5\textwidth}
\begin{tabular}{|l|l|l|l|l|l|}
\hline
\textbf{Tool}       & \begin{tabular}[c]{@{}l@{}}\textbf{Iteration} \\\textbf{Number}\end{tabular}             & \begin{tabular}[c]{@{}l@{}}\textbf{Target} \\ \textbf{TBox}\\ \textbf{Definition}\end{tabular} & \begin{tabular}[c]{@{}l@{}}\textbf{Mapping}\\ \textbf{Generartion}\end{tabular} & \begin{tabular}[c]{@{}l@{}}\textbf{ETL}\\ \textbf{Design}\end{tabular} & \textbf{Total}\\ \hline
\s\   & 1 & 186                                                                & 51                                                    &85  &322                                                                             \\ \hline
\s\   & 2 & 146                                                                & 30                                                    & 65    &241                                                                          \\ \noalign{\hrule height 1.5pt}
\scon\      & 1 & 97                                                                & 46 & 35 & 178                                                                              \\ \hline
\scon\      & 2 & 58                                                                & 36 & 30                                                                             & 124                                                                              \\ \noalign{\hrule height 1.5pt}
\sauto\      & 1 & 97                                                                & 40 & 2                                                                             & 139                                                                              \\ \hline
\sauto\      & 2 & 58                                                                & 34 & 2                                                                             & 94                                                                              \\ \noalign{\hrule height 1.5pt}
\end{tabular}
\end{adjustbox}
\end{table}

\begin{figure*}[h]
\centering
\includegraphics[scale=0.6]{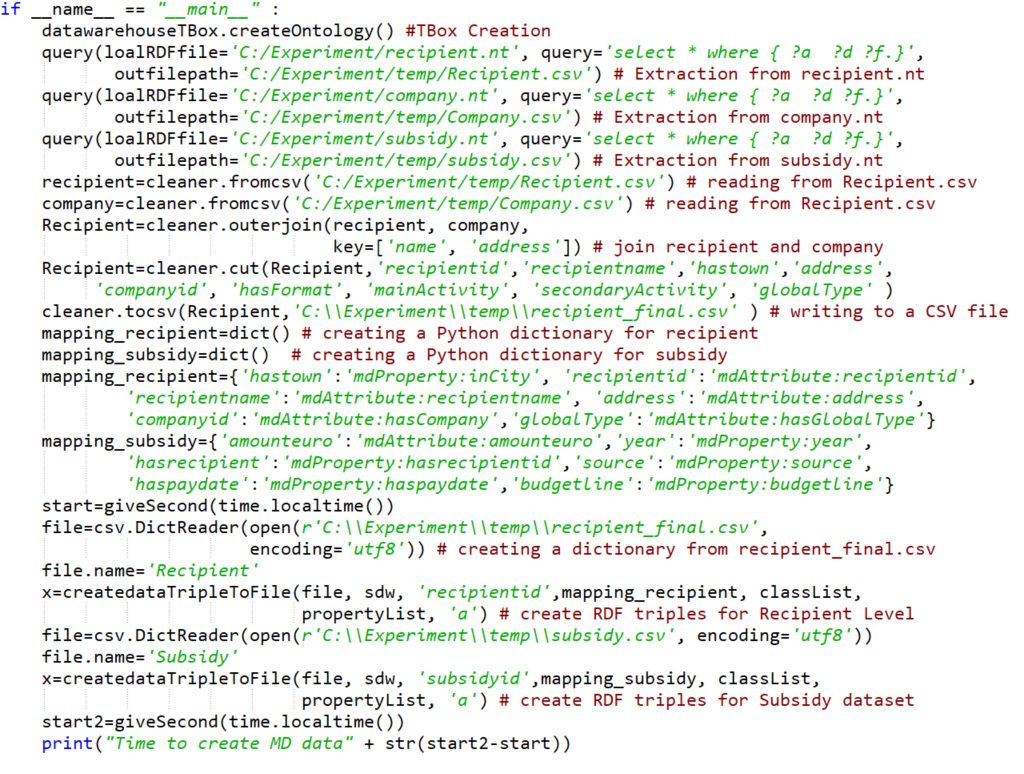}
\centering
\caption{The segment of the main method to populate  the \texttt{sdw:SubsidyMD} dataset and the \texttt{sdw:Recipient} using \s.}
\label{fig:programmable}
\end{figure*}

\begin{figure*}[h]
\centering
\includegraphics[scale=0.4]{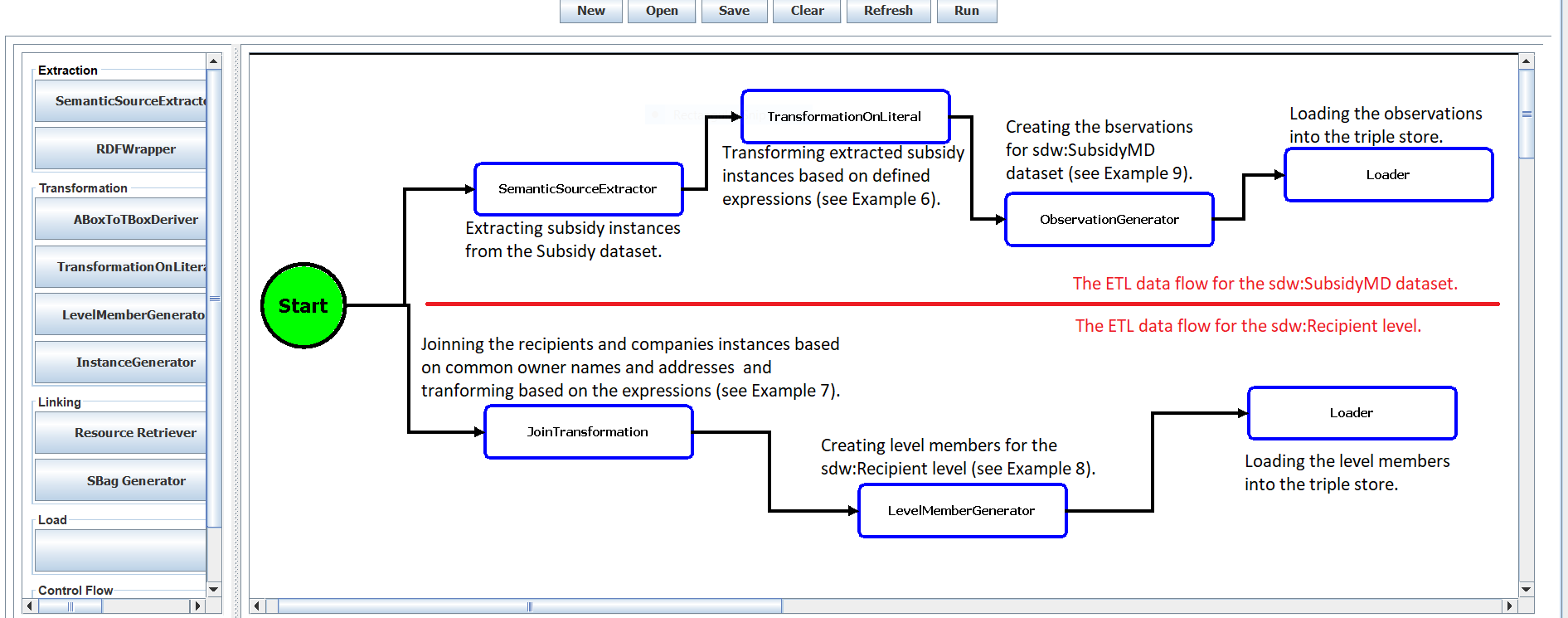}
\centering
\caption{The ETL flows to populate the \texttt{sdw:SubsidyMD} dataset and the \texttt{sdw:Recipient} level using \scon. }
\label{fig:concept}
\end{figure*}

\begin{figure}[h!]
\centering
\includegraphics[width=1\linewidth]{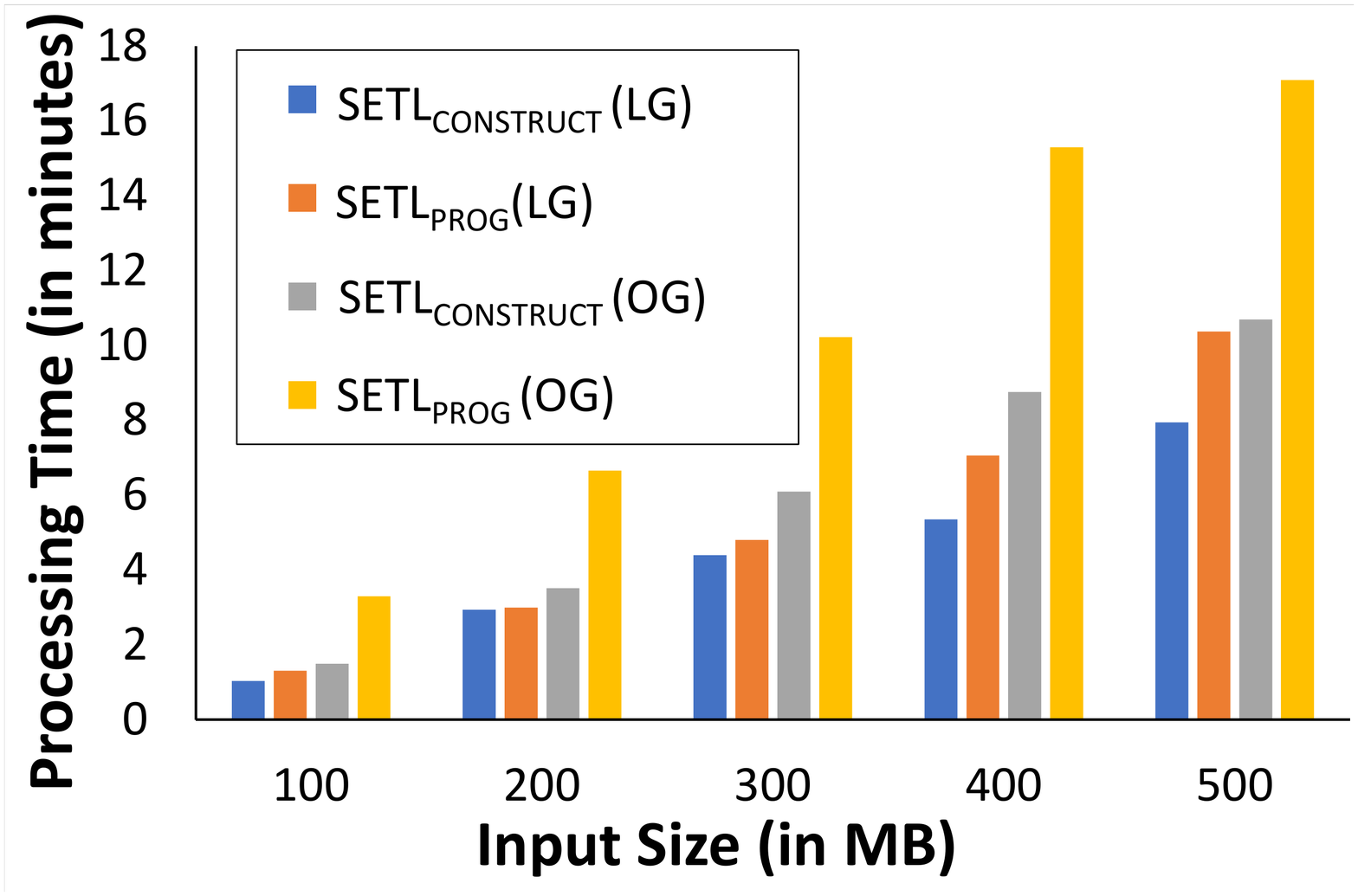}
  \caption{Comparison of \scon\ and \s\ for semantic transformation.Here, LG and OG stand for \textit{LevelMemberGenerator} and \textit{ObservationGenerator}.}
  \label{fig:comparison}
\end{figure}

\begin{figure}[h!]
\centering
  \includegraphics[width=1\linewidth]{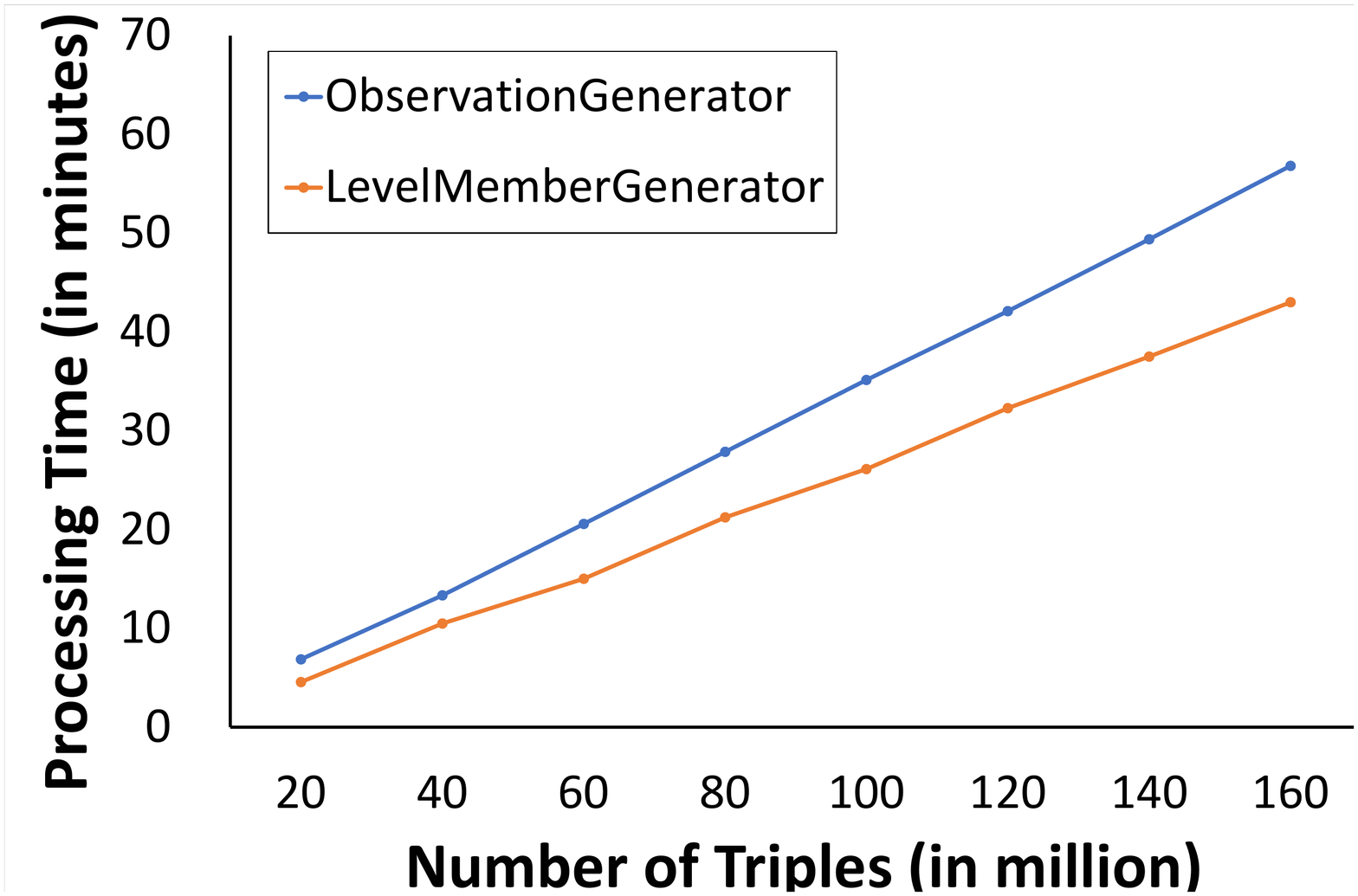}
  \caption{Scalability of \textit{LevelMemberGenerator} and \textit{ObservationGenerator}.}
  \label{fig:scalability}
\end{figure}

\s\ took twice as long as \scon\ and \sauto\ to develop a target TBox. In \s, to create a target TBox construct, for example a level \textit{l}, we need to instantiate the \textit{Concept()} class for \textit{l} and then add its different property values by calling different methods of \textit{Concept()}. \scon\ and \sauto\ create \textit{l} by typing the input or selecting from the suggested items. Thus, \scon\ and \sauto\ also reduce the risk of making mistakes in an error-prone task, such as creating an ETL. In \s, we typed all the source and target properties to create a mapping dictionary for each source and target concept pair. However, to create mappings, \scon\ and \sauto\ select different constructs from a source as well as a target TBox and only need to type when there are expressions and/or filter conditions of queries. Moreover, \sauto\ took less time than \scon\ in mapping generation because we did not need to create mappings for intermediate results. 

To create ETL data flows using \s, we had to write scripts for cleansing, extracting, transforming, and loading. \scon\ creates ETL flows using drag-and-drop options. Note that the mappings in \scon\ and \sauto\ use expressions to deal with cleansing and transforming related tasks; however, in \s\, we cleansed and transformed the data in the ETL design phase. Hence, \s\ took more time in designing ETL compared to \scon. On the other hand, \sauto\ creates the ETL data flows automatically from a given mapping file and the target TBox. Therefore, \sauto\ took only two minutes to create the flows.   In short, \s\ is a programmatic environment, while \scon\ and \sauto\ are drag and drop tools. We exemplify this fact by means of Figures~\ref{fig:programmable} and~\ref{fig:concept}, which showcase the creation of the ETL data flows for the \texttt{sdw:SubsidyMD} dataset and the \texttt{sdw:Recipient} level. To make it more readable and understandable, we add comments at the end of the lines of Figure~\ref{fig:programmable} and in each operation of Figure~\ref{fig:concept}. In summary, using \scon, the development time is cut in almost half (41\% less development time than \s); and using \sauto, it is cut by another 27\%. 

\begin{table}[h!]
\centering
\caption{\label{tab:performance}ETL execution time required for each sub-phase of the ETL processes created using \s\ and \scon.}
\resizebox{0.5\textwidth}{!}{%
\begin{tabular}{|l|l|l|l|l|l|}
\hline
\textbf{\begin{tabular}[c]{@{}l@{}}Performance \\ Metrics\end{tabular}}                          & \textbf{Systems} & \textbf{\begin{tabular}[c]{@{}l@{}}Extraction and\\ Traditional\\ Transformation\end{tabular}} & \textbf{\begin{tabular}[c]{@{}l@{}}Semantic\\ Transformation\end{tabular}} & \textbf{Loading}                                             & \textbf{\begin{tabular}[c]{@{}l@{}}Total\\ Processing\\ Time\end{tabular}} \\ \hline
\multirow{2}{*}{\textbf{\begin{tabular}[c]{@{}l@{}}Processing\\ Time (in minutes)\end{tabular}}} & \s                & 33                                                                                             & 17.86                                                                      & 21                                                           & 71.86                                                                      \\ \cline{2-6} 
                                                                                                 & \scon             & 43.05                                                                                          & 39.42                                                                      & 19                                                           & 101.47                                                                     \\ \hline
\multirow{2}{*}{\textbf{\begin{tabular}[c]{@{}l@{}}Input \\ Size\end{tabular}}}                  & \s                & \begin{tabular}[c]{@{}l@{}}6.2 GB (Jena TDB)\\ + 6.1 GB (N-Triples)\end{tabular}               & \begin{tabular}[c]{@{}l@{}}496 MB\\ (CSV)\end{tabular}                     & \begin{tabular}[c]{@{}l@{}}4.1 GB\\ (N-Triples)\end{tabular} & \multirow{4}{*}{}                                                          \\ \cline{2-5}
                                                                                                 & \scon             & \begin{tabular}[c]{@{}l@{}}6.2 GB (Jena TDB)\\ +6.1 GB (N-Triples)\end{tabular}                & \begin{tabular}[c]{@{}l@{}}6.270 GB\\ (N-Triples)\end{tabular}             & \begin{tabular}[c]{@{}l@{}}4.1 GB\\ (N-Triples)\end{tabular} &                                                                            \\ \cline{1-5}
\multirow{2}{*}{\textbf{\begin{tabular}[c]{@{}l@{}}Output\\ Size\end{tabular}}}                  & \s                & 490 MB (CSV)                                                                                   & \begin{tabular}[c]{@{}l@{}}4.1 GB\\ (N-Tripels)\end{tabular}               & \begin{tabular}[c]{@{}l@{}}3.7 GB\\ (Jena TDB)\end{tabular}  &                                                                            \\ \cline{2-5}
                                                                                                 & \scon             & \begin{tabular}[c]{@{}l@{}}6.270 GB\\ (N-Triples)\end{tabular}                                 & \begin{tabular}[c]{@{}l@{}}4.1 GB\\ (N-Triples)\end{tabular}               & \begin{tabular}[c]{@{}l@{}}3.7 GB\\ (Jena TDB)\end{tabular}  &                                                                            \\ \hline
\end{tabular}%
}
\end{table}

\subsection{Performance}
Since the ETL processes of \scon\ and \sauto\ are the same and only differ in the developer effort needed to create them, this section only compares the performance of \s\ and \scon. We do so by analyzing the time required to create the use case \sd\ by executing the respective ETL processes. To evaluate the performance of similar types of operations, we divide an ETL process into three sub-phases: extraction and traditional transformation, semantic transformation, as well as loading and discuss the time to complete each.

Table~\ref{tab:performance} shows the processing time (in minutes), input and output size of each sub-phase of the ETL processes created by \s\ and \scon. The input and output formats of each sub-phase are shown in parentheses. The extraction and traditional transformation sub-phases in both systems took more time than the other sub-phases. This is because they include time for 1) extracting data from large RDF files, 2) cleansing and filtering the noisy data from the DBD and Subsidy datasets, and 3) joining the DBD and Subsidy datasets. \scon\ took more time than \s\ because its \textit{TransformationOnLiteral} and \textit{JoinTransformation} operations use SPARQL queries to process the input file whereas \s\ uses the methods from the Petl Python library to cleanse the data extracted from the sources. 

\scon\ took more time during the semantic transformation than \s\ because \scon\ introduces two improvements over \s: 1) To guarantee the uniqueness of an IRI, before creating an IRI for a target TBox construct (e.g., a level member, an instance, an observation, or the value of an object or roll-up property), the operations of \scon\ look up the IRI provenance graph to check the availability of an existing IRI for that TBox construct. 2) As input, the operations of \scon\ take RDF (N-triples) files that are larger in size compared to the CSV files (see Table~\ref{tab:performance}), the input format of the semantic transformation process of \s. To ensure our claims, we run an experiment for measuring the performance of the semantic transformation procedures of \s\  and the operations of \scon\ by excluding the additional two features introduced in \scon\ operations (i.e., a \scon\ operation does not lookup the IRI provenance graph before creating IRIs and takes a CSV input). Figure~\ref{fig:comparison} shows the processing time taken by \scon\ operations and \s\ procedures to create level members and observations with increasing input size. In the figure, LG and OG represent level member generator and observation generator operations (in case of \scon) or procedures (in case of \s). 

In summary, to process an input CSV file with 500 MB in size, \scon\ takes 37.4\% less time than \s\ to create observations and 23,4\% less time than \s\ to create level members. The figure also shows that the processing time difference between the corresponding \scon\ operation and the \s\ procedure increases with the size of the input. In order to guarantee scalability when using the Jena library, a \scon\ operation takes the (large) RDF input in the N-triple format, divides the file into several smaller chunks, and processes each chunk separately. Figure~\ref{fig:scalability} shows the processing time taken by \textit{LevelMemberGenerator} and \textit{ObservationGenerator} operations with the increasing number of triples. We shows the scalability of the \textit{LevelMemberGenerator} and \textit{ObservationGenerator} because they creates data with MD semantics. The figure shows that the processing time of both operations increase linearly with the increase in the number of triples, which ensures that both operations are scalable. \scon\ takes less time in loading than \s\ because \s\ uses the Jena TDB loader command to load the data while \scon\ programmatically load the data using the Jena API's method. 

In summary, \s\ and \scon\ have similar performance (29\% difference in total processing time). \scon\ ensures the uniqueness of IRI creation and uses RDF as a canonical model, which makes it more general and powerful than \s. 

Besides the differences of the performances already explained in Table~\ref{tab:performance},  \scon\ also includes an operation to update the members of the SDW levels, which is not included in \s. Since the ETL process for our use case SDW did not include that operation, we scrutinize the performance of this specific operation of \scon\ in the following.

\begin{figure*}[h!]

\begin{subfigure}{.5\textwidth}
  \centering
  \includegraphics[width=1\linewidth]{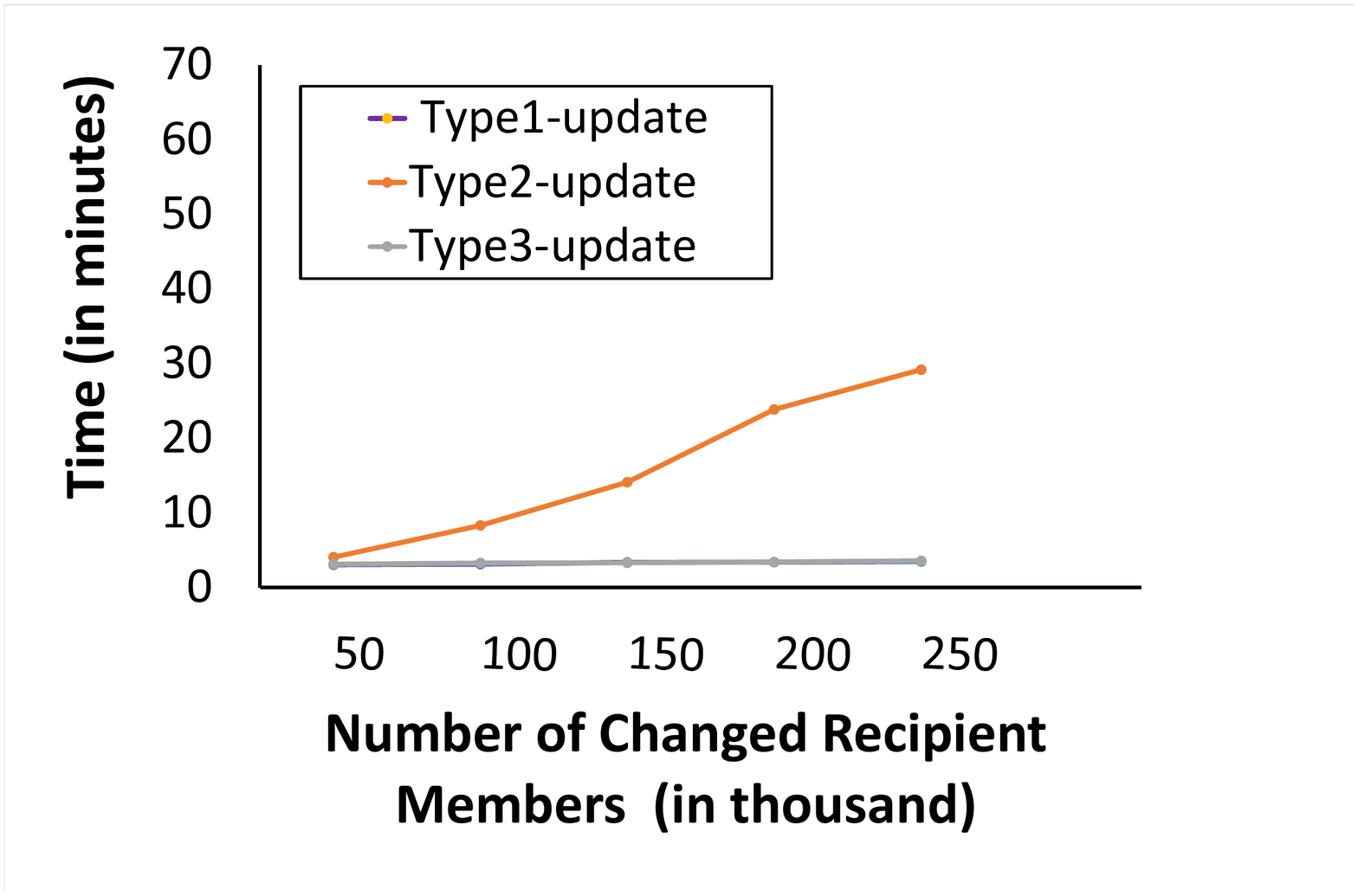}
  \caption{Processing time of increasing changed recipient members.}
  \label{fig:recipienttime1}
\end{subfigure}%
\begin{subfigure}{.5\textwidth}
  \centering
  \includegraphics[width=1\linewidth]{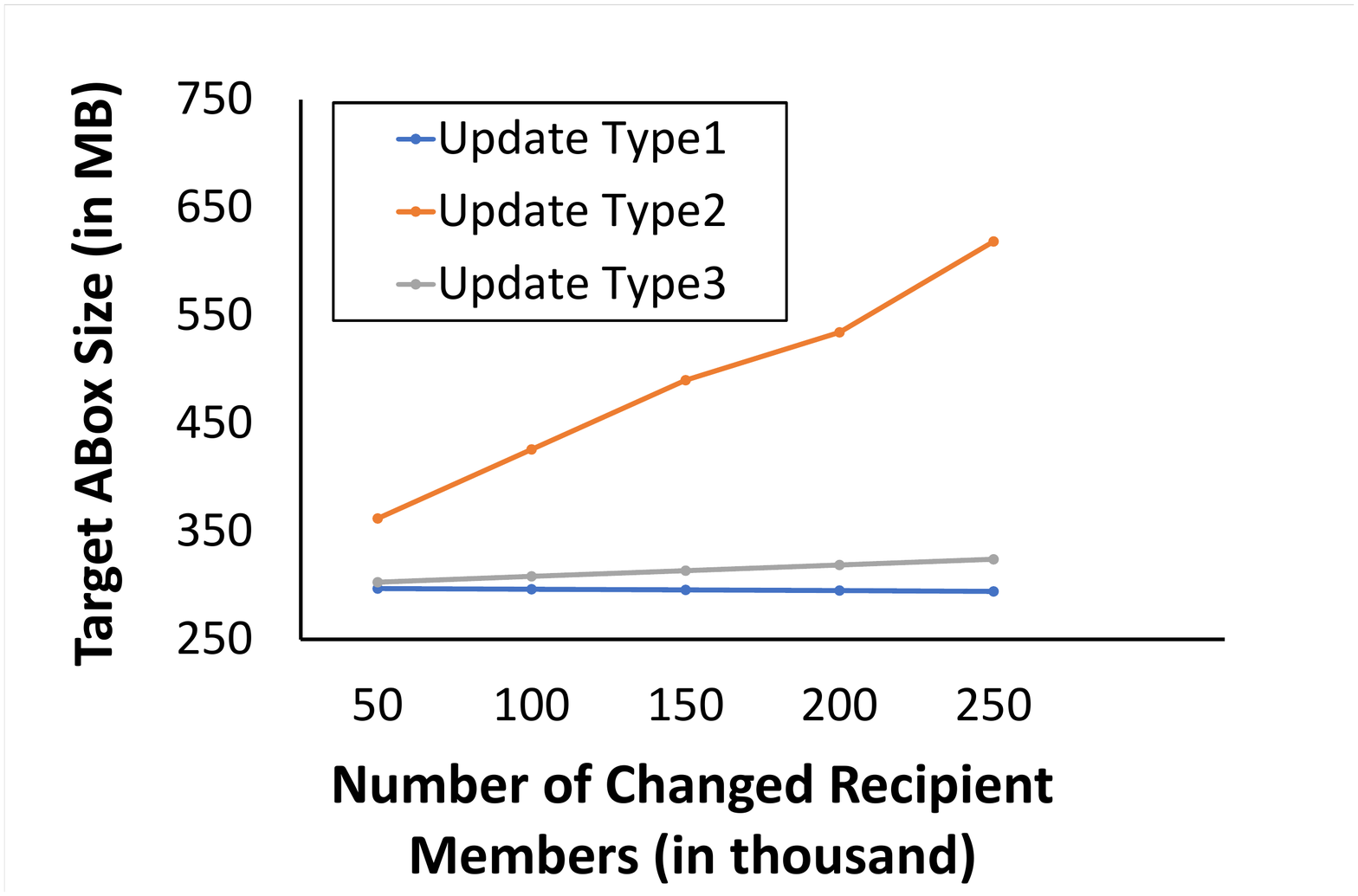}
  \caption{Target size of increasing changed recipient members.}
  \label{fig:recipientsize1}
\end{subfigure}
\begin{subfigure}{.49\textwidth}
  \centering
  \includegraphics[width=1\linewidth]{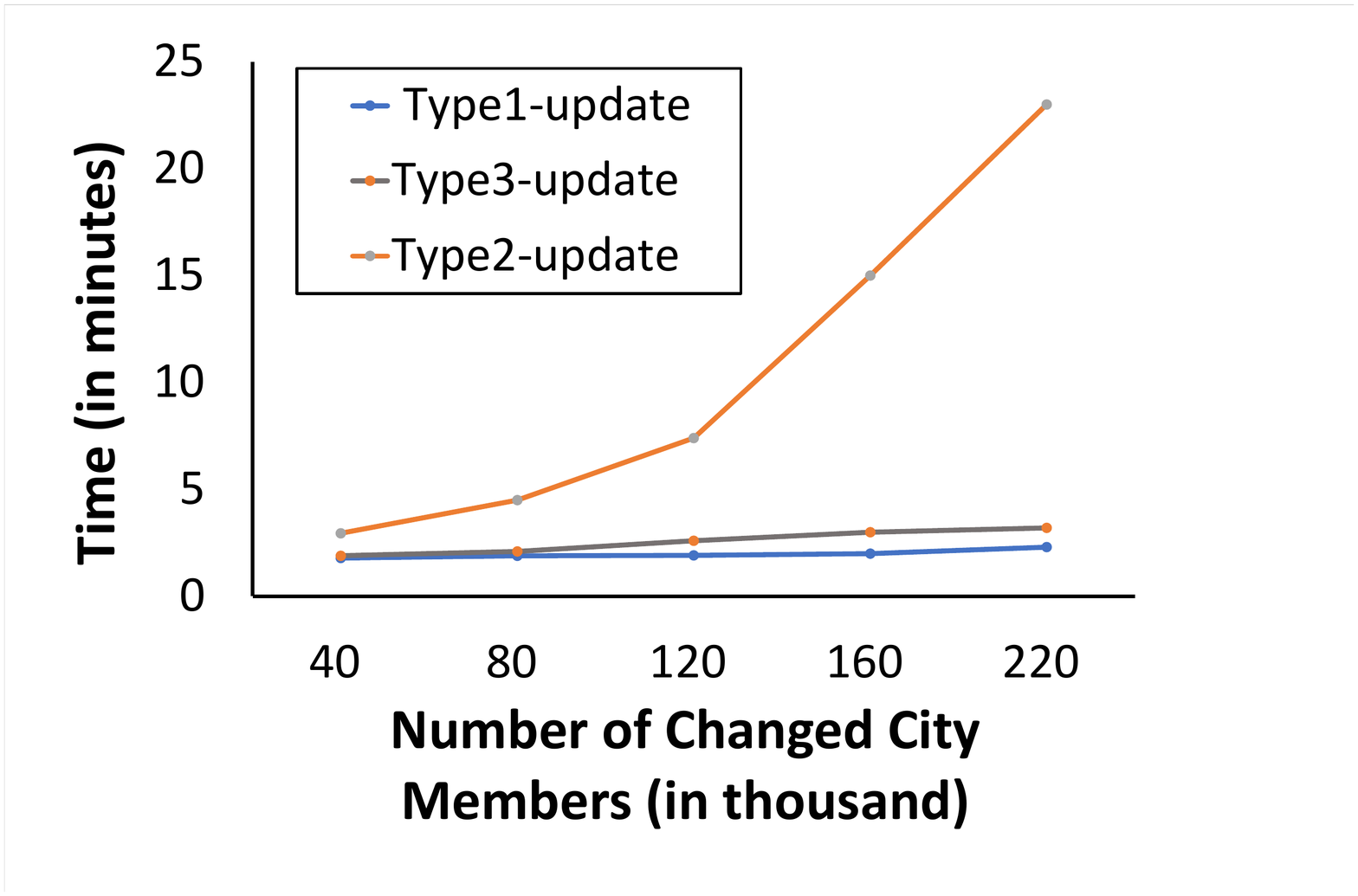}
  \caption{Processing time of increasing changed city members.}
  \label{fig:citytime1}
\end{subfigure}
\begin{subfigure}{.49\textwidth}
  \centering
  \includegraphics[width=1\linewidth]{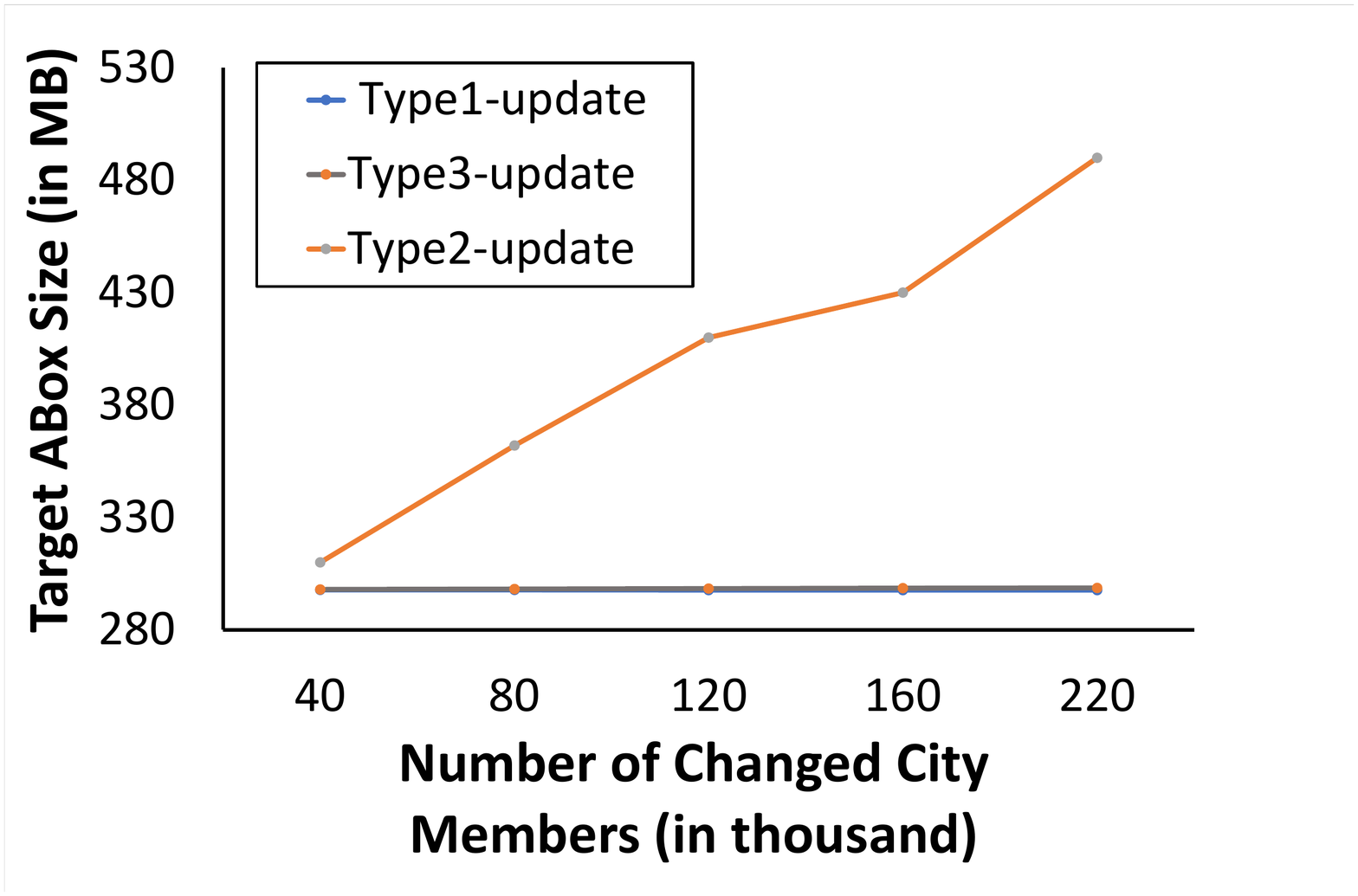}
  \caption{Target size of increasing changed city members.}
  \label{fig:citysize1}
\end{subfigure}

\caption{ Performance of \textit{UpdateLevel} based on processing time and target ABox size with the changing of the \texttt{sdw:Recipient} and \texttt{sdw:City} members.}
\label{fig:updatelevel}
\end{figure*}

\paragraph{Performance analysis of UpdateLevel operation} Figure~\ref{fig:updatelevel} shows the performance of the \textit{UpdateLevel} operation. To evaluate this operation, we consider two levels: \texttt{mdProperty:Recipient} and \texttt{mdPrope-\\rty:City}. We consider these two levels because \texttt{mdProperty:Recipient} is the largest level (in terms of size) in our use case, and \texttt{mdProperty:Ci-\\ty} is the immediate upper level of \texttt{mdProperty:R-\\ecipient} in the \texttt{mdStructure:Address} hierarchy of the dimension \texttt{mdProperty:Beneficiary}. Therefore, we can record how the changes in cities propagate to recipients, especially in Type2-update. The \texttt{sdw:Recipient} level is the lowest granularity level in the \texttt{mdStructure:Address} hierarchy; therefore, changes in a recipient (i.e., a member of \texttt{sdw:Recipient}) only affect that recipient. Figure~\ref{fig:recipienttime1} shows the processing time with the increasing number of recipients. As a Type2-update creates a new version for each changed level member, it takes more time than a Type1-update and a Type3-update. A Type3-update takes more time than a Type1-update because it keeps the record of old property values besides the current ones. Figure~\ref{fig:recipientsize1} shows how the size of the target ABox increases with the increasing number of recipients. The target ABox size increases linearly with the increasing number of recipients (see Figure~\ref{fig:recipientsize1}) for Type2-update and Type-3 updates because they keep additional information. However, the target ABox size decreases with the increasing number of recipients for Type1-updates; this is because the current property-values are smaller than the older ones in size.

Figure~\ref{fig:citytime1} shows the processing time when increasing the number of updated cities. Since \texttt{sdw:City} is the immediate upper level of \texttt{sdw:Recipient} in the \texttt{mdStructure:address} hierarchy, to reflect a changed city in the target ABox, a Type2-update creates new versions for itself as well as for the recipients living in that city. Therefore, a Type2-update takes more processing time in comparison to a Type1-update and a Type-3 update. Figure~\ref{fig:citysize1} shows the target ABox size with the increasing number of cities. The figure shows that the target ABox size for Type2-updates increases slowly within the range from 120 to 160 (X-axis), and it indicates that the changed cities within this range contain fewer recipients than the cities in other ranges.      

In overall, we conclude that \s\ and \textit{SETL\textsubscript{CON-}} \textit{\textsubscript{STRUCT}} have similar performance and \scon\ is a scalable framework. Moreover, \scon\ supports SDWs update.

\section{Related Work}
\label{sec:relatedwork}
Nowadays, combining SW and BI technologies is an emerging research topic as it opens interesting research opportunities. As a DW deals with both internal and (increasingly) external data presented in heterogeneous formats, especially in the RDF format, semantic issues should be considered in the integration process~\cite{abello2014using}. Furthermore, the popularity of SW data gives rise to new requirements for BI tools to enable OLAP-style analysis over this type of data~\cite{jakobsen2015optimizing}. Therefore, the existing research related to semantic ETL is divided into two lines: 1) on the one hand, the use of SW technologies to physically integrate heterogeneous sources and 2) on the other hand, enabling OLAP analysis over SW data. 

One prominent example following the first research line is~\cite{skoutas2007ontology}, which  presents an ontology-based approach to enable the construction of ETL flows.  At first, the schema of both the sources and the DW are defined by a common graph-based model, named the datastore graph. Then, the semantics of the datastore graphs of the data sources and the DW are described by generating an (OWL-based) application ontology, and the mappings between the sources and the target are established through that ontology. In this way this approach addresses heterogeneity issues among the source and target schemata and finally demonstrates how the use of an ontology enables a high degree of automation from the source to the target attributes, along with the appropriate ETL transformations. Nonetheless, computationally complex ETL operations like slowly changing dimensions and the annotation of the application ontology with MD semantics are not addressed in this work. Therefore, OLAP queries cannot be applied on the generated DW. 

Another piece of existing work~\cite{bellatreche2013semantic} aligned to this line of research proposes a methodology describing some important steps required to make an SDW, which enables to integrate data from semantic databases. This approach also misses the annotation of the SDW with MD semantics. \cite{thenmozhi2014ontological} has proposed an approach to support data integration tasks in two steps: 1) constructing ontologies from XML and relational sources and 2) integrating the derived ontologies by means of existing ontology alignment and merging techniques. However, ontology alignment techniques are complex and error-prone. \cite{bansal2014towards} presents a semantic ETL framework at the conceptual level. This approach utilizes the SW technologies to facilitate the integration process and discusses the use of different available tools to perform different steps of the ETL process. \cite{andersen2014publishing} presents a method to spatially integrate a Danish Agricultural dataset and a Danish Business dataset using an ontology. The approach uses SQL views and other manual processes to cleanse the data and Virtuoso for creating and storing integrated RDF data. \cite{knap2018unifiedviews} presents UnifiedViews, an open-source ETL framework that supports management of RDF data. Based on the SPARQL queries, they define four types of Data Processing Units (DPUs): Extractor, Transformer, Loader, and Quality Assessor. However, the DPUs do not support to generate MD RDF data. 

In the second line of research, a prominent paper is ~\cite{nebot2012building}, which outlines a semi-automatic method for incorporating SW data into a traditional MD data management system for OLAP analysis. The proposed method allows an analyst to accomplish the following tasks: 1) designing the MD schema from the TBox of an RDF dataset, 2) extracting the facts from the ABox of the dataset and populating the MD fact table, and 3) producing the dimension hierarchies from instances of the fact table and the TBox to enable MDX queries over the generated DW. However, the generated DW no longer preserves the SW data principles defined in~\cite{heath2011linked}; thus, OLAP analysis directly over SW data is yet to be addressed. To address this issue, \cite{colazzo2014rdf} introduces the notion of a \textit{lens}, called the analytical schema, over an RDF dataset. An analytical schema is a graph of classes and properties, where each node of the schema presents a set of facts that can be analyzed by traversing the reachable nodes. \cite{hilal2017olap} presents a self-service OLAP endpoint for an RDF dataset. This approach first superimposes an MD schema over the RDF dataset. Then, a semantic analysis graph is generated on top of that MD schema, where each node of the graph represents an analysis situation corresponding to an MD query, and an edge indicates a set of OLAP operations. 

Both \cite{colazzo2014rdf} and \cite{hilal2017olap} require either a \textit{lens} or a semantic analysis graph to define MD views over an RDF dataset. Since most published SW data contains facts and figures,  W3C recommends the Data Cube (QB)~\cite{cyganiak2014rdf} vocabulary to standardize the publication of SW data with MD semantics. Although QB is appropriate to publish statistical data and several publishers (e.g.,~\cite{petrou2014publishing}) have already used the vocabulary for publishing statistical datasets, it has limitations to define MD semantics properly. The QB4OLAP~\cite{qb4olap2014modeling} vocabulary enriches QB to support MD semantics by providing constructs to define 1) a cube structure in terms of different level of dimensions, measures, and attaching aggregate functions with measures and 2) a dimension structure in terms of levels, level attributes, relationships, the cardinality of relationships among the levels, and hierarchies of the dimension. Therefore, MD data can be published either by enriching data already published using QB  with dimension levels, level members, dimension hierarchies, and the association of aggregate functions to measures without affecting the existing observations ~\cite{varga2016dimensional} or using QB4OLAP from scratch ~\cite{galarraga2017qboairbase,ibragimov2014towards,nath2017setl}.   

In~\cite{kampgen2012interacting}, the authors present an approach to enable OLAP operations on a single data cube published using the QB vocabulary and shown the applicability of their OLAP-to-SPARQL mapping in answering business questions in the financial domain. However, their OLAP-to-SPARQL mapping may not result in the most efficient SPARQL query and requires additional efforts and a longer time to get the results as they consider that the cube is queried on demand and the DW is not materialized. While some approaches have proposed techniques to optimize execution of OLAP-style SPARQL queries in a federated setting~\cite{ibragimov2015processing}, others have considered view materialization~\cite{ibragimov2016optimizing,galarraga2018answering}. The authors in ~\cite{varga2016dimensional} present a semi-automatic method to enrich the QB dataset with QB4OLAP terms. However, there is no guideline for an ETL process to populate a DW annotated with QB4OLAP terms.

After analyzing the two research lines, we can draw some conclusions. Although each approach described above addresses one or more aspects of a semantic ETL framework, there is no single platform that supports them all (target definition, source-to-target mappings generation, ETL generations, MD target population, and evolution). To solve this problem, we have proposed a Python-based programmable semantic ETL (\s) framework~\cite{nath2017setl} that provides a number of powerful modules, classes, and methods for performing the tasks mentioned above. It facilitates developers by providing a higher abstraction level that lowers the entry barriers. We have experimentally shown that \s\ performs better in terms of programmer productivity, knowledge base quality, and performance, compared to other existing solutions. However, to use it, developers need a programming background. Although \s\ enables to create an ETL flow and provides methods by combining several tasks, there is a lack of a well-defined set of basic semantic ETL constructs that allow users more control in creating their ETL process. Moreover, how to update an SDW to synchronize it with the changes taking place in the sources is not discussed. Further, in a data lake/big data environment, the data may come from heterogeneous formats, and the use of the relational model as the canonical model may generate an overhead. Transforming JSON or XML data to relational data to finally generate RDF can be avoided by using RDF as the canonical model instead. To this end, several works have discussed the appropriateness of knowledge graphs for data integration purposes and specifically, as a canonical data model~\cite{cudreleveraging,rouces2017framebase,rouces2016heuristics}. An additional benefit of using RDF as a canonical model is that it allows adding semantics without being compliant to a fixed schema. The present paper presents the improvements introduced on top of \s\ to remedy its main drawbacks discussed above. As there are available RDF Wrappers (e.g., ~\cite{dimou2014rml,sequeda2013ultrawrap}) to convert another format to RDF, in this paper, we focus on only semantic data and propose an RDF-based two-layered (\dl\ and  \exl) integration process. We also propose a set of high-level ETL constructs (tasks/operations) for each layer, with the aim of overcoming the drawbacks identified above for \s. We also provide an operation to update an SDW based on the changes in source data. On top of that, we characterize the ETL flow in the \dl\ by means of creating an RDF based source-to-target mapping file, which allows to automate the ETL execution flows.

\section{Conclusion and Future Work}
\label{sec:conclusion}
In this paper, we proposed a framework of a set of high-level ETL constructs to integrate semantic data sources. The overall integration process uses the RDF model as canonical model and is divided into the Definition and Execution Layer. In the Definition Layer, ETL developers create the metadata (target and source TBoxes, and source-to-target mappings). We propose a set of high-level ETL operations for semantic data that can be used to create ETL data flows in the Execution Layer. As we characterize the transformations in the Definition Layer in terms of source-to-target mappings at the schema level, we are able to propose an automatic algorithm to generate ETL data flows in the Execution Layer. We developed an end-to-end prototype \scon\ based on the high-level constructs proposed in this paper. We also extended it to enable automatic ETL execution flows generation (and named it \sauto). The experiment shows that 1) \scon\ uses 92\% fewer NOTC than \s, and \sauto\ further reduces NOUC by another 25\%; 2) using\scon, the development time is almost cut in half compared to \s, and is cut by another 27\% using \sauto; 3) \scon\ is scalable and has similar performance compared to \s. 

An extension of this work is to define the formal semantics of the operations using an ontology and mathematically prove the correctness and completeness of the proposed ETL constructs. The source-to-target mappings act as a mediator to generate data according to the semantics encoded in the target. Therefore, we plan to extend our framework from purely physical to also virtual data integration where instead of materializing all source data in the DW, ETL processes will run on demand. When considering virtual data integration, it is important to develop query optimization techniques for OLAP queries on virtual semantic DWs, similar to the ones developed for virtual integration of data cubes and XML data~\cite{pedersen2004integrating,yin2006evaluating,pedersen2002query}. Another interesting work will be to apply this layer-based integration process in a Big Data and Data Lake environment. The metadata produced in the Definition Layer can be used as basis to develop advanced catalog features for Data Lakes. Furthermore, we plan to investigate how to integrate provenance information into the process~\cite{ahlstrom2016towards}. Another aspect of future work is the support of spatial data~\cite{gur2018foundation,gur2017geosemolap}. We will also develop a new implementation with scalable, parallel ETL computing and optimize the operations.


\begin{acks}
This research is partially funded by the European Commission through the Erasmus Mundus Joint Doctorate Information Technologies for Business Intelligence (EM IT4BI-DC), the Poul Due Jensen Foundation, and the Danish Council for Independent Research (DFF) under grant agreement no. DFF-8048-00051B.
\end{acks}

\begin{appendix}
\section{}
\label{sec:operations}
\subsection{Semantics of the \exl\ Operations}\label{sec:exl}

In this section, we provide the detailed semantics of the ETL operations described in Section~\ref{sec:executionLayer}. The  operations depend on some auxiliary functions. Here, we first present the semantics of the auxiliary functions and then the semantics of the operations. We present each function and operation in terms of input Parameters and semantics. To distinguish the auxiliary functions from the operations, we use small capital letters to write an auxiliary function name, while an operation name is written in italics. 

\subsubsection{Auxiliary Functions}
\vspace{0.2cm}
\paragraph{\underline{\textsc{executeQuery}($Q$, $G$, $outputHeader$)}}\label{executequery} This function provides  similar functionality as SPARQL SELECT queries. 

\textbf{Input Parameters}: Let $I$, $B$, $L$, and $V$ be the sets of IRIs, blank nodes, literals, and query variables. We denote the set of RDF terms $(I \cup B \cup L)$ as $T$ for an RDF graph $G$. $V$ is disjoint from $T$. A query variable $v \in V$ is prefixed by the symbol $'?'$. A query pattern $Q$ can be recursively defined as follows\footnote{We follow the same syntax and semantics  used in~\cite{perez2006semantics} for a query pattern.}:
\begin{enumerate}
\item An RDF triple pattern $t_p$ is a query pattern $Q$. A $t_p$\footnote{To make it easily distinguishable, here we use comma to separate the components of a triple pattern and an RDF triple.} allows query variables in any positions of an RDF triple, i.e., $t_p\in(I\cup B \cup V)\times(I\cup V)\times(T \cup V)$.
\item If $Q_1$ and $\mathcal{Q}_2$ are query patterns, then $(Q_1\; AND\\\; Q_2)$, $(Q_1\; OPT\; Q_2)$, and $(Q_1 \;UNION\; Q_2)$ are also query patterns.
\item If $Q$ is a query pattern and $F_c$ is a filter condition then $(Q\;FILTER\; F_c)$ is a query pattern. A filter condition is constructed using elements of the set $(T \cup V)$ and constants, the equality symbol ($=$), inequality symbols ($<, \geq, \leq, >$), logical connectivities ($\neg, \vee, \wedge $), unary predicates like bound, isBlank, and isIRI plus other features described in~\cite{prud2006sparql}.
\end{enumerate}
$G$ is an RDF graph over which $Q$ is evaluated and $outputHeader$ is an output header list, which is a list of query variables and/or expressions over the variables used in $Q$. Here, expressions are  standard SPARQL expressions defined in~\cite{harris2013sparql}.

\textbf{Semantics}: For the query pattern $Q$, let $\mu$ be a partial function that maps $var(Q)$ to $T$, i.e., $\mu: var(Q)\rightarrow T$. The domain of $\mu$, denoted by $dom(\mu)$, is the set of variables occurring in $Q$ for which $\mu$ is defined. We abuse notation and say $\mu (Q)$ is the set of triples obtained by replacing the variables in the triple patterns of $Q$ according to $\mu$. The semantics of the query pattern are recursively described below.
\begin{enumerate}

\item If the query pattern $Q$ is a triple pattern $t_p$, then the evaluation of  $Q$ against $G$ is the set of all mappings that can map $t_p$ to a triple contained in $G$, i.e., 
 $ \llbracket Q \rrbracket _{G} =  \llbracket t_p \rrbracket _{G}=\{ \mu \;|\; dom(\mu)=var(t_p) \; \wedge \; \mu(t_p) \in G\}.       $
 
\item If $Q$ is $(Q_1 \; AND\; Q_2)$, then $  \llbracket Q \rrbracket _{G}$ is the natural join of $\llbracket Q_1 \rrbracket _{G} $ and $ \llbracket Q_2 \rrbracket _{G} $, i.e.,    
$  \llbracket Q \rrbracket _{G} = \llbracket Q_1 \rrbracket _{G} \bowtie \llbracket Q_2 \rrbracket _{G} =\{ \mu _1 \cup \mu _2 \;|\; \mu _1 \in \llbracket Q_1 \rrbracket _{G}, \;\mu _2 \in \llbracket Q_2 \rrbracket _{G},\\ \mu _1 \;and\; \mu _2 \; are \; compatible \; mappings, i.e., \forall ?x \in dom(\mu _1) \cap dom(\mu _2),\mu _1 (?x)= \mu _2 (?x)\}  $. 

 \item If $Q$ is $(Q_1 \; OPT \; Q_2)$, then $  \llbracket Q \rrbracket _{G}$ is the left outer join of $\llbracket Q_1 \rrbracket _{G} $ and $ \llbracket Q_2 \rrbracket _{G} $, i.e., $
  \llbracket Q \rrbracket _{G} = \llbracket Q_1 \rrbracket _{G} \tiny  \louterjoin \llbracket Q_2 \rrbracket _{G}
  		=(\llbracket Q_1 \rrbracket _{G} \bowtie \llbracket Q_2 \rrbracket _{G}) \cup (\llbracket Q_1 \rrbracket _{G} \setminus (\llbracket Q_2 \rrbracket _{G})$.

\item If $Q$ is $(Q_1 \; UNION \; Q_2)$, then $
  \llbracket Q \rrbracket _{G} = \llbracket Q_1 \rrbracket _{G} \cup \llbracket Q_2 \rrbracket _{G}$
   		$=\{ \mu \;|\; \mu \in \llbracket Q_1 \rrbracket _{G} \; or \; \mu \in \llbracket Q_2 \rrbracket _{G}\;\}  $. 

\item If $Q=(Q_1 \; FILTER \; F_c)$, then $\llbracket Q \rrbracket _{G} = \{\mu \;|\; \mu \in \llbracket Q_1 \rrbracket _{G}\; \wedge\; \mu \models F_c \}$, where a mapping $\mu$ satisfies a filter condition $F_c$, denoted by $\mu \models F_c$.



\end{enumerate} 
This function returns a set of tuples where each tuple contains the values of the variables and/or the values derived by solving the expressions defined in $outputHeader$  according to $\mu$.

\vspace{0.2 cm}
\noindent\textit{\underline{\textsc{getPropertiesFromExpressions}(sTBox, exps)}}\label{para:extractpropertiesfromexpression} 
An expression is a combination of one or more properties, operators, and SPARQL functions defined in~\cite{harris2013sparql}. This function returns a subset of properties of a source TBox that are used in the given expressions.

\textbf{Input Parameters}: \textit{sTBox} is a TBox and \textit{exps} is a set of expressions. 

\textbf{Semantics}: For each $exp \in exps$, this function computes the intersection of the set of IRIs used in $exp$ and the set of properties of \textit{sTBox}. Then, it returns the union of all intersection sets. Here, \textit{returnIRIs(exp)} returns the set of IRIs used in $exp$ and $\mathcal{P}(t)$ is defined in Equation~\ref{eq:property}. The semantic is defined as follows:
\begin{multline*}
\textsc{getPropertiesFromExpressions}(sTBox, \\exps)= \bigcup_{exp \in exps} returnIRIs(exp) \cap \mathcal{P}(sTBox).
\end{multline*} 

This function returns a set of properties.

\paragraph{\underline{\textsc{validateExpressions}($exps$,$Q$, flag)}}\label{para:replacexpression} The source expressions/properties (defined by \texttt{map:source4Ta-\\rgetPropertyValue}) in property-mappings contains properties. However, the \textit{outputHeader} parameter of \textsc{executeQuery()} allows only query variables. Therefore to extract data from an RDF graph using \textsc{executeQuery()}, the properties used in the expressions/properties of property-mappings should be replaced by corresponding query variables. On the other hand, to match the expressions/properties used in \textit{outputHeader} of \textsc{executeQuery()}  with the expressions/properties (defined by \texttt{map:source4TargetPro-\\pertyValue}) in property-mappings , the query variables used in expressions/properties need to be replaced by the corresponding properties. Given a set of expressions, a query pattern, and a flag, this function replaces the used term (either IRI or query variables) in expressions with alternative ones from the query pattern based on the flag and returns a list of the validated expressions.

\textbf{Input Parameters}: $exps$ is a list of expressions, $Q$ is a query pattern, and \textit{flag} indicates whether it will replace the IRIs (\textit{flag}=1) or query variables (\textit{flag}=0) used in $exps$.

\textbf{Semantics}: If \textit{flag=1} the function replaces the $exps$ IRIs with the corresponding query variables used in $Q$. For each IRI \textit{iri} used in an expression, \textsc{validateExpressions} replaces \textit{iri} with the object of the triple pattern whose predicate is \textit{iri} in $Q$. Finally, it returns the list of expressions, where each expression does not contain any IRIs. 
 
If \textit{flag=0} the function replaces the $exps$ query variables with the corresponding predicate IRIs used in $Q$. For each query variable \textit{?q} used in an expression, it replaces \textit{?q} with the predicate of the triple pattern whose object is \textit{?q} in $Q$. Finally, it returns the list of expressions, where each expression does not contain any query variables.

\vspace{0.2 cm}
\noindent \textit{\underline{\textsc{mappedSourceInstances}(sc, sTBox, sABox, prop-}} \\\textit{\underline{ertyMappings)}}\label{para:mappedsourceinstances} 
This function returns a dictionary describing the instances of a source concept.

\textbf{Input Parameters}: $sc$ is a source construct, $sTBox$ and $sABox$ are the source TBox and ABox, $property-\\Mappings$ is a set of property-mappings. 

\textbf{Semantics}: At first, this function retrieves instances of $sc$ with their corresponding properties and values. Here, we consider only those properties that are directly mapped to target properties and/or used in source expressions. Both the properties and expressions are defined in \textit{propertyMappings} by \texttt{map:sour-\\ce4TargetPropertyValue}. Then, it creates a dictionary, a set of $(key, value)$ pairs. In a pair, $key$ represents an instance IRI and $value$ in turn represents a set of $(p_i, v_i)$ pairs, where $p_i$ represents a property and $v_i$ represents a value for $p_i$. It is explained as follows:
\begin{multline*} 
\textsc{mappedSourceInstances}(sc,sTBox, sABox, \\propertyMapping)= dictionary(\textsc{executeQu-}\\\textsc{ery}((?i, \texttt{rdf:} \textit{type}\footnote{If the source construct is either a QB level property or dataset, then \texttt{rdf:type} is replaced by either \texttt{qb4o:memberOf} or \texttt{qb4o:dataset}}, sc) AND (?i , ?p, ?v)FILTER\\ (?p \;  \in \; \textsc{getPropertiesFromExpressions}(\\sTBox, mappedExpressions(sc, propertyMap\\pings)))), sABox, (?i, ?p , ?v))). 
\end{multline*}

Here, \textit{mappedExpressions(sc, propertyMappings)} returns the set of source properties/expressions (defined by \texttt{map:source4TargetPropertyValue}) used in \textit{propertyMappings}. The \textit{dictionary((?i, ?p, ?v))} function first groups the input tuples by \textit{?i} and then for each instance $i\in ?i$, it creates a set of $(p_i, v_i)$ pairs, where $p_i$ is a property of $i$ and $v_i$ is the value for $p_i$. Finally, \textsc{mappedSourceInstances(..)} returns the dictionary created.

\vspace{0.2 cm}
\noindent \textit{\underline{\textsc{generateIRI}(sIRI, value, tType, tTBox, iriGraph )}} Every resource in an SDW is uniquely identified by an IRI defined under the namespace of the SDW. This function creates an equivalent target IRI for a source resource. Additionally, it keeps that information in the IRI graph. 

\textbf{Input Parameters}: $sIRI$ is the source IRI, $value$ is the literal to be used to ensure the uniqueness of the generated IRI, $tType$ is the type of the generated IRI in $tTBox$, $tTBox$ is the target TBox, and $iriGraph$ is the IRI graph. The IRI graph is an RDF graph that keeps a triple for each resource in the SDW with their corresponding source IRI.

\textbf{Semantics}: Equation~\ref{eq:generateiri} formulates how to create the IRI. First, the function checks whether there is an equivalent IRI for $sIRI$ in $iriGraph$ using $lookup(sIRI, iriGraph)$. It returns the target IRI if it finds an existing IRI in $iriGraph$; otherwise, it generates a new IRI by concatenating \textit{prefix(tTBox)} and $validate(value)$ for a target concept or property, or creates an instance IRI by concatenating \textit{tType} and $validate(value)$. Here, \textit{prefix(tTBox)} returns the namespace of $tTBox$; $concat()$ concatenates the input strings; $validate(value)$ modifies $value$ according to the naming convention rules of IRIs described in~\cite{richardson2008restful}; $\mathcal{C}(T)$ and $\mathcal{P}(T)$ are defined in Equation~\ref{eq:concept} and \ref{eq:property}. Upon the creation of the IRI, this function adds an  RDF triple (tIRI \texttt{owl:sameAs} sIRI) to $iriGraph$.   

\def\stackalignment{l}
\textit{generateIRI(sIRI, value, tType, tTBox, iriGraph )=}
\begin{equation}\label{eq:generateiri}
\scriptsize\begin{cases}
 lookup(sIRI,iriGraph) 
&if \;lookup(sIRI,iriGraph)!= \\[-3 pt]
   &NULL\\  

  concat(tType,``\#", \\ validate(value))
  &if \;lookup(value,iriGraph)=\\[-3 pt]
   & NULL\; \land
\;   tType \in (\mathcal{C}(tTBox) \\[-3 pt]
   &\cup \mathcal{P}(tTBox)) \\
   
  concat(prefix(tTBox),\\``\#", validate(value))
  &if lookup(sIRI,iriGraph)= \\[-3 pt]
   &NULL\; \land
   \; tType \notin (\mathcal{C}(tTBox)\\[-3 pt]
   & \cup \mathcal{P}(tTBox)) 
    
\end{cases}
\end{equation}

This function returns an IRI.  
\vspace{0.2 cm}

\noindent \textit{\underline{\textsc{tuplesToTriples}(T, tc, Q, propertyMappings,}} \\\textit{\underline{ (i, $exp_1,.., exp_n$))}} As the canonical model of our integration process is RDF model, we need to convert the instance descriptions given in the tuple format into equivalent RDF triples. This function returns a set of RDF triples from the given set of tuples.  

\textbf{Input Parameters}: $T$ is a set of tuples (1st normal form tabular format), $tc$ is a target construct, $Q$ is a query pattern, $propertyMappings$ is a set of property-mappings, and  $(i, exp_1,.., exp_n)$ is the tuple format. 

\textbf{Semantics}: Equation~\ref{eq:transformeddata} shows the semantics of this function. Each tuple of $T$ represents an instance-to-be of the target construct $tc$, and each expression ($exp_i$) and the value of the expression ($val(exp_i)$) represent the corresponding target property and its value of the instance. Here, $val(exp_i)$ is the value of an expression in a source tuple. As an expression can contain query variables and expressions used in $propertyMappings$ do not use query variables, \textsc{validateExpressions}\textit{($exp_i$, Q, 0)} replaces the query variables with equivalent properties from $Q$, and \textit{return(x, propertyMappings)} returns the target construct mapped to source construct $x$ from $propertyMappings$.

\begin{multline}
\label{eq:transformeddata}
\textsc{tuplesToTriples}(T, tc, Q, propertyMappings,\\(i, exp_1,.., exp_n))=\\\bigcup_{(val(i), val(exp_1),.., val(exp_n)) \in T} \{ \{(val(i),\texttt{rdf:type}, tc\\))\}\;\cup\;   \bigcup_{p=1}^n\{(val(i), return(\textsc{validateExpressions}\\(exp_p,Q,0), propertyMappings), val(exp_i))\} \}
\end{multline}

This function returns a set of RDF triples. 
\subsubsection{\exl\ Operations}
This section gives the semantics of each operation category-wise.

\noindent \textbf{Extraction Operations}

\vspace{0.2 cm}
\paragraph{\underline{GraphExtractor(\textit{Q}, \textit{G}, \textit{outputPattern}, \textit{tABox})}}\label{PatternMatcher} This operation is functionally equivalent to SPARQL CONSTRUCT queries to extract data from sources. 

\textbf{Input Parameters}: $Q$ is a query pattern as defined in \textsc{ExecuteQuery()}. $G$ is an RDF graph over which $Q$ is evaluated. The parameter $outputPattern$ is a set of triple patterns from  $(I\:\cup B\:\cup\: var(Q))\times (I\:\cup \: var(Q)\times (T\: \cup\: var(Q))$, where $var(Q)\subseteq V$ is the set of query variables occurring in $Q$. Here, $I, B, T,$ and $V$ are the sets defined in \textsc{ExecuteQuery()}. The parameter \textit{tABox} is the location where the output of this operation is stored.   

\textbf{Semantics}: This operation generates the output in two steps: 1) First, $Q$ is matched against $G$ to obtain a set of bindings  for the variables in $Q$ as discussed in \textsc{ExecuteQuery()}. 2) Then, for each variable binding ($\mu$), it instantiates the triple patterns in $outputPattern$ to create new RDF triples. The result of the query is a merged graph, including all the created triples for all variable bindings, which will be stored in $tABox$. Equation~\ref{eq:graphextractor} formally defines it.  

\textit{GraphExtractor(Q, G, outputPattern, tABox) =}
\begin{multline}\label{eq:graphextractor}
   \bigcup_{t_h \in outputPattern}\{ \mu(t_h) \;|\; \mu \in \llbracket Q \rrbracket _{G} \; \wedge \;\mu(t_h)\; is\;a \;\\ well-formed\; RDF \; triple \}.       
\end{multline}

\noindent \textbf{Transformation Operations}

\vspace{0.2 cm}
\noindent \textit{\underline{TransformationOnLiteral(sConstruct, tConstruct,}}\\ \textit{\underline{ sTBox, sABox, propertyMappings, tABox)}} This operation creates a target ABox from a source ABox based on the expressions defined in property-mappings.
 
\textbf{Input Parameters}: \textit{sConstruct} and \textit{tConstruct} are a source and target TBox construct, \textit{sTBox} and \textit{sABox} are the source TBox and ABox, \textit{propertyMappings} is a set of property-mappings, and \textit{tABox} is the output location.

\textbf{Semantics}: First, we retrieve the instances of \textit{sConstruct} \textit{ins(c)} from \textit{sABox} using the \textsc{executeQuery()} function, which is formally described as follows: 
  
\textit{ins(c)=\textsc{executeQuery}(q(c), sABox,(?i, \textsc{validateExpressions}(list(cElements),q(c),1))}.

\noindent Here:
\begin{itemize}[noitemsep,nolistsep, topsep=0pt]
\item $c=sConstruct$. 

\item \textit{q(c)=(((?i, }\texttt{rdf:type},\textit{ c) AND (?i,?p,?v)) FILTER (?p $\in$ cProperties))} is a query pattern. 

\item \textit{cProperties= \textsc{getPropertiesFromExpressions}(sTBox, cElements)} is the set of source properties used in source expressions defined in \textit{propertyMappings}.

\item \textit{cElements=\textsc{executeQuery}((?pm, \texttt{map:sour-}\\\texttt{ce4TargetPropertyValue}, ?sp), propertyMappings, ?sp)}  is the set of source expressions in \textit{propertyMappings} defined by \texttt{map:source4-\\TargetPropertyValue}.

\item \textit{\textsc{validateExpressions}(list(cElements),q(c),1)} replaces the source properties used in \textit{cElements} with the corresponding query variables from \textit{q(c)} as \textit{outputHeader} parameter \textsc{executeQuery()} does not allows any properties. Since \textsc{validateExpressions} takes a list of expressions as a parameter, list(cElement) creates a list for \textit{scElements}.
\end{itemize} 
\noindent Now, we transform all tuples of $ins(c)$ into equivalent RDF triples to get $tABox$, i.e.,

$output=\textsc{tuplesToTriples}(ins(c), tConstruct,\\ q(c), (?i,exp_1, ,..,exp_n))$. 

The output of this operation is $output$, a set of RDF triples, which will be stored in $tABox$. 
\vspace{0.2 cm}

\noindent\textit{\underline{JoinTransformation(sConstruct, tConstruct, sTBox,}}\\ \textit{\underline{ tTBox, sABox, tABox, comProp, propertyMappings)}} This operation joins and transforms the instances of source and target based on property-mappings.

\textbf{Input Parameters}: \textit{sConstruct} and \textit{tConstruct} are a source and target\footnote{The term target here does not mean target schema but in the sense it is defined in concept-mappings. Source and target both can be either source concepts or intermediate concepts (from an intermediate result).} TBox construct, $sTBox$ and $tTBox$ are the source and target TBoxes;; $sABox$ and $tABox$ are the source and target ABoxes; $comProp$ is a set of common properties between \textit{tConstruct} and \textit{sConstruct} and $propertyMapping$ is the set of property-mappings. 

\textbf{Semantics}: At first,  we create an ABox by taking the set union of instances (including their properties and values) of both \textit{sConstruct} and \textit{tConstruct} and apply a query on the ABox using the \textsc{executeQuery()} function. The query pattern of the function joins the instances of two concepts based on their common properties and finally the function returns a set of target instances \textit{ins(sc,tc)} with the values of the source expressions (defined by \texttt{map:source4TargetPropertyValue}) in \textit{propertyMappings}. As the query pattern includes triple patterns from both \textit{sConstruct} and \textit{tConstruct}, the source expression can be composed of both source and target properties. This is described as follows:

\textit{ins(tc,sc)=\textsc{executeQuery} ((q(tc) \textit{OPT} q(sc)),uni-\\on(extractIns(tc,tABox), extractIns(sc,sABox)), (?i, \textsc{validateExpressions}(scElements, (q(tc) \textit{OPT} q(sc)),1)).}

\noindent Here:
\begin{itemize}[noitemsep,nolistsep, topsep=0pt]

\item \textit{tc=tConstruct} and \textit{sc=sConstruct}. 

\item \textit{q(tc)=(($?i_{tc}$,} \texttt{rdf:type}, \textit{tc) AND ($?i_{tc}$,$?p_{tc},?v_{tc}$) $tp_{com}(?i_{tc}$,``target", comProp) FILTER($?p_{tc} \in$ tcProperties))} is a query pattern.
\begin{itemize}[noitemsep,nolistsep, topsep=0pt]
\item[-] $tp_{com}(?i_{tc}, target, comProp)$ is a function that joins a triple pattern ( i.e., $\;AND\;(?i_{tc}, \\scom_i, ?scom_i)$) in $q(tc)$ for each pair $(scom_i,\\ tcom_i)\in comProp$.
\item[-]\textit{tcProperties= \textsc{getPropertiesFromExpressions}(tTBox, scElements)} represents the set of source properties used in source expressions.
\end{itemize}
\item \textit{q(sc)=(($?i_{sc}$,} \texttt{rdf:type}, \textit{sc) AND ($?i_{sc},?p_{sc},?v_{sc}$) $sp_{com}(?i_{sc}$,``source", comProp) FILTER($?p_{sc} \in$ scProperties))} is the query pattern.
\begin{itemize}[noitemsep,nolistsep, topsep=0pt]
\item[-] $sp_{com}(?i_{sc}, ``source", comProp)$ is a function that joins a triple pattern ( i.e., $\;AND\;(?i_{sc}, \\tcom_i, ?scom_i)$) in $q(sc)$ for each pair $(scom_i,\\ tcom_i)\in comProp$ .
\item[-] \textit{scProperties= \textsc{getPropertiesFromExpressions}(sTBox, scElements)} represents the set of source properties used in source expressions.
\end{itemize}


\item \textit{scElements=\textsc{executeQuery}((?pm, \texttt{map:sour-}\\\texttt{ce4TargetPropertyValue}, ?sp), propertyMappings, ?sp)} is the set of source expressions in \textit{propertyMappings} defined by \texttt{map:source4-\\TargetPropertyValue}.

\item \textit{extractInstance(c, abox)=\textsc{executeQuery}(((?i, rdf:type, c) AND (?i,?p,?v)), abox,(?i, ?p, ?v))} retrieves the instances (with their linked properties and values) of the concept $c$ from the given ABox $abox$.

\item $union(s1,s2)$ returns the set union of two given sets $s1$ and $s2$.

\item \textit{\textsc{validateExpressions}(list(scElements),q(sc,\\tc),1)} replaces the source properties used in \textit{scElements} with the corresponding query variables from \textit{q(tc,sc)} as the \textit{outputHeader} parameter of \textsc{executeQuery()} does not allow any properties. Since \textsc{validateExpressions} takes a list of expressions as a parameter, list(cElement) creates a list for \textit{scElements}.
\end{itemize}
Now, we transform all tuples of $ins(sc,tc)$ into equivalent RDF triples to get the transformed $tABox$, i.e.,

\textit{output=\textsc{tuplesToTriples}(ins(tc,sc), tConstruct, q(tc,sc), ($?i,exp_1, ,..,exp_n$))}. 

The output of this operation is $output$, a set of RDF triples, which will be stored in $tABox$.

\vspace{0.2 cm}
\noindent\textit{\underline {LevelMemberGenerator(sConstruct, level, sTBox, }}\\ \textit{\underline {sABox, tTBox, iriValue, iriGraph, propertyMappings,}} \\ \textit{\underline {tABox)}} The objective of this operation is to create QB4OLAP-complaint level members.

\textbf{Input Parameters}: \textit{sConstruct} is the source construct, \textit{level} is a target level, \textit{sTBox} and \textit{sABox} are the source TBox and ABox, \textit{iriValue} is the rule of creating level members' IRIs, \textit{iriGraph} is the IRI graph, \textit{proeprtyMappings} is a set of property-mappings, and \textit{tABox} is the output location. 

 \textbf{Semantics}: First, we retrieve the instances of \textit{sConstruct} with their properties and values, i.e., 
\begin{multline}\label{eq:retrieveinstance}
Ins=\textsc{mappedSourceInstances}(sConstruct,\\ sTBox, sABox, propertyMappings)
\end{multline} 
 
To enrich \textit{level} with the dictionary \textit{Ins}, we define a set of triples \textit{LM} by taking the union of \textit{IdentityTriples} (for each \textit{(i,pv)} pair in \textit{Ins}) and the union of \textit{DescriptionTriples} (for each $(p_i,v_i)$ pair in $pv$). Equation~\ref{eq:lm} defines \textit{LM}.  
\begin{multline}\label{eq:lm}
\noindent LM=\bigcup_{(i,pv)\in Ins} (IdentityTriples \;\cup \\[-3 pt]\bigcup_{(p_i,v_i)\in pv}DescriptionTriple)
\end{multline}

\noindent Here: 

-$IdentityTriples= $
\vspace{-0.5 cm}
\begin{multline}\label{eq:lmidentity}
\{(lmIRI, \texttt{rdf:type},\texttt{qb4o:LevelMember}), \\[-3 pt] (lmIRI, 
\texttt{qb4o:memberOf},level)\} 
\end{multline}
-$lmIRI =$
\vspace{-0.5 cm}
\begin{align}\label{eq:lmiri}
\scriptsize \begin{cases}
i &if\; iriValue=``sameAs\\[-3 pt]
  &SourceIRI"\\
\textsc{generateIRI}(i,resolve(i,pv,iriValue),\\[-3 pt] range(level, tTBox, iriGraph), tTBox) \\[-3 pt]
  &if \;range(level, tTBox)!\\[-3 pt]&= NULL\\
\textsc{generateIRI}(i, resolve(i, pv, iriValue),\\[-3 pt]level, tTBox, iriGraph) \\[-3 pt]
  &if \;range(level, tTBox)\\[-3 pt]&= NULL\\
\end{cases}
\end{align}
-$DescriptionTriple=$
\vspace{-0.5cm}
\begin{align}\label{eq:lmdescription}
\noindent\scriptsize \begin{cases}
\{(lmIRI, return(p_i, property-\\[-3 pt]Mappings), \;
  \textsc{generateIRI}(v_i, \\[-3 pt]iriValue(v_i),range((return(\\[-3 pt]v_i, propertyMappings),\\[-3 pt] tTBox), tTBox, iriGraph ))\}
  &if \;targetType(return(v_i,\\[-3 pt]
   & propertyMappings),   tTBox)\\[-3 pt]
   &\in \{ rollupProperty, ObjectP-\\[-3 pt]
   &roperty   \}  \\
 \{(lmIRI, 
  return(p_i, \\[-3 pt]propertyMappings),v_i)\}
  &if\; targetType(p_i, propertyMap-\\[-3 pt]
   &pings,  tTBox)= levelAttribute     
\end{cases}
\end{align}

Each instance of $sConstruct$ is a \texttt{qb4o:LevelMem\\ber} and a member of \textit{level}; therefore, for each instance of $sConstruct$, $LM$ in Equation~\ref{eq:lm} includes two identity triples, described in Equation~\ref{eq:lmidentity}. %
Equation~\ref{eq:lmiri} describes how to create an IRI for a level member. As we described in Section~\ref{para:s2t}, the rule for creating IRIs can be different types. If \textit{iriValue} is "sameAsSourceIRI" then the function returns the source IRI; otherwise it resolves value of $iriValue$ using the \textit{resolve(i, pv, iriValue)} function---this function returns either the value of a property/expression or next incremental value---, and finally it creates a new IRI by calling the \textsc{generateIRI()} function. As some datasets  (e.g.,~\cite{galarraga2017qboairbase}) use the IRI of the level to create the IRIs of its members, whereas others (Eurostat (\url{https://ec.europa.eu/eurostat/data/database}), Greece (\url{http://linked-statistics.gr/}) linked datasets) use the IRI of the range of the level, we generalize it using Equation~\ref{eq:lmiri}. If a range exists for the target level, the IRI of that range is used as a target type--- the type of the resources for which  IRIs will be created--- for creating IRIs for the members of that level (second case in Equation~\ref{eq:lmiri}); otherwise, the level's IRI is used as a target type (third case in Equation~\ref{eq:lmiri}). Here, $range(level, tTBox)$ returns the range of $level$ from $tTBox$.

Equation~\ref{eq:lmdescription} creates a triple for each $(p_i, v_i)$ pair of \textit{Ins} (in Equation~\ref{eq:retrieveinstance}). If $p_i$ corresponds to either an object property or a rollup property in $tTBox$, then a new IRI is created for $v_i$ and the target type of the IRI is the range of the target construct that is mapped to $p_i$ (first case in Equation~\ref{eq:lmdescription}). Here, $return(x, mappings)$ function returns the target construct that is mapped to $x$ from the set of property-mappings $mappings$; $targetType(x,tTBox)$ returns the type of a target construct $x$ from \textit{tTBox}; and $iriValue(v)$ retrieves the value to be used to create the IRI. If $v$ is a literal, it simply returns it, otherwise, it splits $v$ by either ``/" or ``\#" and returns the last portion of $v$. If $p_i$ corresponds to a level attribute (i.e., datatype property), then the object of the triple generated for the level member will be the same as $v$ (second case in Equation~\ref{eq:lmdescription}). 

The output of this operation is $LM$ and the operation stores the output in $tABox$.


\vspace{0.2cm}
\noindent\textit{\underline {ObservationGenerator(sConstruct, dataset, sTBox, }}\\ \textit{\underline {sABox, tTBox iriValue, iriGraph, propertyMappings,}} \\ \textit{\underline {tABox)}}: This operation creates  QB4OLAP-complaint observations from the source data.

\textbf{Input Parameters}: $sConstruct$ is the source construct, $dataset$ is a target QB dataset, $sTBox$ and $sABox$ are the source TBox and ABox, \textit{iriValue} is the rule of creating level members' IRIs, $iriGraph$ is the IRI graph, $proeprtyMappings$ is a set of property-mappings, and $tABox$ is the output location. 

\textbf{Semantics}: First, we retrieve the instances of \textit{sConstruct} with their properties and values using Equation~\ref{eq:retrieveinstance}. To populate $dataset$ with the dictionary $Ins$, we define a set of triples $OB$ which is equivalent to $LM$ in Equation~\ref{eq:lm}.

{\setlength{\mathindent}{0cm}
\begin{multline}\label{eq:obidentity}
\scriptsize IdentityTriples= \{(oIRI, \texttt{rdf:type},\texttt{qb:Observ} \\[-3 pt]\scriptsize \texttt{ation}), (oIRI, 
\texttt{qb:dataset}, dataset)\} 
\end{multline}
}
\noindent $oIRI =$
\begin{align}\label{eq:obiri}
\scriptsize \begin{cases}
i &if\; iriValue=``sameAs\\[-3 pt]
  &SourceIRI"\\
\textsc{generateIRI}(i, resolve(i, pv,\\[-3 pt] iriValue),dataset, tTBox,iriGraph) 
  &otherwise\\
\end{cases}
\end{align}

\noindent $DescriptionTriple=$
\begin{align}\label{eq:obdescription}
\scriptsize \begin{cases}
\{(oIRI,
  return(p_i, \\[-3 pt]propertyMappings), \;
  lmIRI_{o})\}
  &if \;targetType(return(p_i, \\[-3 pt] &propertyMappings),tTBox)\\[-3 pt]&= LevelProperty \\
\{(oIRI,
  return(p_i,\\[-3 pt] propertyMappings), \;  v_i)\}
  &if \;targetType(return(p_i, \\[-3 pt] &propertyMappings), tTBox)\\[-3 pt]&= MeasureProperty
\end{cases}
\end{align}

\noindent $lmIRI_{o}=$
\begin{align}
\label{eq:lmirici}
\scriptsize \begin{cases} 
\textsc{generateIRI}(v_i, iriValue(v_i),\\[-3 pt] range(return(p_i, property-\\[-3 pt]Mappings), tTBox),\\[-3 pt] tTBox, iriGraph)
&if \;range(return (p_i, propertyM\\[-3 pt]&appings),tTBox))!= NULL\\
generateIRI(v_i, iriValue(v_i),\\[-3 pt] return(p_i,propertyMappings)\\[-3 pt],tTBox, iriGraph)
&if \;range(return (p_i, propertyM\\[-3 pt]&appings),tTBox))= NULL
\end{cases}
\end{align}

Each instance of $sConstruct$ is a \texttt{qb:Observation} and the QB dataset of that observation is $dataset$; therefore, for each instance of $sConstruct$, $OB$ in Equation~\ref{eq:lm} includes two identity triples. Equation~\ref{eq:obidentity} redefines Equation~\ref{eq:lmidentity} for $OB$. Equation~\ref{eq:obiri} describes how to create an IRI for an observation. If \textit{iriValue} is ``sameAsSourceIRI" then the function returns the source IRI; otherwise it generates an observation IRI by calling the \textsc{generateIRI()} function. Equation~\ref{eq:obdescription} redefines Equation~\ref{eq:lmdescription} for $OB$. For each $(p_i,v_i)$ in the dictionary, the equation creates a triple. If $p_i$ represents a level in the target, the predicate of the triple is the corresponding level of $p_i$ and the object of the triple is a level member of that level. Equation~\ref{eq:lmirici} shows how to create the IRI for the object (level member) of that triple. If $p_i$ represents a measure property in the target, the predicate of the triple is the corresponding target measure property of $tTBox$ and the object of the target triple is similar to $v_i$.

The output of this operation is $OB$, which will be stored in $tABox$.

\paragraph{\underline{ChangedDataCapture(nABox, oABox, flag)}} This operation triggers either the SDW evolution (i.e., enriched with new instances) or update (i.e., reflect the changes in the existing instances). 

\textbf{Input Parameters}: \textit{nABox} and \textit{oABox} are the sets of triples that represent the new source data and the old source data, \textit{flag} indicates the type of difference the operation will produced. If \textit{flag= 0}, the operation produces the new instances, and if \textit{flag= 1}, the operation produces the set of updated triples (i.e., the triples in $oABox$ updated in $nABox$). 

\textbf{Semantics}: Here, \textit{nABox} and \textit{oABox} both are a set of triples where first element of each triple corresponds to a level member in the target, second element represents a property (either a level attribute or a rollup property) and the third element represents a value of the property for that instance. First, we retrieve the sets of instances  from \textit{nABox} and \textit{oABox} using the \textsc{executeQuery()} function, shown as follows:
 $Ins_{nABox}= \textsc{executeQuery}((?i, \texttt{rdf:type}, ?v),\\ nABox, ?i)$\\
$Ins_{oABox}=\textsc{executeQuery}((?i, \texttt{rdf:type}, ?v),\\ oABox, ?i)$
 
 The set difference of $Ins_{nABox}$ and $Ins_{oABox}$ gets the new instances, i.e., $Ins_{new}= Ins_{nABox} - Ins_{oABox}$.  We get the description (properties and their values) of new instances using Equation~\ref{eq:newinsdes}. 
\begin{multline}\label{eq:newinsdes}
InsDes_{new}= \bigcup_{i \in Ins_{new}} \textsc{executeQuery}((?s,?p,?v)\\ FILTER (?s=i)),  nABox, (?s,?p,?v))
\end{multline} 
 
 To get the triples that are changed over time, we first remove the description of new instances (derived in Eq.~\ref{eq:newinsdes}) from \textit{newABox} i.e., $InsDes_{old}=newABox - InsDes_{new}$; then get the set of changed triples by taking the set difference of $InsDes_{old}$ and $oldABox$, i.e., 
 
$ChangedTriples = InsDes_{old}- oldABox$.   

 If (\textit{flag=0}) then the output is \textit{InsDes}$_{new}$, else\\ \textit{ChangedTriples}. The output overwrites $oABox$. 


\vspace{0.2 cm}

\noindent \textit{\underline{updateLevel(level, updatedTriples, sABox, tTBox, tABox,}}\\ \textit{ \underline{ propertyMappings, iriGraph)}} This operation reflect the changes in source to the SDW. 

\textbf{Input Parameters}: $level$ is a target level, \textit{updatedTriples} is the set of updated triples (generated by the \textit{ChangedDataCapture} operation), $sABox$ is the source data of $level$, $tTBox$ and $tABox$ are the target TBox and ABox, $propertyMappings$ is the set of property-mappings, and $iriGraph$ is the IRI graph. 



\textbf{Semantics}: As we consider that the changes only occur at the instance level, not at the schema level, only the 3rd elements (objects) of the triples of \textit{sABox} and \textit{updatedTriples} can be different. For the simplicity of calculation, we define \textit{sABox}, and \textit{updatedTriples} as the following relations: 

\textit{sABox = (instance, property, oldVal)}

\textit{updatedTriples = (instance, property, newVal)}


This operation updates $tABox$ by deleting invalid triples from $tABox$ and inserting new triples that reflect the changes. As SPARQL does not support any (SQL-like) update statement, we define two sets \textit{DeleteTriples} and \textit{InsertTriples} that contain the triples to be deleted from $tABox$ and inserted to $tABox$, respectively. To get the final $tABox$, we first take the set difference between $tABox$ and $DeleteTriples$ using Equation~\ref{eq:aboxafterdeletion} and then, take the union of $tABox$ and $InsertTriples$ using Equation~\ref{eq:aboxafterinsertion}. 
 
 \begin{equation}\label{eq:aboxafterdeletion}
 tABox= tABox - DeleteTriples
\end{equation} 
\vspace{-1 cm}
\begin{equation}\label{eq:aboxafterinsertion}
 tABox = tABox \cup InsertTriples
\end{equation} 

To get the updated objects and old objects of triples for source instances, we take the natural join between $updateddTriples$ and $sABox$, i.e., 

\begin{equation}\label{eq:typejoin}
 NewOldValues=updatedTriples \Join sABox
\end{equation}

We define \textit{DeleteTriples} and \textit{InsertTriples} as

\begin{equation}\label{eq:delete}
 DeleteTriples =\bigcup_{(i, p ,nV ) \in \; updatedTriples} del (i, p ,nV )
\end{equation}
\vspace{-1 cm}
\begin{equation}\label{eq:insert}
  InsertTriples =\bigcup_{(i,p,nV,oV) \in \; NewOldValues} in(i,p,nV,oV)
\end{equation}

The $del (i, p ,nV )$ and $in(i,p,nV)$ functions depends on the update type of the level attribute corresponding to $p$. We retrieve the update type of the level attribute from $tTBox$ using Equation~\ref{eq:updatetype}. The \textit{updateType(prop,tTBox)} function returns the update type of a level attribute \textit{prop} from \textit{tTBox} (see the blue-colored rectangle of Figure~\ref{fig:qb13}). 
\begin{multline}\label{eq:updatetype}
updateType= (updateType(return(p,\\ propertyMappings), tTBox)) 
\end{multline}

In the following, we describe how the functions, $del (i, p ,nV)$ and $in(i,p,nV,oV)$ are defined for different values of  \textit{updateType}.

\underline{If \textit{updateType}= Type-1 update}

If \textit{updateType} is \textit{Type-1 update}, the new value of the property will replace the old one. Therefore, the triple holding the old value for the property will be deleted from $tABox$ and a new triple with the updated value of the property will be inserted into $tABox$. Equation~\ref{eq:type1delete} defines $del (i, p ,nV )$ for the Type-1 update. For an instance $i$ in the source, the existing equivalent IRI in the target can be retrieved from the IRI graph $iriGraph$, i.e., $IRI_i= lookup(i,iriGraph)$. \textit{return(p, propertyMappings)} in Equation~\ref{eq:type1delete} returns the target level attribute that is mapped to $p$ from  \textit{propertyMappings}.
\begin{multline}\label{eq:type1delete}
del(i, p, nV)=\textsc{executeQuery}(((?i, ?p,?val) \\[-3 pt]
FILTER\; (?i=IRI_i\; \&\&\; ?p=return(p, property\\[-3 pt]Mappings))) , 
tABox, (?i \;?p \; ?val))
\end{multline}

Equation~\ref{eq:type1insert} describes how to create an RDF triple for the Type-1 update. First, we retrieve the equivalent value of $oV$ in the target using Equation~\ref{eq:objectretrieval}. As the retrieved value can be either a literal or an IRI, we create the object of the triple by replacing the $oV$ portion of that value with $nV$.  The \textit{replace(org\_str, search\_pattern, replace\_pattern)} function updates the string $org\_str$ by replacing the substring \textit{search\_pattern} with \textit{replace\_pattern}. The subject of the output RDF triple is the level member equivalent to $i$ and the predicate is the target property equivalent to $p$. 
\begin{multline}
\label{eq:type1insert}
in (i,p,nV,oV)=  \{(IRI_i, return(p, mapping), \\[-3 pt]replace(targetValue(i,p), oV, nV)\}
\end{multline}
where,\vspace{-0.5 cm}
\begin{multline}\label{eq:objectretrieval}
targetValue(i,p)= \textsc{executeQuery}((IRI_i, \\return(p,mapping),?v), tABox, ?v)
\end{multline}

\underline{If \textit{updateType}= Type-3 update}

Like the Type-1 update, Type-3 update also replaces the old property value with the current one. Besides, it also keeps the old property value as the latest old property value. Therefore, the triples that contain  the current and old property value should be removed from the target. Equation~\ref{eq:type3delete} defines \textit{del(i,p,nV)} for the Type-3 update, which retrieves the triples containing both current and old property values. 

As besides new value Type-3 keeps the latest old value of the property by adding another property, we define \textit{in(i,p,nV,oV)} in Equation~\ref{eq:type3 insert} which creates a triple for the new value and a triple for the old one. For a property $property$,  \textit{concat(property, ``\_oldValue")} creates a property to contain the old value of the $property$ by  concatenating the $``\_oldValue"$ with $property$. The function $targetValue(i,p)$ returns the objects of triple whose subject and predicate are corresponds to $i$ and $p$, defined in Equation~\ref{eq:objectretrieval}.

\begin{multline}\label{eq:type3delete}
del(i, p, nV)=\textsc{executeQuery}(((?i,?p,?val) \\
FILTER(?i=IRI_i\; \&\&\; ?p \;\in(return(p, property\\Mappings), concat(return(p, propertyMappings),\\``\_oldValue"))), 
tABox, (?i,?p,?val))
\end{multline}
\vspace{-1 cm}
\begin{multline}
\label{eq:type3 insert}
in(i,p,nV,oV)= \{(IRI_i, return(p, mapping), \\replace(targetValue(i,p), oV, nV),
(IRI_i, concat\\(return(p, propertyMappings), ``\_oldValue"), \\targetValue(i,p))
\}
\end{multline}

\underline{If \textit{updateType}= Type-2 update}

In Type-2 update, a new version for the level member is created (i.e., it keeps the previous version and creates a new updated version). Since the validity interval (defined by \texttt{type2:toDate}) and  status (defined by \texttt{type2:status}) of the previous version need to be updated, triples describing the validity interval and status of the previous version should be deleted from $tABox$. Equation~\ref{eq:type2delete} defines $del(i,p,nV)$ for Type-2 update. The first operand of the union in Equation~\ref{eq:type2delete} retrieves the expired triples of $IRI_i$ (the level member corresponding to $i$) from $tABox$. As the level of $IRI_i$ can be an upper level in a hierarchy, the validity intervals and the status of the level members of lower levels referring to $IRI_i$ need to be updated, too. Therefore, the current validity intervals and status of the associated level members will also be deleted. 
The second operand of union in Equation~\ref{eq:type2delete} $getExpiredTriplesAll(IRI_i)$ returns the set of expired triples of the members of all lower levels referring to  $IRI_i$ (described in Equation~\ref{eq:expiredall}). The function $getTriplesImmediate(IRI_i)$  returns the set of expired triples of the members of immediate child level referring to $IRI_i$ (described in Equation~\ref{eq:expiredintermediate}).   
\begin{multline}\label{eq:type2delete}
 del(i,p,nV)=\textsc{executeQuery}(((IRI_i, ?p,?val)\\ FILTER \; (?p\; \in \;(\texttt{type2:toDate},\\\texttt{type2:status}))), tABox, (IRI_i, ?p, ?val))\\ \cup\;  getExpiredTriplesAll(IRI_i)
\end{multline}
\vspace{-1 cm}
\begin{multline}\label{eq:expiredall}
\scriptsize getExpiredTriplesAll(IRI_i)= \\ \bigcup_{(i_c \;p \;v) \;\in\; getExpiredTriplesImmediate(IRI_i)} \\getExpiredTriplesAll(i_c)
\end{multline}
\vspace{-1 cm}
\begin{multline}\label{eq:expiredintermediate}
\scriptsize getExpiredTriplesImmediate(IRI_i)=\scriptsize \\\textsc{executeQuery}((?i_c,?p,IRI_i)\;AND\; (?i_c,\\  \texttt{rdf:type}, \texttt{qb4o:LevelMember})\;AND\; (?i_c,\\?p,?v)  \;FILTER \;( ?p \;\in\;(\texttt{type2:toDate},\\ \texttt{type2:status}))), tABox, (?i_c,?p,?val)) 
\end{multline}
 To reflect the changes in the target, a new version of the level member (with the changed and unchanged property values) is created. Equation~\ref{eq:type2insert} defines  \textit{in(i,p,nV,oV)} for Type-2 update. The IRI for the new version of $IRI_i$ is \textit{$newIRI_i$ = updateIRI($IRI_i$, iriGraph)}. The \textit{updateIRI(iri, iriGraph)} function updates the existing IRI $iri$ by appending the current date with it and returns the updated one. An RDF triple is generated for $nV$, where the object is composed by replacing the old value with the new value, the predicate is the target level attribute equivalent to $p$, and the subject is $newIRI_i$ (first operand of the union operation in Equation~\ref{eq:type2insert}). The call of \textit{update2Insert($iri_o, iri_n$,cp)} in Equation~\ref{eq:type2insert} returns the triples required to reflect the changes in $tABox$. Here, $iri_o$ and $iri_n$ are the old and new version (IRI) of a level member and $cp$ is the predicate of the updated triple (i.e., the property whose value has been changed). The values of \texttt{type2:status} for new and existing versions are set to ``Current" and ``Expired", respectively. The validity interval of existing version is ended on \textit{sysdate()-1} and the validity interval of new version is from \textit{sysdate()} to ``9999-12-31". The \textit{sysdate()} returns the current date in \textit{year-month-day} format as we consider that at most one update operation can be applied on the SDW in a day. The value ``9999-12-31" in the \texttt{type2:toDate} indicates that the instance is still valid; this is a usual notation in temporal databases~\cite{vaisman2014data}. As the new version of the level member also contains the triples for its unchanged properties, we retrieve the triples of a level member's old version $iri_o$ with unchanged properties using \textit{retrieveTriples($iri_o, cp$)} which returns all triples of $iri_o$ except the triples with the properties \textit{cp} (changed property), \texttt{type2:toDate, type2:fromDate,} and \texttt{type2:status} (described in Equation~\ref{eq:retrievetriple}). We replace the IRI of old version with the new one using the \textit{replace()} function. To propagate the current version to its descendants (i.e., the level members those are associated to $iri_o$); the function \textit{updateAssociates($iri_o, iri_n$)} creates new versions for the associated members of all lower levels in the hierarchy so that they refer to the correct version (described in Eq.~\ref{eq:connectedinstanceall}). The function \textit{getAssociates($iri_o$)} returns the set of triples that describe the members that are connected to $iri_o$ with a property. For each member connected to $iri_o$, we create a new version of it and also recursively update the members dependent on it using Equation~\ref{eq:deriveA}.


\begin{multline}
\label{eq:type2insert}
in(i,p,nV,oV)=\\\{(newIRI_i, return(p, propertyMappings), \\replace(targetValue(i,p),oV, nV)\} \;\cup\\ update2Insert(IRI_i, newIRI_i, return(p, \\propertyMappings))
\end{multline}
\vspace{-1 cm}
\begin{multline}\label{eq:update2insert}
update2Insert(iri_o, iri_n, cp)=\\
\{
(iri_o, \texttt{type2:status}, ``Expired"),\\
(iri_o,  \texttt{type2:toDate}, sysdate()-1),\\
 (iri_n , \texttt{type2:fromDate}, sysdate()), \\
 (iri_n , \texttt{type2:toDate}, ``9999-12-31"),\\
(iri_n, \texttt{type2:status}, ``Current")
\}\\ \cup \; replace(retrieveTriples(iri_o, cp), iri_o, iri_n))\\ \cup\; 
updateAssociates(iri_o, iri_n)
\end{multline}
\vspace{-1 cm}
\begin{multline}
\label{eq:retrievetriple}
retrieveTriples(lm,cp)=\textsc{executeQuery}(((lm, \\?p, ?v)\;FILTER\; (?p\; \notin \;(cp, \texttt{type2:toDate}, \\ \texttt{type2:fromDate}, \texttt{type2:status}))),\\ tABox, (lm, ?p, ?v))
\end{multline}
\vspace{-1 cm}
\begin{multline}\label{eq:connectedinstanceall}
updateAssociates(iri_o, iri_n)=\\ \bigcup_{(i_c, p)\in getAssociates(iri_o)}  \scriptsize recursiveUpdate(i_c, p)
\end{multline}
\vspace{-1 cm}
\begin{multline}\label{eq:connectedinstancesintermediate}
getAssociates(iri_o)= \textsc{executeQuery}\\(
((?i_c,?p,iri_o)\; AND\; (?i_c,\texttt{rdf:type}, \texttt{qb4o:}\\\texttt{LevelMember})), 
tABox, (?i_c, ?p)
)
\end{multline}
\vspace{-1 cm}
\begin{multline}\label{eq:deriveA}
 recursiveUpdate(i_c, p) =\\
\{(updateIRI(i_c,iriGraph), p, i_c)\} \;\cup \\ update2Insert(i_c, updateIRI(i_c, iriGraph), p)
\end{multline}

This operation updates the target ABox, $tABox$.

We refer to~\cite{nath2017setl} for the semantics of the \textit{ExternalLinking} operation and to~\cite{harth2014linked} for the semantics of the \textit{MaterializeInference} operation.
\end{appendix}

\bibliographystyle{ios1}

\bibliography{references}

\begin{thebibliography}{56}
\ifx \bisbn   \undefined \def \bisbn  #1{ISBN #1}\fi
\ifx \binits  \undefined \def \binits#1{#1} \fi
\ifx \bauthor  \undefined \def \bauthor#1{#1} \fi
\ifx \bjtitle  \undefined \def \bjtitle#1{\textit{#1}}\fi
\ifx \batitle  \undefined \def \batitle#1{#1} \fi
\ifx \bctitle  \undefined \def \bctitle#1{#1} \fi
\ifx \bvolume  \undefined \def \bvolume#1{\textbf{#1}}\fi
\ifx \byear  \undefined \def \byear#1{#1} \fi
\ifx \bissue  \undefined \def \bissue#1{#1} \fi
\ifx \bfpage  \undefined \def \bfpage#1{#1} \fi
\ifx \blpage  \undefined \def \blpage #1{#1} \fi
\ifx \burl  \undefined \def \burl#1{#1} \fi
\ifx \doiurl  \undefined \def \doiurl#1{#1} \fi
\ifx \betal  \undefined \def \betal{et al.} \fi
\ifx \binstitute  \undefined \def \binstitute#1{#1} \fi
\ifx \beditor  \undefined \def \beditor#1{#1} \fi
\ifx \bpublisher  \undefined \def \bpublisher#1{#1} \fi
\ifx \bbtitle  \undefined \def \bbtitle#1{\textit{#1}} \fi
\ifx \bedition  \undefined \def \bedition#1{#1} \fi
\ifx \bseriesno  \undefined \def \bseriesno#1{#1} \fi
\ifx \blocation  \undefined \def \blocation#1{#1} \fi
\ifx \bsertitle  \undefined \def \bsertitle#1{#1} \fi
\ifx \bsnm \undefined \def \bsnm#1{#1} \fi
\ifx \bsuffix \undefined \def \bsuffix#1{#1} \fi
\ifx \bparticle \undefined \def \bparticle#1{#1} \fi
\ifx \barticle \undefined \def \barticle#1{#1} \fi
\ifx \botherref \undefined \def \botherref #1{#1} \fi
\ifx \url \undefined \def \url#1{#1} \fi
\ifx \bchapter \undefined \def \bchapter#1{#1} \fi
\ifx \bbook \undefined \def \bbook#1{#1} \fi
\ifx \bcomment \undefined \def \bcomment#1{#1} \fi
\ifx \oauthor \undefined \def \oauthor#1{#1} \fi
\ifx \citeauthoryear \undefined \def \citeauthoryear#1{#1} \fi
\ifx \texttildelow  \undefined \def \texttildelow{\symbol{126}} \fi
\def \endbibitem {}
\ifx \bconflocation  \undefined \def \bconflocation#1{#1} \fi

\bibitem{hartig2019linked}
\begin{botherref}
\oauthor{\binits{O.}~\bsnm{Hartig}},
\oauthor{\binits{K.}~\bsnm{Hose}} and
\oauthor{\binits{J.F.}~\bsnm{Sequeda}},
{Linked Data Management.},
2019.
\end{botherref}
\endbibitem

\bibitem{deb2020setlbi}
\begin{bchapter}
\bauthor{\binits{R.P.}~\bsnm{Deb~Nath}},
\bauthor{\binits{K.}~\bsnm{Hose}},
\bauthor{\binits{T.B.}~\bsnm{Pedersen}},
\bauthor{\binits{O.}~\bsnm{Romero}} and
\bauthor{\binits{A.}~\bsnm{Bhattacharjee}},
\bctitle{{SETL$_{BI}$: An Integrated Platform for Semantic Business
  Intelligence}},
in: \bbtitle{Companion Proceedings of the Web Conference 2020},
\byear{2020},
pp.~\bfpage{167}--\blpage{171}.
\end{bchapter}
\endbibitem

\bibitem{kalampokis2015challenges}
\begin{bchapter}
\bauthor{\binits{E.}~\bsnm{Kalampokis}},
\bauthor{\binits{B.}~\bsnm{Roberts}},
\bauthor{\binits{A.}~\bsnm{Karamanou}},
\bauthor{\binits{E.}~\bsnm{Tambouris}} and
\bauthor{\binits{K.A.}~\bsnm{Tarabanis}},
\bctitle{{Challenges on Developing Tools for Exploiting Linked Open Data
  Cubes.}},
in: \bbtitle{SemStats@ ISWC},
\byear{2015}.
\end{bchapter}
\endbibitem

\bibitem{nebot2012building}
\begin{barticle}
\bauthor{\binits{V.}~\bsnm{Nebot}} and
\bauthor{\binits{R.}~\bsnm{Berlanga}},
\batitle{{Building Data Warehouses with Semantic Web Data}},
\bjtitle{Decision Support Systems}
\bvolume{52}(\bissue{4})
(\byear{2012}),
\bfpage{853}--\blpage{868}.
\end{barticle}
\endbibitem

\bibitem{heath2011linked}
\begin{barticle}
\bauthor{\binits{T.}~\bsnm{Heath}} and
\bauthor{\binits{C.}~\bsnm{Bizer}},
\batitle{{Linked Data: Evolving the Web into a Global Data Space}},
\bjtitle{Synthesis lectures on the semantic web: theory and technology}
\bvolume{1}(\bissue{1})
(\byear{2011}),
\bfpage{1}--\blpage{136}.
\end{barticle}
\endbibitem

\bibitem{berlanga2011semantic}
\begin{botherref}
\oauthor{\binits{R.}~\bsnm{Berlanga}},
\oauthor{\binits{O.}~\bsnm{Romero}},
\oauthor{\binits{A.}~\bsnm{Simitsis}},
\oauthor{\binits{V.}~\bsnm{Nebot}},
\oauthor{\binits{T.}~\bsnm{Pedersen}},
\oauthor{\binits{A.}~\bsnm{Abell{\'o}~Gamazo}} and
\oauthor{\binits{M.J.}~\bsnm{Aramburu}},
{Semantic Web Technologies for Business Intelligence}
(2011).
\end{botherref}
\endbibitem

\bibitem{varga2016dimensional}
\begin{barticle}
\bauthor{\binits{J.}~\bsnm{Varga}},
\bauthor{\binits{A.A.}~\bsnm{Vaisman}},
\bauthor{\binits{O.}~\bsnm{Romero}},
\bauthor{\binits{L.}~\bsnm{Etcheverry}},
\bauthor{\binits{T.B.}~\bsnm{Pedersen}} and
\bauthor{\binits{C.}~\bsnm{Thomsen}},
\batitle{{Dimensional Enrichment of Statistical Linked Open Data}},
\bjtitle{Journal of Web Semantics}
\bvolume{40}
(\byear{2016}),
\bfpage{22}--\blpage{51}.
\end{barticle}
\endbibitem

\bibitem{ciferri2013cube}
\begin{barticle}
\bauthor{\binits{C.}~\bsnm{Ciferri}},
\bauthor{\binits{R.}~\bsnm{Ciferri}},
\bauthor{\binits{L.}~\bsnm{G{\'o}mez}},
\bauthor{\binits{M.}~\bsnm{Schneider}},
\bauthor{\binits{A.}~\bsnm{Vaisman}} and
\bauthor{\binits{E.}~\bsnm{Zim{\'a}nyi}},
\batitle{{Cube Algebra: A Generic User-centric Model and Query Language for
  OLAP Cubes}},
\bjtitle{International Journal of Data Warehousing and Mining (IJDWM)}
\bvolume{9}(\bissue{2})
(\byear{2013}),
\bfpage{39}--\blpage{65}.
\end{barticle}
\endbibitem

\bibitem{etcheverry2015modeling}
\begin{botherref}
\oauthor{\binits{L.}~\bsnm{Etcheverry}},
\oauthor{\binits{S.S.}~\bsnm{Gomez}} and
\oauthor{\binits{A.}~\bsnm{Vaisman}},
{Modeling and Querying Data Cubes on the Semantic Web},
\textit{arXiv preprint arXiv:1512.06080}
(2015).
\end{botherref}
\endbibitem

\bibitem{nath2017setl}
\begin{barticle}
\bauthor{\binits{R.P.D.}~\bsnm{Nath}},
\bauthor{\binits{K.}~\bsnm{Hose}},
\bauthor{\binits{T.B.}~\bsnm{Pedersen}} and
\bauthor{\binits{O.}~\bsnm{Romero}},
\batitle{{SETL: A Programmable Semantic Extract-Transform-Load Framework for
  Semantic Data Warehouses}},
\bjtitle{Information Systems}
\bvolume{68}
(\byear{2017}),
\bfpage{17}--\blpage{43}.
\end{barticle}
\endbibitem

\bibitem{harth2014linked}
\begin{bbook}
\bauthor{\binits{A.}~\bsnm{Harth}},
\bauthor{\binits{K.}~\bsnm{Hose}} and
\bauthor{\binits{R.}~\bsnm{Schenkel}},
\bbtitle{{Linked Data Management}},
\bpublisher{CRC Press},
\byear{2014}.
\end{bbook}
\endbibitem

\bibitem{baader2003description}
\begin{bbook}
\bauthor{\binits{F.}~\bsnm{Baader}},
\bauthor{\binits{D.}~\bsnm{Calvanese}},
\bauthor{\binits{D.}~\bsnm{McGuinness}},
\bauthor{\binits{P.}~\bsnm{Patel-Schneider}} and
\bauthor{\binits{D.}~\bsnm{Nardi}},
\bbtitle{{The Description Logic Handbook: Theory, Implementation and
  Applications}},
\bpublisher{Cambridge university press},
\byear{2003}.
\end{bbook}
\endbibitem

\bibitem{cyganiak2014rdf}
\begin{botherref}
\oauthor{\binits{R.}~\bsnm{Cyganiak}},
\oauthor{\binits{D.}~\bsnm{Reynolds}} and
\oauthor{\binits{J.}~\bsnm{Tennison}},
{The RDF Data Cube Vocabulary},
\textit{W3C Recommendation, W3C (Jan. 2014)}
(2014).
\end{botherref}
\endbibitem

\bibitem{kimball1996data}
\begin{bbook}
\bauthor{\binits{R.}~\bsnm{Kimball}},
\bbtitle{{The Data Warehouse Toolkit: Practical Techniques for Building
  Dimensional Data Warehouses}},
\bpublisher{John Wiley \& Sons, Inc.},
\byear{1996}.
\end{bbook}
\endbibitem

\bibitem{nath2015towards}
\begin{bchapter}
\bauthor{\binits{R.P.D.}~\bsnm{Nath}},
\bauthor{\binits{K.}~\bsnm{Hose}} and
\bauthor{\binits{T.B.}~\bsnm{Pedersen}},
\bctitle{{Towards a Programmable Semantic Extract-Transform-Load Framework for
  Semantic Data Warehouses}},
in: \bbtitle{Acm Eighteenth International Workshop on Data Warehousing and Olap
  (dolap 2015)},
\binstitute{Association for Computing Machinery},
\byear{2015}.
\end{bchapter}
\endbibitem

\bibitem{vaisman2014data}
\begin{bbook}
\bauthor{\binits{A.}~\bsnm{Vaisman}} and
\bauthor{\binits{E.}~\bsnm{Zim{\'a}nyi}},
\bbtitle{{Data Warehouse Systems: Design and Implementation}},
\bpublisher{Springer},
\byear{2014}.
\end{bbook}
\endbibitem

\bibitem{li2008rimom}
\begin{barticle}
\bauthor{\binits{J.}~\bsnm{Li}},
\bauthor{\binits{J.}~\bsnm{Tang}},
\bauthor{\binits{Y.}~\bsnm{Li}} and
\bauthor{\binits{Q.}~\bsnm{Luo}},
\batitle{{Rimom: A Dynamic Multistrategy Ontology Alignment Framework}},
\bjtitle{IEEE Transactions on Knowledge and data Engineering}
\bvolume{21}(\bissue{8})
(\byear{2008}),
\bfpage{1218}--\blpage{1232}.
\end{barticle}
\endbibitem

\bibitem{harris2013sparql}
\begin{botherref}
\oauthor{\binits{S.}~\bsnm{Harris}},
\oauthor{\binits{A.}~\bsnm{Seaborne}} and
\oauthor{\binits{E.}~\bsnm{Prud’hommeaux}},
{SPARQL 1.1 Query Language},
\textit{W3C recommendation}
\textbf{21}(10)
(2013).
\end{botherref}
\endbibitem

\bibitem{kostylev2015construct}
\begin{bchapter}
\bauthor{\binits{E.V.}~\bsnm{Kostylev}},
\bauthor{\binits{J.L.}~\bsnm{Reutter}} and
\bauthor{\binits{M.}~\bsnm{Ugarte}},
\bctitle{{CONSTRUCT Queries in SPARQL}},
in: \bbtitle{18th International Conference on Database Theory (ICDT 2015)},
\binstitute{Schloss Dagstuhl-Leibniz-Zentrum fuer Informatik},
\byear{2015}.
\end{bchapter}
\endbibitem

\bibitem{seddiqui2011augmentation}
\begin{bchapter}
\bauthor{\binits{M.H.}~\bsnm{Seddiqui}},
\bauthor{\binits{S.}~\bsnm{Das}},
\bauthor{\binits{I.}~\bsnm{Ahmed}},
\bauthor{\binits{R.P.D.}~\bsnm{Nath}} and
\bauthor{\binits{M.}~\bsnm{Aono}},
\bctitle{{Augmentation of Ontology Instance Matching by Automatic Weight
  Generation}},
in: \bbtitle{Information and Communication Technologies (WICT), 2011 World
  Congress on},
\binstitute{IEEE},
\byear{2011},
pp.~\bfpage{1390}--\blpage{1395}.
\end{bchapter}
\endbibitem

\bibitem{nath2012resolving}
\begin{bchapter}
\bauthor{\binits{R.P.D.}~\bsnm{Nath}},
\bauthor{\binits{H.}~\bsnm{Seddiqui}} and
\bauthor{\binits{M.}~\bsnm{Aono}},
\bctitle{{Resolving Scalability Issue to Ontology Instance Matching in Semantic
  Web}},
in: \bbtitle{2012 15th International Conference on Computer and Information
  Technology (ICCIT)},
\binstitute{IEEE},
\byear{2012},
pp.~\bfpage{396}--\blpage{404}.
\end{bchapter}
\endbibitem

\bibitem{nath2014efficient}
\begin{barticle}
\bauthor{\binits{R.P.D.}~\bsnm{Nath}},
\bauthor{\binits{M.H.}~\bsnm{Seddiqui}} and
\bauthor{\binits{M.}~\bsnm{Aono}},
\batitle{{An Efficient and Scalable Approach for Ontology Instance Matching}},
\bjtitle{JCP}
\bvolume{9}(\bissue{8})
(\byear{2014}),
\bfpage{1755}--\blpage{1768}.
\end{barticle}
\endbibitem

\bibitem{polleres2013rdfs}
\begin{bchapter}
\bauthor{\binits{A.}~\bsnm{Polleres}},
\bauthor{\binits{A.}~\bsnm{Hogan}},
\bauthor{\binits{R.}~\bsnm{Delbru}} and
\bauthor{\binits{J.}~\bsnm{Umbrich}},
\bctitle{{RDFS and OWL Reasoning for Linked Data}},
in: \bbtitle{Reasoning Web. Semantic Technologies for Intelligent Data Access},
\bpublisher{Springer},
\byear{2013},
pp.~\bfpage{91}--\blpage{149}.
\end{bchapter}
\endbibitem

\bibitem{casters2010pentaho}
\begin{bbook}
\bauthor{\binits{M.}~\bsnm{Casters}},
\bauthor{\binits{R.}~\bsnm{Bouman}} and
\bauthor{\binits{J.}~\bsnm{Van~Dongen}},
\bbtitle{{Pentaho Kettle Solutions: Building Open Source ETL Solutions with
  Pentaho Data Integration}},
\bpublisher{John Wiley \& Sons},
\byear{2010}.
\end{bbook}
\endbibitem

\bibitem{andersen2014publishing}
\begin{bchapter}
\bauthor{\binits{A.B.}~\bsnm{Andersen}},
\bauthor{\binits{N.}~\bsnm{G{\"u}r}},
\bauthor{\binits{K.}~\bsnm{Hose}},
\bauthor{\binits{K.A.}~\bsnm{Jakobsen}} and
\bauthor{\binits{T.B.}~\bsnm{Pedersen}},
\bctitle{{Publishing Danish Agricultural Government Data as Semantic Web
  Data}},
in: \bbtitle{Joint International Semantic Technology Conference},
\binstitute{Springer},
\byear{2014},
pp.~\bfpage{178}--\blpage{186}.
\end{bchapter}
\endbibitem

\bibitem{abello2014using}
\begin{barticle}
\bauthor{\binits{A.}~\bsnm{Abell{\'o}}},
\bauthor{\binits{O.}~\bsnm{Romero}},
\bauthor{\binits{T.B.}~\bsnm{Pedersen}},
\bauthor{\binits{R.}~\bsnm{Berlanga}},
\bauthor{\binits{V.}~\bsnm{Nebot}},
\bauthor{\binits{M.J.}~\bsnm{Aramburu}} and
\bauthor{\binits{A.}~\bsnm{Simitsis}},
\batitle{{Using Semantic Web Technologies for Exploratory OLAP: A Survey}},
\bjtitle{IEEE transactions on knowledge and data engineering}
\bvolume{27}(\bissue{2})
(\byear{2014}),
\bfpage{571}--\blpage{588}.
\end{barticle}
\endbibitem

\bibitem{jakobsen2015optimizing}
\begin{bchapter}
\bauthor{\binits{K.A.}~\bsnm{Jakobsen}},
\bauthor{\binits{A.B.}~\bsnm{Andersen}},
\bauthor{\binits{K.}~\bsnm{Hose}} and
\bauthor{\binits{T.B.}~\bsnm{Pedersen}},
\bctitle{Optimizing RDF Data Cubes for Efficient Processing of Analytical
  Queries.},
in: \bbtitle{COLD},
\byear{2015}.
\end{bchapter}
\endbibitem

\bibitem{skoutas2007ontology}
\begin{barticle}
\bauthor{\binits{D.}~\bsnm{Skoutas}} and
\bauthor{\binits{A.}~\bsnm{Simitsis}},
\batitle{{Ontology-Based Conceptual Design of ETL Processes for both Structured
  and Semi-structured Data}},
\bjtitle{IJSWIS}
\bvolume{3}(\bissue{4})
(\byear{2007}),
\bfpage{1}--\blpage{24}.
\end{barticle}
\endbibitem

\bibitem{bellatreche2013semantic}
\begin{bchapter}
\bauthor{\binits{L.}~\bsnm{Bellatreche}},
\bauthor{\binits{S.}~\bsnm{Khouri}} and
\bauthor{\binits{N.}~\bsnm{Berkani}},
\bctitle{{Semantic Data Warehouse Design: From ETL to Eeployment A La Carte}},
in: \bbtitle{International Conference on Database Systems for Advanced
  Applications},
\binstitute{Springer},
\byear{2013},
pp.~\bfpage{64}--\blpage{83}.
\end{bchapter}
\endbibitem

\bibitem{thenmozhi2014ontological}
\begin{barticle}
\bauthor{\binits{M.}~\bsnm{Thenmozhi}} and
\bauthor{\binits{K.}~\bsnm{Vivekanandan}},
\batitle{{An Ontological Approach to Handle Multidimensional Schema Evolution
  for Data Warehouse}},
\bjtitle{International Journal of Database Management Systems}
\bvolume{6}(\bissue{3})
(\byear{2014}),
\bfpage{33}.
\end{barticle}
\endbibitem

\bibitem{bansal2014towards}
\begin{bchapter}
\bauthor{\binits{S.K.}~\bsnm{Bansal}},
\bctitle{{Towards a Semantic Extract-Transform-Load (ETL) Framework for Big
  Data Integration}},
in: \bbtitle{Big Data},
\byear{2014},
pp.~\bfpage{522}--\blpage{529}.
\end{bchapter}
\endbibitem

\bibitem{knap2018unifiedviews}
\begin{barticle}
\bauthor{\binits{T.}~\bsnm{Knap}},
\bauthor{\binits{P.}~\bsnm{Hane{\v{c}}{\'a}k}},
\bauthor{\binits{J.}~\bsnm{Kl{\'\i}mek}},
\bauthor{\binits{C.}~\bsnm{Mader}},
\bauthor{\binits{M.}~\bsnm{Ne{\v{c}}ask{\`y}}},
\bauthor{\binits{B.}~\bsnm{Van~Nuffelen}} and
\bauthor{\binits{P.}~\bsnm{{\v{S}}koda}},
\batitle{{UnifiedViews: An ETL Tool for RDF Data Management}},
\bjtitle{Semantic Web}
\bvolume{9}(\bissue{5})
(\byear{2018}),
\bfpage{661}--\blpage{676}.
\end{barticle}
\endbibitem

\bibitem{colazzo2014rdf}
\begin{bchapter}
\bauthor{\binits{D.}~\bsnm{Colazzo}},
\bauthor{\binits{F.}~\bsnm{Goasdou{\'e}}},
\bauthor{\binits{I.}~\bsnm{Manolescu}} and
\bauthor{\binits{A.}~\bsnm{Roati{\c{s}}}},
\bctitle{{RDF Aanalytics: Lenses over Semantic Graphs}},
in: \bbtitle{WWW},
\binstitute{ACM},
\byear{2014},
pp.~\bfpage{467}--\blpage{478}.
\end{bchapter}
\endbibitem

\bibitem{hilal2017olap}
\begin{bchapter}
\bauthor{\binits{M.}~\bsnm{Hilal}},
\bauthor{\binits{C.G.}~\bsnm{Schuetz}} and
\bauthor{\binits{M.}~\bsnm{Schrefl}},
\bctitle{{An OLAP Endpoint for RDF Data Analysis Using Analysis Graphs.}},
in: \bbtitle{ISWC},
\byear{2017}.
\end{bchapter}
\endbibitem

\bibitem{petrou2014publishing}
\begin{botherref}
\oauthor{\binits{I.}~\bsnm{Petrou}} and
\oauthor{\binits{G.}~\bsnm{Papastefanatos}},
{Publishing Greek Census Data as linked open data},
\textit{ERCIM News}
\textbf{96}
(2014).
\end{botherref}
\endbibitem

\bibitem{qb4olap2014modeling}
\begin{bchapter}
\bauthor{\binits{L.}~\bsnm{Etcheverry}},
\bauthor{\binits{A.}~\bsnm{Vaisman}} and
\bauthor{\binits{E.}~\bsnm{Zim{\'a}nyi}},
\bctitle{{Modeling and Querying Data Warehouses on the Semantic Web Using
  QB4OLAP}},
in: \bbtitle{DaWak},
\byear{2014},
pp.~\bfpage{45}--\blpage{56}.
\end{bchapter}
\endbibitem

\bibitem{galarraga2017qboairbase}
\begin{bchapter}
\bauthor{\binits{L.}~\bsnm{Gal{\'a}rraga}},
\bauthor{\binits{K.A.M.}~\bsnm{Mathiassen}} and
\bauthor{\binits{K.}~\bsnm{Hose}},
\bctitle{{QBOAirbase: The European Air Quality Database as an RDF Cube}},
in: \bbtitle{International Semantic Web Conference (Posters, Demos \& Industry
  Tracks)},
\byear{2017}.
\end{bchapter}
\endbibitem

\bibitem{ibragimov2014towards}
\begin{bchapter}
\bauthor{\binits{D.}~\bsnm{Ibragimov}},
\bauthor{\binits{K.}~\bsnm{Hose}},
\bauthor{\binits{T.B.}~\bsnm{Pedersen}} and
\bauthor{\binits{E.}~\bsnm{Zim{\'a}nyi}},
\bctitle{{Towards Exploratory OLAP over Linked Open Data--A Case Study}},
in: \bbtitle{Enabling Real-Time Business Intelligence},
\bpublisher{Springer},
\byear{2014},
pp.~\bfpage{114}--\blpage{132}.
\end{bchapter}
\endbibitem

\bibitem{kampgen2012interacting}
\begin{bchapter}
\bauthor{\binits{B.}~\bsnm{K{\"a}mpgen}},
\bauthor{\binits{S.}~\bsnm{O’Riain}} and
\bauthor{\binits{A.}~\bsnm{Harth}},
\bctitle{{Interacting with Statistical Linked Data via OLAP Operations}},
in: \bbtitle{Extended Semantic Web Conference},
\binstitute{Springer},
\byear{2012},
pp.~\bfpage{87}--\blpage{101}.
\end{bchapter}
\endbibitem

\bibitem{ibragimov2015processing}
\begin{bchapter}
\bauthor{\binits{D.}~\bsnm{Ibragimov}},
\bauthor{\binits{K.}~\bsnm{Hose}},
\bauthor{\binits{T.B.}~\bsnm{Pedersen}} and
\bauthor{\binits{E.}~\bsnm{Zim{\'a}nyi}},
\bctitle{Processing aggregate queries in a federation of SPARQL endpoints},
in: \bbtitle{European Semantic Web Conference},
\binstitute{Springer},
\byear{2015},
pp.~\bfpage{269}--\blpage{285}.
\end{bchapter}
\endbibitem

\bibitem{ibragimov2016optimizing}
\begin{bchapter}
\bauthor{\binits{D.}~\bsnm{Ibragimov}},
\bauthor{\binits{K.}~\bsnm{Hose}},
\bauthor{\binits{T.B.}~\bsnm{Pedersen}} and
\bauthor{\binits{E.}~\bsnm{Zim{\'a}nyi}},
\bctitle{Optimizing aggregate SPARQL queries using materialized RDF views},
in: \bbtitle{International Semantic Web Conference},
\binstitute{Springer},
\byear{2016},
pp.~\bfpage{341}--\blpage{359}.
\end{bchapter}
\endbibitem

\bibitem{galarraga2018answering}
\begin{bchapter}
\bauthor{\binits{L.}~\bsnm{Gal{\'a}rraga}},
\bauthor{\binits{K.}~\bsnm{Ahlstr{\o}m}},
\bauthor{\binits{K.}~\bsnm{Hose}} and
\bauthor{\binits{T.B.}~\bsnm{Pedersen}},
\bctitle{{Answering Provenance-Aware Queries on RDF Data Cubes under Memory
  Budgets}},
in: \bbtitle{International Semantic Web Conference},
\binstitute{Springer},
\byear{2018},
pp.~\bfpage{547}--\blpage{565}.
\end{bchapter}
\endbibitem

\bibitem{cudreleveraging}
\begin{botherref}
\oauthor{\binits{P.}~\bsnm{Cudr{\'e}-Mauroux}},
{Leveraging Knowledge Graphs for Big Data Integration: the XI Pipeline},

\textit{Semantic Web},
1--5.
\end{botherref}
\endbibitem

\bibitem{rouces2017framebase}
\begin{barticle}
\bauthor{\binits{J.}~\bsnm{Rouces}},
\bauthor{\binits{G.}~\bsnm{De~Melo}} and
\bauthor{\binits{K.}~\bsnm{Hose}},
\batitle{{FrameBase: Enabling Integration of Heterogeneous Knowledge}},
\bjtitle{Semantic Web}
\bvolume{8}(\bissue{6})
(\byear{2017}),
\bfpage{817}--\blpage{850}.
\end{barticle}
\endbibitem

\bibitem{rouces2016heuristics}
\begin{bchapter}
\bauthor{\binits{J.}~\bsnm{Rouces}},
\bauthor{\binits{G.}~\bsnm{De~Melo}} and
\bauthor{\binits{K.}~\bsnm{Hose}},
\bctitle{{Heuristics for Connecting Heterogeneous Knowledge via FrameBase}},
in: \bbtitle{European Semantic Web Conference},
\binstitute{Springer},
\byear{2016},
pp.~\bfpage{20}--\blpage{35}.
\end{bchapter}
\endbibitem

\bibitem{dimou2014rml}
\begin{botherref}
\oauthor{\binits{A.}~\bsnm{Dimou}},
\oauthor{\binits{M.}~\bsnm{Vander~Sande}},
\oauthor{\binits{P.}~\bsnm{Colpaert}},
\oauthor{\binits{R.}~\bsnm{Verborgh}},
\oauthor{\binits{E.}~\bsnm{Mannens}} and
\oauthor{\binits{R.}~\bsnm{Van~de Walle}},
{RML: a generic language for integrated RDF mappings of heterogeneous data}
(2014).
\end{botherref}
\endbibitem

\bibitem{sequeda2013ultrawrap}
\begin{barticle}
\bauthor{\binits{J.F.}~\bsnm{Sequeda}} and
\bauthor{\binits{D.P.}~\bsnm{Miranker}},
\batitle{{Ultrawrap: SPARQL execution on relational data}},
\bjtitle{Journal of Web Semantics}
\bvolume{22}
(\byear{2013}),
\bfpage{19}--\blpage{39}.
\end{barticle}
\endbibitem

\bibitem{pedersen2004integrating}
\begin{bchapter}
\bauthor{\binits{D.}~\bsnm{Pedersen}},
\bauthor{\binits{J.}~\bsnm{Pedersen}} and
\bauthor{\binits{T.B.}~\bsnm{Pedersen}},
\bctitle{{Integrating XML Data in the TARGIT OLAP System}},
in: \bbtitle{Proceedings. 20th International Conference on Data Engineering},
\binstitute{IEEE},
\byear{2004},
pp.~\bfpage{778}--\blpage{781}.
\end{bchapter}
\endbibitem

\bibitem{yin2006evaluating}
\begin{barticle}
\bauthor{\binits{X.}~\bsnm{Yin}} and
\bauthor{\binits{T.B.}~\bsnm{Pedersen}},
\batitle{{Evaluating XML-Extended OLAP Queries Based on Physical Algebra}},
\bjtitle{Journal of Database Management (JDM)}
\bvolume{17}(\bissue{2})
(\byear{2006}),
\bfpage{85}--\blpage{116}.
\end{barticle}
\endbibitem

\bibitem{pedersen2002query}
\begin{bchapter}
\bauthor{\binits{D.}~\bsnm{Pedersen}},
\bauthor{\binits{K.}~\bsnm{Riis}} and
\bauthor{\binits{T.B.}~\bsnm{Pedersen}},
\bctitle{{Query Optimization for OLAP-XML Federations}},
in: \bbtitle{Proceedings of the 5th ACM international workshop on Data
  Warehousing and OLAP},
\byear{2002},
pp.~\bfpage{57}--\blpage{64}.
\end{bchapter}
\endbibitem

\bibitem{ahlstrom2016towards}
\begin{bchapter}
\bauthor{\binits{K.}~\bsnm{Ahlstr{\o}m}},
\bauthor{\binits{K.}~\bsnm{Hose}} and
\bauthor{\binits{T.B.}~\bsnm{Pedersen}},
\bctitle{{Towards Answering Provenance-Enabled SPARQL Queries Over RDF Data
  Cubes}},
in: \bbtitle{Joint International Semantic Technology Conference},
\binstitute{Springer},
\byear{2016},
pp.~\bfpage{186}--\blpage{203}.
\end{bchapter}
\endbibitem

\bibitem{gur2018foundation}
\begin{barticle}
\bauthor{\binits{N.}~\bsnm{G{\"u}r}},
\bauthor{\binits{T.B.}~\bsnm{Pedersen}},
\bauthor{\binits{E.}~\bsnm{Zim{\'a}nyi}} and
\bauthor{\binits{K.}~\bsnm{Hose}},
\batitle{{A Foundation for Spatial Data Warehouses on the Semantic web}},
\bjtitle{Semantic Web}
\bvolume{9}(\bissue{5})
(\byear{2018}),
\bfpage{557}--\blpage{587}.
\end{barticle}
\endbibitem

\bibitem{gur2017geosemolap}
\begin{bchapter}
\bauthor{\binits{N.}~\bsnm{G{\"u}r}},
\bauthor{\binits{J.}~\bsnm{Nielsen}},
\bauthor{\binits{K.}~\bsnm{Hose}} and
\bauthor{\binits{T.B.}~\bsnm{Pedersen}},
\bctitle{{GeoSemOLAP: Geospatial OLAP on the Semantic Web made easy}},
in: \bbtitle{Proceedings of the 26th International Conference on World Wide Web
  Companion},
\byear{2017},
pp.~\bfpage{213}--\blpage{217}.
\end{bchapter}
\endbibitem

\bibitem{perez2006semantics}
\begin{bchapter}
\bauthor{\binits{J.}~\bsnm{P{\'e}rez}},
\bauthor{\binits{M.}~\bsnm{Arenas}} and
\bauthor{\binits{C.}~\bsnm{Gutierrez}},
\bctitle{{Semantics and Complexity of SPARQL}},
in: \bbtitle{International semantic web conference},
Vol.~\bseriesno{4273},
\binstitute{Springer},
\byear{2006},
pp.~\bfpage{30}--\blpage{43}.
\end{bchapter}
\endbibitem

\bibitem{prud2006sparql}
\begin{botherref}
\oauthor{\binits{E.}~\bsnm{Prud}},
\oauthor{\binits{A.}~\bsnm{Seaborne}} \betal,
{SPARQL Query Language for RDF}
(2006).
\end{botherref}
\endbibitem

\bibitem{richardson2008restful}
\begin{bbook}
\bauthor{\binits{L.}~\bsnm{Richardson}} and
\bauthor{\binits{S.}~\bsnm{Ruby}},
\bbtitle{{RESTful Web Services}},
\bpublisher{" O'Reilly Media, Inc."},
\byear{2008}.
\end{bbook}
\endbibitem

\end{thebibliography}

\end{document}